%% file: B2G-19-002_temp.tex
\pdfoutput=1
\documentclass[11pt,twoside,a4paper,cmspaper,final,collab]{cms-tdr}
\def\svnVersion{df0b521}\def\svnDate{2021/12/14}\def\cmsCernNoTag{CERN-EP-2021-159}\def\cmsCernDate{\today}\def\cmsMessage{Published in Physical Review D as \href{http://dx.doi.org/10.1103/PhysRevD.105.032008}{\doi{10.1103/PhysRevD.105.032008}.}}
\begin{document}\cmsNoteHeader{B2G-19-002}

\ifthenelse{\boolean{cms@external}}{\providecommand{\cmsTable}[1]{#1}}{\providecommand{\cmsTable}[1]{\resizebox{\textwidth}{!}{#1}}}
\ifthenelse{\boolean{cms@external}}{\providecommand{\cmsLeft}{upper\xspace}}{\providecommand{\cmsLeft}{left\xspace}}
\ifthenelse{\boolean{cms@external}}{\providecommand{\cmsRight}{lower\xspace}}{\providecommand{\cmsRight}{right\xspace}}
\ifthenelse{\boolean{cms@external}}{\providecommand{\NA}{\ensuremath{\cdots}\xspace}}{\providecommand{\NA}{\ensuremath{\text{---}}\xspace}}
\ifthenelse{\boolean{cms@external}}{\providecommand{\CL}{C.L.\xspace}}{\providecommand{\CL}{CL\xspace}}
\ifthenelse{\boolean{cms@external}}{\providecommand{\CLnp}{C.L\xspace}}{\providecommand{\CLnp}{CL\xspace}}
\newcommand{\qqbarpr}{\ensuremath{\PQq\PAQq^{(\prime)}}\xspace}
\renewcommand{\bbbar}{\ensuremath{\PQb{}\PAQb}\xspace}
\newcommand{\nobbbar}{no-\bbbar}
\newcommand{\LUMISIXTEEN}{35.9\fbinv}
\newcommand{\LUMISEVENTEEN}{41.5\fbinv}
\newcommand{\LUMIEIGHTEEN}{59.7\fbinv}
\newcommand{\LUMIFULL}{137\fbinv}
\newcommand{\WW}{\ensuremath{\PW\PW}\xspace}
\newcommand{\ZZ}{\ensuremath{\PZ\PZ}\xspace}
\newcommand{\WZ}{\ensuremath{\PW\PZ}\xspace}
\newcommand{\WV}{\ensuremath{\PW\PV}\xspace}
\newcommand{\WH}{\ensuremath{\PW\PH}\xspace}
\newcommand{\tW}{\ensuremath{\PQt\PW}\xspace}
\newcommand{\GBulk}{\ensuremath{\PXXG_\text{bulk}}\xspace}
\newcommand{\ktilde}{\ensuremath{\tilde{k}}\xspace}
\newcommand{\MPl}{\ensuremath{\overline{M}_\text{Pl}}\xspace}

\newcommand{\LambdaR}{\ensuremath{\Lambda_\text{R}}\xspace}
\newcommand{\Wprpm}{\ensuremath{\PWpr^\pm}\xspace}
\newcommand{\gV}{\ensuremath{g_\text{V}}\xspace}
\newcommand{\cH}{\ensuremath{c_\text{H}}\xspace}
\newcommand{\cF}{\ensuremath{c_\text{F}}\xspace}
\newcommand{\XtoWW}{\ensuremath{\text{X}\to\WW}\xspace}
\newcommand{\GbuToWW}{\ensuremath{\GBulk\to\WW}\xspace}
\newcommand{\ZprToWW}{\ensuremath{\PZpr\to\WW}\xspace}
\newcommand{\WprToWZ}{\ensuremath{\PWpr\to\WZ}\xspace}
\newcommand{\WprToWH}{\ensuremath{\PWpr\to\WH}\xspace}
\newcommand{\Wtolnu}{\ensuremath{\PW\!\to\!\Pell\PGn}\xspace}
\newcommand{\Vtoqqbarpr}{\ensuremath{\PV\!\to\!\qqbarpr}\xspace}
\newcommand{\Wtoqqbarpr}{\ensuremath{\PW\!\to\!\PQq\PAQq^{\prime}}\xspace}
\newcommand{\Htobbbar}{\ensuremath{\PH\!\to\!\bbbar}\xspace}
\newcommand{\Ztobbbar}{\ensuremath{\PZ\!\to\!\bbbar}\xspace}
\newcommand{\WWtolnuqqbarpr}{\ensuremath{\WW\!\to\!\Pell\PGn\qqbar^{\prime}}\xspace}
\newcommand{\WZtolnuqqbar}{\ensuremath{\WZ\!\to\!\Pell\PGn\qqbar}\xspace}
\newcommand{\WHtolnubbbar}{\ensuremath{\WH\!\to\!\Pell\PGn\bbbar}\xspace}
\newcommand{\Wlep}{\ensuremath{\PW_\text{lep}}\xspace}
\newcommand{\Vhad}{\ensuremath{\PV_\text{had}}\xspace}
\newcommand{\mWV}{\ensuremath{m_{\WV}}\xspace}
\newcommand{\mWVpart}{\ensuremath{\mWV^\text{part.}}\xspace}
\newcommand{\mjet}{\ensuremath{m_\text{jet}}\xspace}
\newcommand{\genjetpt}{\ensuremath{p_\text{T,jet}^\text{gen.}}\xspace}
\newcommand{\mX}{\ensuremath{m_\text{X}}\xspace}
\newcommand{\GbuToWWexcl}{1.8\TeV}
\newcommand{\RadToWWexcl}{3.1\TeV}
\newcommand{\ZprToWWexcl}{4.0\TeV}
\newcommand{\WprToWZexcl}{3.9\TeV}
\newcommand{\WprToWHexcl}{4.0\TeV}
\newcommand{\Wplusjets}{\ensuremath{\PW+\text{jets}}\xspace}
\newcommand{\Wtolnuplusjets}{\ensuremath{\PW(\to\!\Pell\PGn)+\text{jets}}\xspace}
\newcommand{\nonRes}{\ensuremath{\PW\!+\!\text{jets}}\xspace}
\newcommand{\res}{\ensuremath{\PW\!+\!\PV\!/\PQt}\xspace}
\newcommand{\nsubj}{\ensuremath{\tau_{21}}\xspace}
\newcommand{\nsubjDDT}{\ensuremath{\tau_{21}^\text{DDT}}\xspace}
\newcommand{\doubleB}{double-\PQb\xspace}
\newcommand{\DoubleB}{Double-\PQb\xspace}
\newcommand{\Dphi}{\ensuremath{\abs{\Delta\phi}}\xspace}
\newcommand{\Dy}{\ensuremath{\abs{\Delta y}}\xspace}
\newcommand{\DetaVBF}{\ensuremath{\abs{\Delta\eta_\text{jj}^\text{VBF}}}\xspace}
\newcommand{\mjjVBF}{\ensuremath{m_\text{jj}^\text{VBF}}\xspace}
\newlength\cmsTabSkip\setlength{\cmsTabSkip}{1ex}

\cmsNoteHeader{B2G-19-002} 
\title{Search for heavy resonances decaying to \texorpdfstring{\WW}{WW}, \texorpdfstring{\WZ}{WZ}, or \texorpdfstring{\WH}{WH} boson pairs in a final state consisting of a lepton and a large-radius jet in proton-proton collisions at \texorpdfstring{$\sqrt{s} = 13\TeV$}{sqrt(s) = 13 TeV}}

\date{\today}

\abstract{A search for new heavy resonances decaying to pairs of bosons (\WW, \WZ, or \WH) is presented. The analysis uses data from proton-proton collisions collected with the CMS detector at a center-of-mass energy of 13\TeV, corresponding to an integrated luminosity of 137\fbinv. One of the bosons is required to be a W boson decaying to an electron or muon and a neutrino, while the other boson is required to be reconstructed as a single jet with mass and substructure compatible with a quark pair from a \PW, \PZ, or Higgs  boson decay. The search is performed in the resonance mass range between 1.0 and 4.5\TeV and includes a specific search for resonances produced via vector boson fusion. The signal is extracted using a two-dimensional maximum likelihood fit to the jet mass and the diboson invariant mass distributions. No significant excess is observed above the estimated background. Model-independent upper limits on the production cross sections of spin-0, spin-1, and spin-2 heavy resonances are derived as functions of the resonance mass and are interpreted in the context of bulk radion, heavy vector triplet, and bulk graviton models. The reported bounds are the most stringent to date.}

\hypersetup{
pdfauthor={CMS Collaboration},
pdftitle={Search for heavy resonances decaying to WW, WZ, or WH boson pairs in a final state consisting of a lepton and a large-radius jet in proton collisions at sqrt(s) = 13 TeV},
pdfsubject={CMS},
pdfkeywords={CMS,  bulk graviton, bulk radion, diboson, resonances, HVT}}

\maketitle

\section{Introduction}
\label{sec:Introduction}

The standard model (SM) of particle physics~\cite{Glashow:1961tr,Salam:1964ry,Weinberg:1967tq} has successfully accommodated a multitude of experimental observations, culminating in the discovery of a Higgs boson (\PH)~\cite{Aad:2012tfa,Chatrchyan:2012xdj,Chatrchyan:2013lba}.
Yet, the SM falls short of addressing several outstanding issues, such as the hierarchy problem, 
{\ie}, explaining the large difference between the Higgs boson mass and the largest scale in the SM, that are necessary components of a consistent theory of nature up to the Planck scale. 
These shortcomings are addressed by a variety of theoretical extensions to the SM, several of which predict the existence of new heavy particles with masses near the \TeVns scale that couple to \PW, \PZ, or Higgs bosons and could be produced in proton-proton ($\Pp\Pp$) collisions at the CERN Large Hadron Collider (LHC).
Models studied in the relevant literature include the bulk scenario of the Randall--Sundrum (RS) model with warped extra dimensions~\cite{Randall:1999ee,Randall:1999vf} and examples of the heavy vector triplet (HVT) framework~\cite{Pappadopulo:2014qza}, which generically represents a number of models that predict additional gauge bosons, such as composite Higgs~\cite{Bellazzini:2014yua,Contino:2011np,Marzocca:2012zn,Greco:2014aza,Lane:2016kvg} and little Higgs~\cite{Schmaltz:2005ky,ArkaniHamed:2002qy} models.

In this paper, a search is presented for a heavy resonance \PX with mass between 1.0 and 4.5\TeV decaying to a pair of bosons, using $\Pp\Pp$ collision data at a center of mass energy of 13\TeV, collected with the CMS detector from 2016 to 2018. The final state considered targets the scenario where one of the two bosons is required to be a \PW boson decaying to an electron or muon and a neutrino, while the other boson is detected as a large-radius jet formed from the merged hadronic decay products of the boson, either a quark pair (\qqbarpr) from a \PW or \PZ boson (collectively referred to as \PV) or a bottom quark pair (\bbbar) from a Higgs boson.

\begin{figure*}[!htb]
        \centering
        \includegraphics[width=0.32\linewidth]{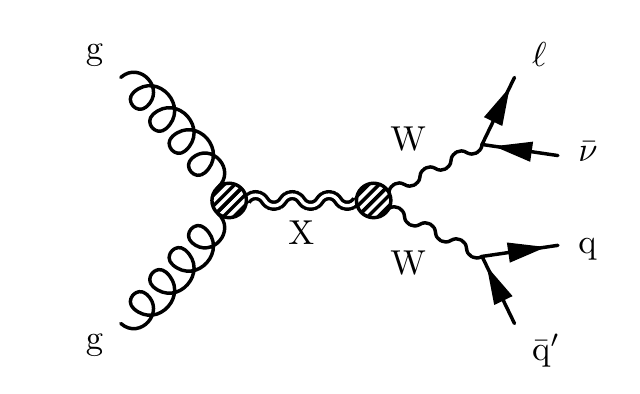}
        \includegraphics[width=0.32\linewidth]{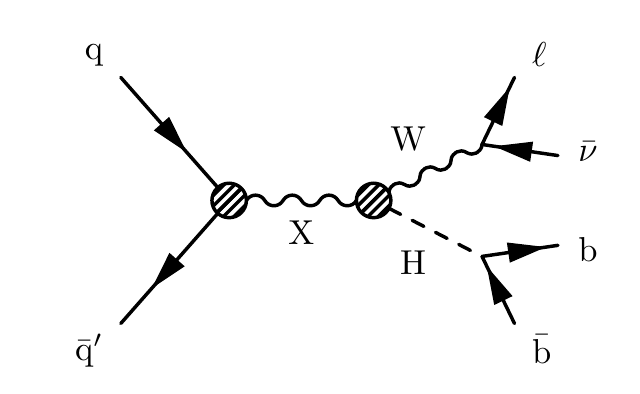}
        \includegraphics[width=0.32\linewidth]{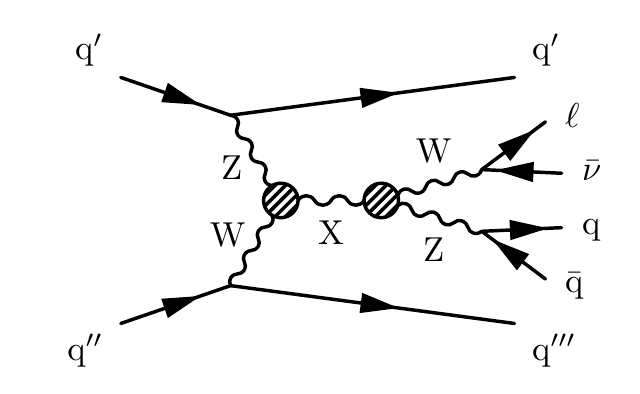}
        \caption{Feynman diagrams for three of the processes studied in this paper: (left) ggF-produced, spin-2 resonance decaying to \WWtolnuqqbarpr; (center) DY-like, charged spin-1 resonance decaying to \WHtolnubbbar; (right) VBF-produced, charged spin-1 resonance decaying to \WZtolnuqqbar. }
        \label{fig:FeynDiag}
\end{figure*}

A boosted \PV or Higgs boson with transverse momentum  $\pt \approx 250\GeV$ and mass $m \approx 100\GeV$ decaying to quarks is expected to have its decay products within a cone defined by an angular separation of  $\DR=\sqrt{\smash[b]{(\Delta\eta)^2+(\Delta\phi)^2}} \approx 2 m/\pt \approx 0.8$, where $\eta$ is the pseudorapidity and $\phi$ is the azimuthal angle. Therefore, the lower bound of 1 \TeV for the resonance mass is appropriate for the requirement that the hadronically decaying boson appears as a single broad massive jet.   
The search targets resonance production via gluon-gluon fusion (ggF) and Drell--Yan-like quark-antiquark annihilation (DY) processes, where no other decay products are expected, as well as production via vector boson fusion (VBF), where the final state contains two additional quark-induced jets in the forward and backward regions of the detector.
Example Feynman diagrams for three representative combinations of production mechanisms and final states studied in this paper are shown in Fig.~\ref{fig:FeynDiag}.

Previous searches for heavy \WW and \WZ resonances in semileptonic final states by the ATLAS~\cite{Aad:2015ufa,Aaboud:2016okv,Aaboud:2017fgj} and CMS~\cite{Khachatryan:2014gha,Sirunyan:2016cao,Sirunyan:2018iff} Collaborations using LHC data collected in 2012, 2015, and 2016, as well as the recent ATLAS search with the complete 2015--2018 data set~\cite{ATLAS:2020fry}, have not observed any statistically significant deviations from the SM background expectation.
In parallel, searches for semileptonically decaying \WH resonances have been reported by ATLAS~\cite{Aad:2015yza,Aaboud:2016lwx,Aaboud:2017cxo} and CMS~\cite{Khachatryan:2016yji,Khachatryan:2016cfx,Sirunyan:2018qob} in a separate series of publications and have yielded similar outcomes.

The analysis is performed by initially selecting events with a well reconstructed \PW boson that decays to an electron or muon and a neutrino, and a large radius jet. The preselected events are then split into 24 categories based on lepton flavor, compatibility of the large radius jet substructure with a vector boson or a Higgs boson, VBF tagging, and compatibility of the event kinematic variables with the spin of the hypothetical resonance. Sensitivity to the spin of the new resonance is introduced by categorizing the events based on the rapidity separation between the \PW boson and the large-radius jet. The distribution of the rapidity separation is different for signals of different spins produced with different production mechanisms and for background events, improving further the sensitivity of the search.

A two-dimensional (2D) fit is then applied in the 24 categories to extract the signal production cross section. The fit is performed  in the plane whose coordinates are defined by the invariant mass of the reconstructed diboson system and by the mass of its \Vtoqqbarpr or \Htobbbar jet component. The fit includes one signal template and two background classes based on the compatibility of the jet mass with a vector boson or with a quark- or gluon-initiated jet. Systematic uncertainties are encoded as nuisance parameters in the fit and are measured in disjoint regions of phase space (control samples). Systematic uncertainties affecting background shapes and yields are constrained further using the data during the likelihood minimization process. Systematic uncertainties affecting the signal are measured precisely in control regions using events with similar particle content and kinematic configuration.

The 2D fit strategy improves the search sensitivity by constraining the backgrounds in regions in the 2D space that are dominated by one of the two background classes.
In addition, the choice of the signal shape in both jet mass and resonance mass in the 2D likelihood enables a simultaneous search for \WW, \WZ, and \WH resonances using the same background estimation procedure for all final states and production mechanisms.

Compared to the previous CMS search for semileptonic \WW and \WZ resonances with 2016 data~\cite{Sirunyan:2018iff}, this analysis still employs the 2D fit, but extends the search sensitivity to \WH decays and VBF production modes by employing \bbbar tagging and VBF tagging. In addition, sensitivity to the spin of the resonance through rapidity separation is introduced for the first time, boosting the reach of the search beyond the improvement expected with the larger data sample.

The paper is organized as follows:
Section~\ref{sec:Detector} describes the CMS detector and the event reconstruction, while Section~\ref{sec:Samples} provides information on the simulation and data samples used. Section~\ref{sec:EventSel} describes the event selection and the categorization of the data in different classes. Section~\ref{sec:Templates} provides details on the 2D signal extraction and Section~\ref{sec:SystUnc} describes the systematic uncertainties. Section \ref{sec:Results} presents the results of the search. Finally, the analysis is summarized in Section~\ref{sec:Summary}.

\section{The CMS detector and event reconstruction}
\label{sec:Detector}
\label{sec:EventReco}

The central feature of the CMS apparatus is a superconducting solenoid of an internal diameter of 6\unit{m} that provides a magnetic field of 3.8\unit{T}. 
Within the solenoid volume are a silicon pixel and strip tracker, a lead tungstate crystal electromagnetic calorimeter (ECAL), and a brass and scintillator hadron calorimeter (HCAL), each composed of a barrel and two endcap sections. 
Forward calorimeters extend the pseudorapidity coverage provided by the barrel and endcap detectors. 
Muons are detected in gas-ionization chambers embedded in the steel flux-return yoke outside the solenoid.  
A more detailed description of the CMS detector, together with a definition of the coordinate system used and the relevant kinematic variables, can be found in Ref.~\cite{Chatrchyan:2008zzk}. 

Event reconstruction relies on the particle-flow (PF) algorithm~\cite{CMS-PRF-14-001}, which aims to identify each individual particle with an optimized combination of information from the various elements of the CMS detector. 
The energy of photons is obtained from the ECAL measurement. 
The energy of electrons is determined from a combination of the electron momentum at the primary interaction vertex as determined by the tracker, the energy of the corresponding ECAL cluster, and the energy sum of all bremsstrahlung photons spatially compatible with originating from the electron track. 
The momentum of muons is obtained from the combined curvature of the corresponding track in both silicon tracker and the muon system. 
The energy of charged hadrons is determined from a combination of their momentum measured in the tracker and the matching ECAL and HCAL energy deposits, corrected for the response function of the calorimeters to hadronic showers. 
Finally, the energy of neutral hadrons is obtained from the corresponding corrected ECAL and HCAL energies.
The missing transverse momentum vector \ptvecmiss is computed as the negative vector sum of the transverse momenta of all the PF candidates in an event, and its magnitude is denoted as \ptmiss~\cite{Sirunyan:2019kia}.

For each event, hadronic jets are clustered from these reconstructed particles using the infrared- and collinear-safe anti-\kt algorithm~\cite{Cacciari:2008gp, Cacciari:2011ma}. 
The jet momentum is determined as the vector sum of all particle momenta in the jet and is found from simulation to be, on average, within 5 to 10\% of the true momentum over the entire \pt spectrum and detector acceptance.
Jet energy corrections are derived from simulation studies so that the average measured response of jets becomes identical to that of particle-level jets~\cite{Khachatryan:2016kdb}.

Additional pp interactions within the same or nearby bunch crossings (pileup) can contribute additional tracks and calorimetric energy depositions, increasing the apparent jet momentum. 
To mitigate this effect, tracks identified to be originating from pileup vertices are discarded and an offset correction is applied to correct for remaining contributions~\cite{CMS-PRF-14-001, Cacciari:2011ma}. 
In the computation of jet substructure variables, a different Pileup-Per Particle Identification (PUPPI) algorithm~\cite{Sirunyan:2020foa,Bertolini:2014bba}, which uses local shape information of charged pileup to rescale the momentum of each particle based on its compatibility with the primary interaction vertex, is employed.

Events of interest are initially selected using a two-tiered trigger system. 
The first level, composed of custom hardware processors, uses information from the calorimeters and muon detectors to select events at a rate of around 100\unit{kHz} within a fixed latency of about 4\mus~\cite{Sirunyan:2020zal}. 
The second level, known as the high-level trigger (HLT), consists of a farm of processors running a version of the full event reconstruction software optimized for fast processing and reduces the event rate to around 1\unit{kHz} before data storage~\cite{Khachatryan:2016bia}.

\section{Data and simulated samples}
\label{sec:Samples}

This search uses data samples of pp collisions collected by the CMS experiment at the LHC at a center-of-mass energy of 13\TeV in 2016, 2017, and 2018.
The performance of the detector on the variables of interest was very similar in the different periods of data taking, therefore they are treated as one single data set, with a total integrated luminosity of \LUMIFULL.
Collision events are selected mainly by HLT algorithms that either require the reconstruction of an electron within $\abs{\eta}<2.5$ or a muon within $\abs{\eta}<2.4$. 

Several electron triggers are combined. In 2016,  \pt thresholds of 27, 55, and 115\GeV are used in association with tight, loose, or no isolation criteria, respectively, while in 2017 and 2018, \pt thresholds of 32 and 115\GeV are used with tight or no isolation criteria, respectively.
Muon triggers have a \pt threshold of 50\GeV.
To further increase the trigger efficiency, another algorithm selects events with $\ptmiss>120\GeV$, exploiting the presence of the high-\pt neutrino in the \Wtolnu decay.
The overall HLT efficiency is larger than 99.7\% for signal events passing the offline selection described in Section~\ref{sec:EventSel}.

Several signal benchmark scenarios are used to interpret the results of the search, focusing on relevant models probed in earlier searches by the ATLAS and CMS Collaborations.
Spin-0 radions~\cite{Goldberger:1999uk,Csaki:1999mp,Csaki:2000zn} and spin-2 gravitons~\cite{Agashe:2007zd, Fitzpatrick:2007qr, Antipin:2007pi} decaying to \WW are generated for the bulk scenario of the RS model of warped extra dimensions~\cite{Randall:1999ee,Randall:1999vf}.
For bulk gravitons, denoted as \GBulk, the ratio \ktilde of the unknown curvature scale of the extra dimension $k$ and the reduced Planck mass \MPl is set to $\ktilde=0.5$, which ensures that the natural width of the graviton is negligible with respect to the experimental resolution~\cite{Oliveira:2014kla}.
For bulk radions, we consider a scenario with $k r_\text{c} \pi = 35$ and $\LambdaR=3\TeV$, where $r_\text{c}$ is the compactification radius and \LambdaR is the ultraviolet cutoff of the theory~\cite{Oliveira:2014kla}.
Spin-1 resonances decaying to \WW, \WZ, or \WH are studied within the HVT framework using benchmark models from Ref.~\cite{Pappadopulo:2014qza}: model B for DY production and model C for VBF.
The HVT framework introduces a triplet of heavy vector bosons with similar masses, of which one is neutral (\PZpr) and two are electrically charged (\Wprpm).
HVT benchmark models are expressed in terms of a few parameters: the strength \cF of the couplings to fermions, the strength \cH of the couplings to the Higgs boson and to longitudinally polarized SM vector bosons, and the interaction strength \gV of the new vector boson. 
In HVT model B ($\gV=3$, $\cH=-0.98$, $\cF=1.02$)~\cite{Pappadopulo:2014qza}, the new resonances are narrow and have large branching fractions to vector boson pairs and suppressed couplings to fermions. 
In model C ($\gV \approx 1$, $\cH \approx 1$, $\cF = 0$), the fermionic couplings are zero, and the resonances are produced only through VBF and decay exclusively to pairs of SM bosons. 
Monte Carlo (MC) simulated samples for bulk radions, bulk gravitons, and resonances of the HVT models are generated at leading order (LO) in quantum chromodynamics (QCD) with \MGvATNLO versions 2.2.2 and 2.4.2~\cite{Alwall:2014hca}. 
For each model, resonance masses in the range 1.0--4.5\TeV are considered, and the resonance width is set to 0.1\% of the resonance mass, ensuring that the width of the signal distribution is dominated by the detector resolution.

Simulated samples for SM background processes are used to optimize the search and to build background templates, as described in Section~\ref{sec:BkgdTemplates}.
The \Wtolnuplusjets process is produced with \MGvATNLO at LO in QCD.
The background from top quark pair events (\ttbar) is generated with \POWHEG v2~\cite{Nason:2004rx,Frixione:2007vw,Alioli:2010xd,Alioli:2011as} at next-to-LO (NLO). 
Single top quark events are generated in the \PQt channel and associated \tW channel at NLO with \POWHEG v2~\cite{Alioli:2009je,Re:2010bp}, while SM diboson processes are generated at NLO with \MGvATNLO using the FxFx merging scheme~\cite{Frederix:2012ps} for \WZ and \ZZ, and with \POWHEG v2 for \WW~\cite{Nason:2013ydw}. 

Parton showering and hadronization are performed with \PYTHIA 8.205 (8.230)~\cite{Sjostrand:2014zea} for 2016 (2017 and 2018) detector conditions. 
The NNPDF 3.0~\cite{Ball:2014uwa} parton distribution functions (PDFs) are used together with the CUETP8M1~\cite{Khachatryan:2015pea} underlying event tune for 2016 conditions (except for \ttbar samples, which use CUETP8M2~\cite{CMS-PAS-TOP-16-021}), while the NNPDF 3.1~\cite{Ball:2017nwa} PDFs and the CP5~\cite{Sirunyan:2019dfx} tune are used for 2017 and 2018 conditions.
To simulate the effect of pileup, 
additional minimum bias interactions are superimposed on the hard-scattering process, and the events are then weighted to match the distributions of the number of pileup interactions observed in 2016, 2017, and 2018 data separately.
All samples are processed through a simulation of the CMS detector based on \GEANTfour~\cite{Agostinelli:2002hh} and are reconstructed using the same algorithms used for collision data.
Simulated events are also reweighted to correct for differences between data and simulation in the efficiencies of the trigger, lepton identification, and \PQb tagging algorithms described in Section~\ref{sec:EventSel}.
 
\section{Event selection and categorization}
\label{sec:EventSel}
\label{sec:EventCateg}

The event selection is designed to isolate events containing a boosted topology consistent with the semileptonic decay of a \WW, \WZ, or \WH pair, involving one energetic electron or muon, large \ptmiss, and a so-called large-radius jet corresponding to a \PW, \PZ, or Higgs boson candidate. 
\begin{figure*}[!htb]
  \centering
  \includegraphics[width=0.4\textwidth]{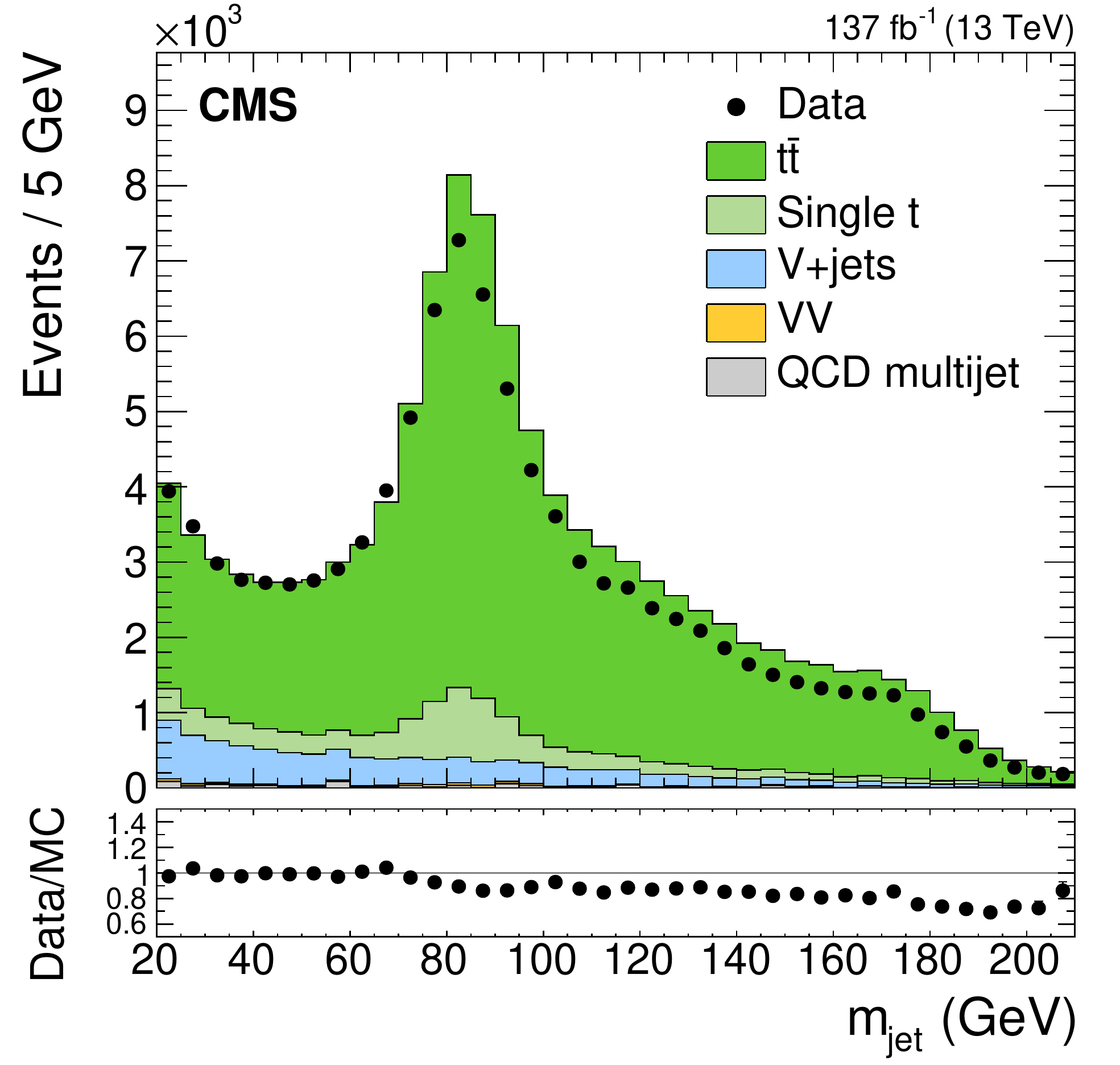}
  \includegraphics[width=0.4\textwidth]{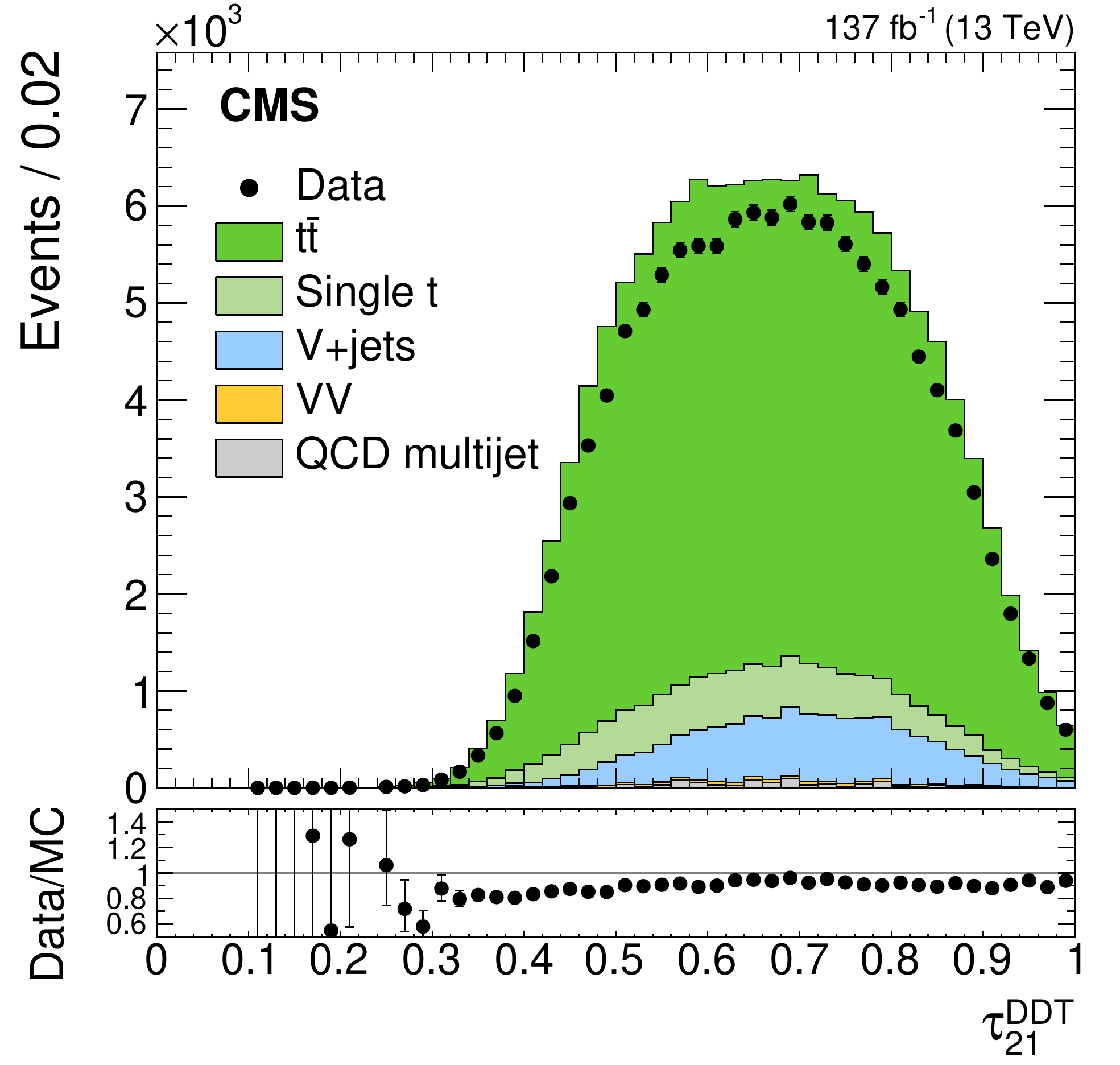}
  \includegraphics[width=0.4\textwidth]{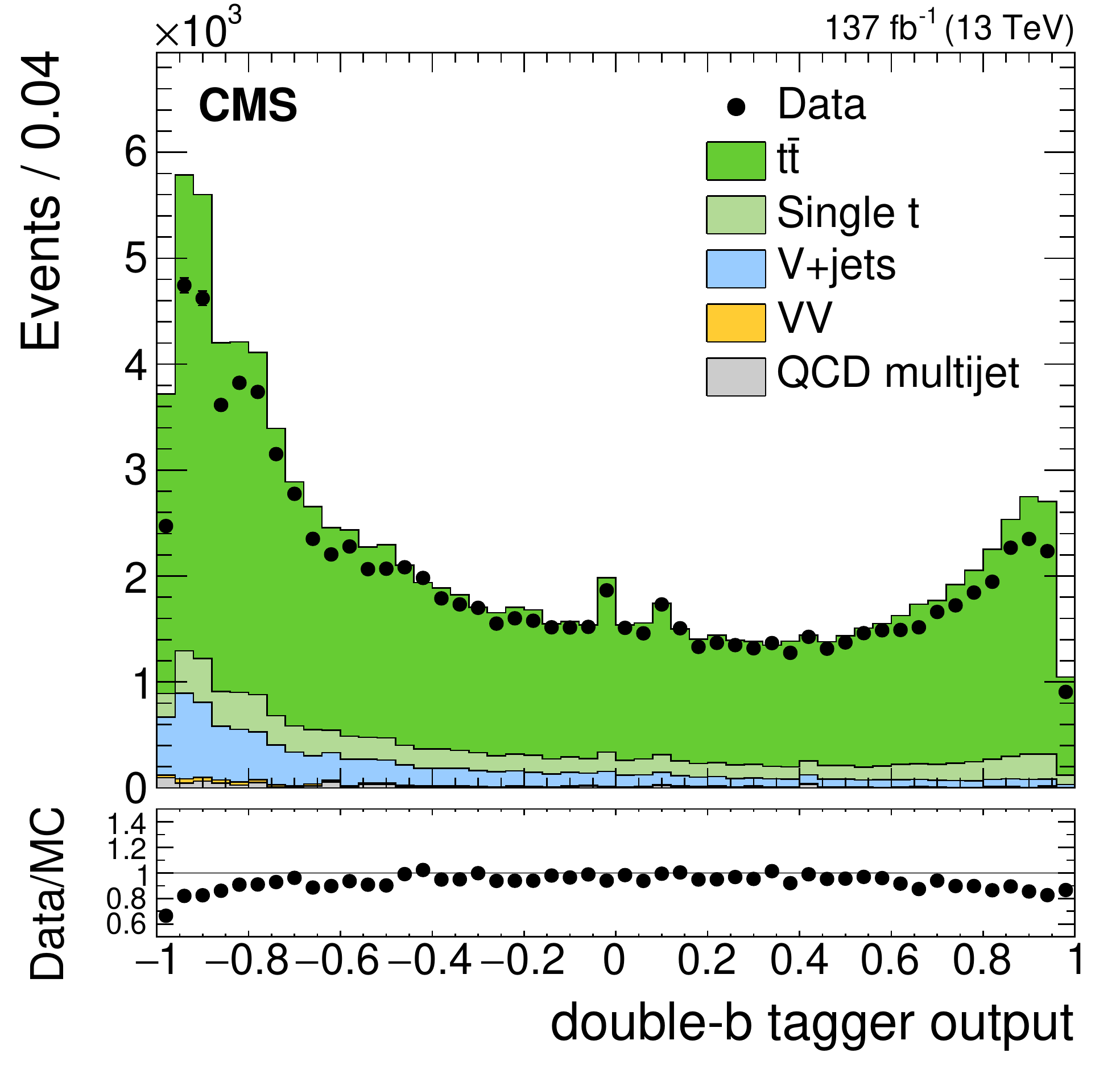}
  \includegraphics[width=0.4\textwidth]{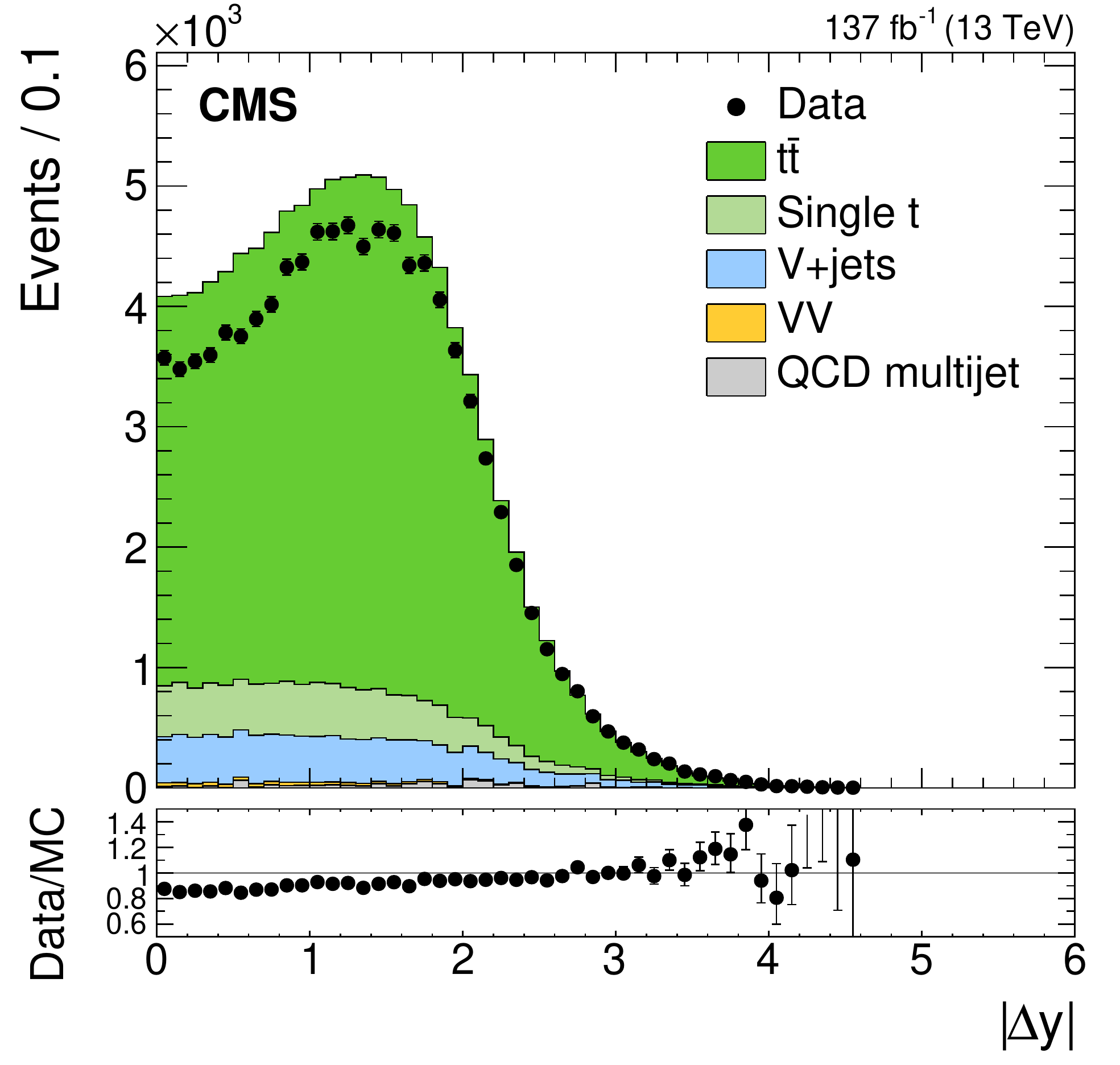}
  \caption{Uncorrected distributions of the soft-drop jet mass \mjet (upper left), mass-decorrelated $N$-subjettiness ratio \nsubjDDT (upper right), \doubleB tagger output (lower left), and difference in rapidity \Dy between the reconstructed bosons (lower right), for data and simulated events in the top quark enriched control region. 
  The lower panels show the ratio of the data to the simulation.
    No event categorization is applied. The events with $\nsubjDDT > 0.80$ are not shown in any distribution other than \nsubjDDT itself, since they are not part of the signal region. The vertical bars correspond to the statistical uncertainties of the data. }
  \label{fig:controlplots}
\end{figure*}

Electron and muon candidates are considered if they satisfy $\pt>55\GeV$, the same $\eta$ acceptance requirements as at the HLT, and a set of lepton reconstruction quality and lepton identification requirements optimized to maintain a large efficiency for high-momentum leptons~\cite{EGM-17-001,MUO-17-001}.
The electrons must satisfy requirements on the ratio of the energies deposited in the HCAL and ECAL, the distribution of the ECAL deposits, their geometrical matching with reconstructed tracks, and the number of reconstructed hits in the silicon tracker. For muons, a track is required that is reconstructed in both the silicon tracker and the muon system. Identification criteria are imposed on the track quality and the number of matched muon hits.  
To reject backgrounds from bottom and charm decays, decays in flight, and misidentified leptons inside jets, muons and electrons are required to be isolated in the detector in a region defined by a cone of $\DR < 0.3$ around the respective lepton.
The muon isolation variable, defined as the \pt sum of all particles within $\DR = 0.3$ of the muon direction, subtracting the muon itself and pileup contributions, is required to be less than 5\% of the muon \pt, 
while electron isolation is defined by applying separate requirements suitable for high energy electrons using the tracker and the calorimeter deposits~\cite{EGM-17-001}.
Additional requirements on the impact parameters of electron and muon tracks with respect to the primary interaction vertex are applied to suppress the contributions from secondary decays and pileup interactions. 
The lepton selection requirements used establish an identification efficiency of about 90\% while ensuring negligible contributions from SM events composed uniquely of jets produced through the strong interaction, referred to as QCD multijet events.

Large-radius jets are clustered using a distance parameter $R=0.8$ in the anti-\kt algorithm and are required to have $\pt>200\GeV$.
Since the signal is expected to be produced centrally, large-radius jets are required to be in the tracker acceptance ($\abs{\eta}<2.5$) so as to exploit the maximum granularity of the CMS detector. 
Another collection of jets is clustered using $R=0.4$ and referred to as standard jets, with $\pt>30\GeV$ and $\abs{\eta}<4.7$.
Both sets of jets are required to pass tight identification requirements~\cite{CMS-PAS-JME-16-003} to remove jets originating from calorimetric noise and track misreconstruction in the silicon tracker.
Large-radius jets located within $\DR<1.0$ of a selected lepton are discarded, as are standard jets located within $\DR<0.8$ of a large-radius jet or within $\DR<0.4$ of a selected lepton.

To identify large-radius jets as hadronic decays of boosted \PW, \PZ, and Higgs bosons, jet substructure techniques are employed.
First, to perform jet grooming, \ie to remove soft, wide-angle radiation from the jet, the modified mass-drop algorithm~\cite{Dasgupta:2013ihk,Butterworth:2008iy} known as ``soft drop'' is used with angular exponent $\beta=0$, soft cutoff threshold $z_\text{cut} = 0.1$, and characteristic radius $R_0=0.8$~\cite{Larkoski:2014wba}.
The invariant mass of the remaining jet constituents is called the soft-drop jet mass, denoted as \mjet, and is one of the two observables of the final 2D fit.
Second, a \PV tagging algorithm is defined based on the $N$-subjettiness ratio between 2-subjettiness and 1-subjettiness~\cite{Thaler:2010tr}, $\nsubj = \tau_{2}/\tau_{1}$, which takes lower values for jets coming from two-prong \PW, \PZ, or \PH decays than for one-prong jets from dominant SM backgrounds.
However, the selection on \nsubj is found to sculpt the distribution of \mjet, affecting the monotonically falling behavior of the \Wplusjets background distributions.
Therefore, to decorrelate \nsubj from the jet mass, the ``designing decorrelated taggers'' (DDT) procedure~\cite{ddt} is followed,
leading to the definition of the mass-decorrelated $N$-subjettiness ratio $\nsubjDDT \equiv \nsubj - M \log{(\mjet^2 /(\pt \mu))}$, with $M=-0.08$ and $\mu=1\GeV$ and using the \pt of the original jet.
In the computation of both \mjet and \nsubjDDT, pileup mitigation relies on the PUPPI algorithm.
Third, to discriminate \Htobbbar and \Ztobbbar decays from other signals and backgrounds that involve light-flavor jets, the so-called \doubleB tagger is used~\cite{BTV-16-002}, which is a multivariate discriminant that combines information from displaced tracks, secondary vertices (SV), and the two-SV system within the Higgs or \PZ boson jet candidate.

The collection of standard jets serves two purposes.
First, the b jet identification algorithm known as DeepCSV~\cite{BTV-16-002}, which relies on a deep neural network that uses track and SV information is applied to standard jets found within $\abs{\eta}<2.5$.
The medium operating point of this algorithm has an efficiency of about 68\% for correctly identifying b jets in simulated \ttbar events, and a misidentification probability of about 1\% for light-flavor jets. 
Events where at least one standard jet passes this operating point are discarded from the signal region, thereby reducing the background from events involving top quark decays, and are used to define a top quark enriched control region.
Second, events containing at least two standard jets are used to make the search sensitive to VBF-produced resonances.
A VBF tagging criterion is defined as $\mjjVBF>500\GeV$ and $\DetaVBF>4$, where \mjjVBF and \DetaVBF are the invariant mass and pseudorapidity separation of the two highest \pt standard jets. The selection requirements were chosen to reject backgrounds from additional jets present in energetic top events and \PW production in association with multiple jets. This criterion is used to define additional categories.

\begin{table*}[tbp]
  \centering
  \topcaption{Summary of the categorization scheme in the analysis. The 24 analysis categories are defined by all possible combinations of the criteria defined in each column.}
  \begin{scotch}{llcc}
   {lepton}  & {purity}  & {\bbbar/VBF tagging} & {spin}\\  \hline\\[-1.8ex]
    muon          & HP: $\nsubjDDT \leq 0.5$      & VBF: $\mjjVBF>500\GeV$ and $\DetaVBF>4$  & LDy: $\Dy \leq 1$ \\ 
    electron      & LP: $0.5 <\nsubjDDT \leq 0.8$ & \nobbbar: no VBF and  $\text{\doubleB tagger} \leq 0.8$ & HDy: $\Dy > 1$ \\ 
                 &                               & \bbbar: no VBF and   $\text{\doubleB tagger} > 0.8$ & \\
  \end{scotch}
  \label{tab:categorization}
\end{table*}

The event selection requires the presence of exactly one identified electron or muon, and events that contain additional electrons (muons) passing $\pt>35\,(20)\GeV$ and otherwise identical requirements are discarded.
The \ptmiss in the event is required to be greater than 80 \GeV if the selected lepton is an electron and greater than 40 \GeV if the selected lepton is a muon. Muons have lower QCD background and are not included in the \ptmiss calculation in the trigger, resulting in the reconstruction of the whole boosted leptonic \PW decay as \ptmiss, achieving higher efficiency at lower offline thresholds.
To reconstruct a \Wtolnu boson candidate, the \ptvecmiss is taken as an estimate of the \ptvec of the neutrino, and the longitudinal component $p_z$ of the neutrino momentum is estimated by imposing a \PW boson mass constraint to the lepton+neutrino system.
This leads to a quadratic equation, of which the solution with smallest magnitude of the neutrino $p_z$ is chosen.
When no real solution is found, only the real part of the two complex solutions is considered.
Besides this leptonically decaying \PW boson candidate, hereafter referred to as \Wlep, the hadronically decaying \PW, \PZ, or Higgs boson candidate is defined as the most energetic large-radius jet in the event and referred to as \Vhad.
The \Vhad is required to pass $\nsubjDDT \leq 0.8$, and the \Wlep and \Vhad are both required to have $\pt>200\GeV$.
They are then combined to form a \WW, \WZ, or \WH diboson candidate, whose invariant mass is denoted as \mWV and is the second observable used in the 2D fit.

Angular criteria are applied in order to select a diboson-like topology:
the angular distance between the selected lepton and the \Vhad is required to be $\DR>\pi/2$,
while the difference in azimuthal angle between the \Vhad and both the \ptvecmiss and the \Wlep directions is required to be $\Dphi > 2$.
The difference in rapidity between the \Wlep and the \Vhad is denoted as \Dy and is used later for event categorization to exploit the fact that signal models investigated in this search tend to have lower values of \Dy compared to backgrounds, except for spin-1 and spin-2 VBF-produced resonances, which significantly populate the $\Dy>1$ region.

The signal region for the 2D fit is defined by two final requirements on the diboson reconstructed mass and soft-drop jet mass, namely $0.7<\mWV<6\TeV$ and $20<\mjet<210\GeV$.
The lower bound on \mWV ensures that the backgrounds have a falling spectrum while allowing a search for resonances with masses greater than $1\TeV$, and the $6\TeV$ upper bound ensures that all observed events are included.
The use of a large window for \mjet allows the selection of background events containing $\PV$ jets as well as top quark jet candidates, while retaining sizeable low- and high-mass sidebands to constrain shapes and normalizations.
The overall signal selection efficiency times acceptance ranges from 22 to 79\%, depending on the benchmark model and increasing with resonance mass.

To enhance the analysis sensitivity to all signals under consideration, each event of the signal region is eventually assigned to one of 24 mutually exclusive search categories, based on a combination of four criteria.
First, the event sample is split according to lepton flavor, distinguishing the electron and muons channels, which helps account for the differences in lepton reconstruction and selection. 
Second, the \PV tagging information is used to split each channel into a high-purity (HP) and a low-purity (LP) subchannel, which correspond to values of the mass-decorrelated $N$-subjettiness ratio in the ranges $\nsubjDDT \leq 0.5$ and $0.5 < \nsubjDDT \leq 0.8$, respectively. 
Third, each subchannel is further divided into three regions, referred to as VBF-tagged for events that satisfy the aforementioned VBF tagging criterion, \doubleB-tagged (\bbbar) for non-VBF-tagged events for which the \doubleB tagger output is larger than 0.8, and non-\doubleB-tagged (\nobbbar) for the remaining events.
Fourth, each of the twelve resulting regions is split into two event categories: LDy, corresponding to a difference in rapidity between the reconstructed bosons of $\Dy \leq 1$; and HDy, which corresponds to $\Dy > 1$. 
The categorization requirements are summarized in Table~\ref{tab:categorization}.

Besides the signal region, a disjoint event sample enriched in \ttbar events with similar kinematic distributions is defined by requiring the presence of a b-tagged standard jet instead of vetoing it. 
This control region is used to compare data and simulation for the main selection variables, to correct the top background yields and \mjet shapes in the signal region, and to compute efficiency scale factors for the \nsubjDDT selection. 
Figure~\ref{fig:controlplots} shows the distributions of \mjet, \nsubjDDT, the \doubleB tagger, and \Dy in this top quark-enriched sample before the aforementioned corrections and scale factors are applied.

\section{Two-dimensional templates}
\label{sec:Templates}

A similar signal extraction strategy is followed as in the previous CMS search for semileptonic \WV resonances with 2016 data~\cite{Sirunyan:2018iff}, using a simultaneous maximum likelihood fit to the (\mWV, \mjet) data distributions in the 24 search categories.
Signal and background templates are constructed using simulated events, after applying corrections to the simulation.
Analytical shapes are used to model the signal, while binned templates are used for background processes. The binning was optimized to maximize the number of bins while ensuring smooth templates in all categories. This process resulted in two binning schemes, one for higher statistics and one for lower statistics categories, respectively. 
Particular care is devoted to constructing smooth background templates, modifying the strategy to accommodate the larger 2D signal region and the fact that new categorization criteria such as VBF tagging and \doubleB tagging cause some categories to be sparsely populated by simulated events.

\subsection{Signal modeling}
\label{sec:SignalTemplates}

The 2D probability density function (pdf) for signal events in the (\mWV, \mjet) plane is described as the product of two one-dimensional (1D) resonant pdfs: 
\begin{equation*}
P_\mathrm{sig}(\mWV,\mjet) = P(\mWV|\mX,\theta) \, P(\mjet|\mX,\theta),
\end{equation*}
where $\theta$ represents sets of nuisance parameters affecting the shape, which we describe in Section~\ref{sec:SignalSystUnc}.
Each factor in this formula is constructed by fitting a double-sided Crystal Ball (dCB) function~\cite{Oreglia:1980cs}, composed of a Gaussian core and asymmetric power-law tails, to the corresponding distribution of simulated signal events for eleven different values of the resonance mass \mX from 1.0 to 4.5\TeV.
Such a model neglects the mild correlation between \mWV and \mjet, which results in a small rotation in the 2D \mWV and \mjet plane that is negligible compared to the experimental resolution uncertainties.
Since the  modeling of the lepton momentum scale and resolution has negligible impact in the shape of the invariant mass of the system compared to the impact of jet and \ptmiss reconstruction,
 the electron and muon channels are merged to gain statistics for the fit in the simulation.
In LP categories, an exponential function is added to the fit model of the \mjet dimension over its entire range, to model properly the low-mass tail.
Each function parameter is interpolated for other values of \mX using a polynomial function.
Separate shape models are built for each studied signal benchmark scenario.
\begin{figure}[!htb]
  \centering
  \includegraphics[width=0.45\textwidth]{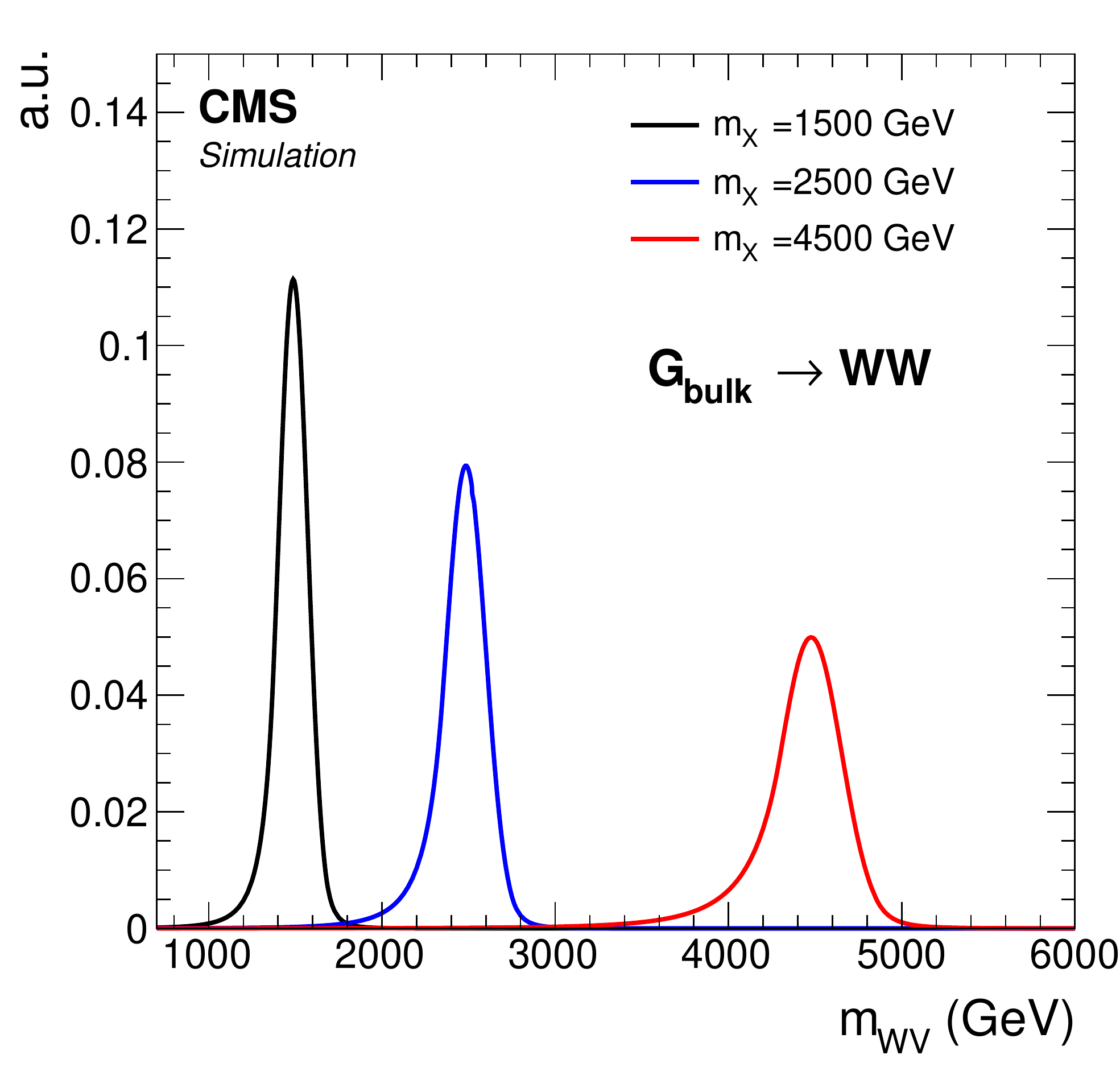}
  \includegraphics[width=0.45\textwidth]{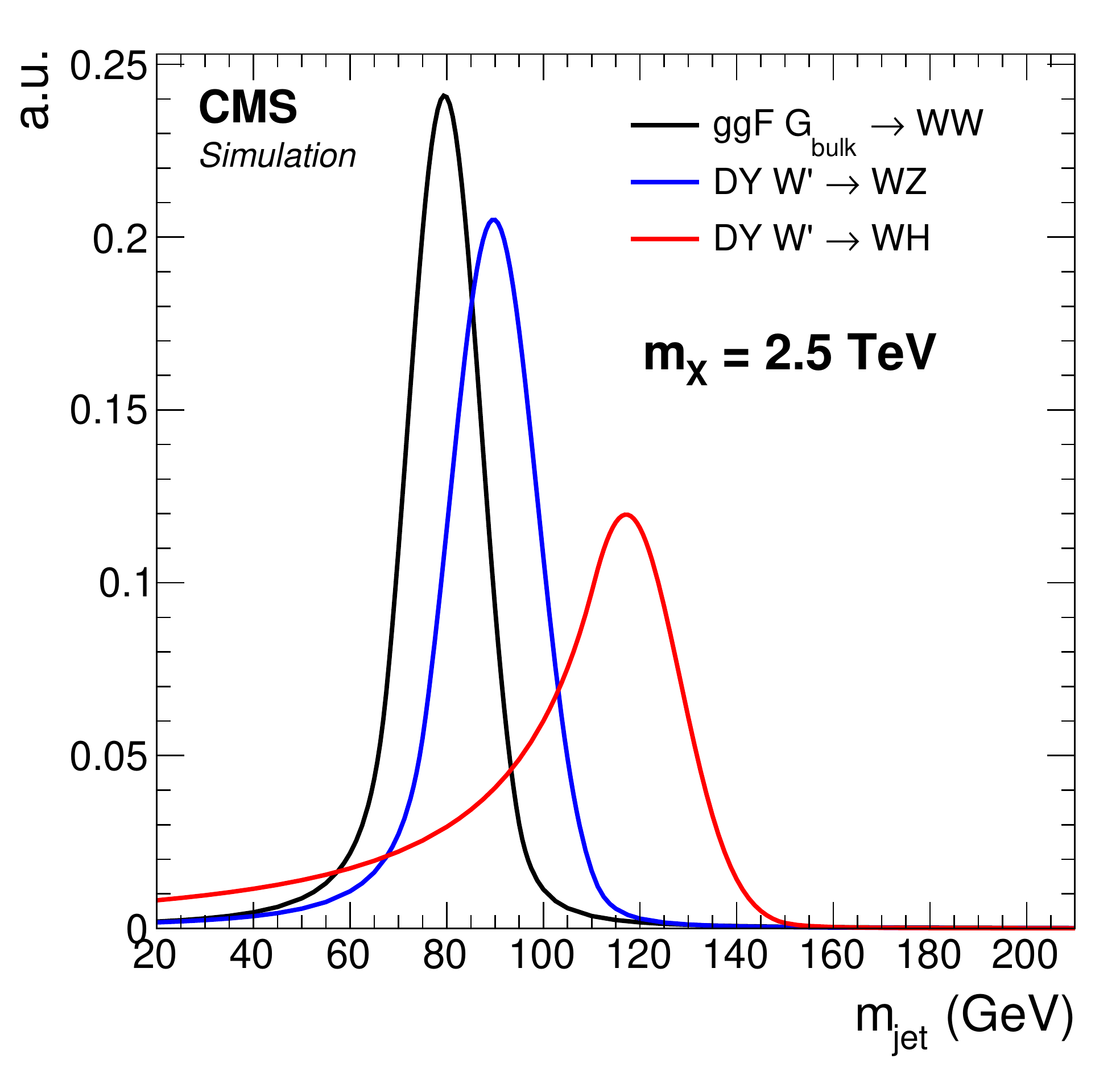}
  \caption{Projections of the 2D signal likelihood along the \mWV dimension (\cmsLeft) and the \mjet dimension (\cmsRight). The \mWV projections are shown for different mass hypotheses of 1.5, 2.5, and 4.5\TeV for a \GBulk signal decaying to \WW. The \mjet projections are shown for \GbuToWW, \WprToWZ, and \WprToWH  for $\mX =  2.5 \TeV$. All distributions are normalized to the same area.}
\label{fig:signalTemplates}
\end{figure}

Figure~\ref{fig:signalTemplates} shows the projections of the 2D likelihood along the \mWV and \mjet dimensions, respectively. The \mWV projections are shown for the \GbuToWW signal for mass hypotheses of 1.5, 2.5, and 4.5\TeV. The distributions are very similar for other spin hypotheses, production mechanisms, and decay modes. The \mjet projections are shown for \GbuToWW, \WprToWZ, and \WprToWH signals for a mass of $\mX= 2.5 \TeV$, demonstrating the sensitivity obtained by CMS reconstruction and jet substructure techniques to \PW, \PZ, and Higgs jet hypotheses. While the \PW and \PZ invariant mass distributions peak near the expected masses, the Higgs mass peak is slightly shifted, since the presence of neutrinos in \PQb quark decays is not accounted for in the calibration of the jet mass.
The experimental resolution for \mjet is found to be of the order of 10\%, whereas that of \mWV ranges from 6\% at 1\TeV to 4\% at 4.5\TeV.
In addition, the expected signal yield in every search category is also parameterized as a function of the collected integrated luminosity so that the resonance production cross section can later be extracted from the fit to data.

\subsection{Background modeling}
\label{sec:BkgdTemplates}

Background events are classified into two classes in the fit, each of which is described by a different pdf:
\begin{enumerate}
\item A background called \res, for which the \mjet shape has two peaks, one near the \PW/\PZ boson masses and the other near the top quark mass, while \mWV has a falling spectrum.
  This resonant background is dominated by \ttbar production, with subdominant contributions from SM diboson and single top quark production, and is defined in the simulation by requiring that both generated quarks from a hadronic \PV decay be located within $\DR=0.8$ of the selected large-radius jet. 
  The \mjet shape structure is thus due to the selection of a partially or fully merged top quark jet or a \PV jet.
\item A background called \nonRes, for which \mjet does not have a peak structure, and \mWV again has a falling spectrum.
  This nonresonant background is dominated by \Wtolnuplusjets events, where the selected jet is produced by the hadronization of one or more partons not originating from a vector boson but is mistagged as a \PV jet.
  In addition, this background also includes \ttbar events where the selected large-radius jet corresponds to a random combination of jets in the event, instead of a \PW boson or top quark hadronic decay.
\end{enumerate}

The \nonRes and \res background shapes are both described as the product of a conditional pdf of \mWV as a function of \mjet, and a \mjet pdf:
\begin{equation*}
P_{\text{bkg}}(\mWV,\mjet)= P(\mWV|\mjet,\theta) \, P(\mjet|\theta),
\end{equation*}
where $\theta$ again represents sets of nuisance parameters, described in Section~\ref{sec:BkgdSystUnc}.
The conditional \mWV pdf is constructed with a similar strategy for both classes of backgrounds that employs a robust kernel density estimation technique that is designed to provide smooth 2D templates in all subcategories. 
For each event in the simulated background sample, particle-level jets are clustered from stable particles using the same algorithms employed in event reconstruction. 
A diboson mass \mWVpart is then defined by combining the reconstructed leptonically decaying W boson and the generated large-radius jet.
A detector response model is derived for the scale and resolution of the diboson mass as a function of the transverse momentum \genjetpt of the generated jet, by comparing the reconstructed and generated variables \mWV and \mWVpart. 
The signal region is divided into slices of the soft-drop jet mass \mjet: 16 slices for \nonRes, and one or two slices for \res in the HP and LP categories, respectively. 
Each simulated event $i$ in a given \mjet slice then contributes to the template \mWV distribution in that slice by adding the following 1D Gaussian distribution:
\begin{linenomath}
\ifthenelse{\boolean{cms@external}} {
\begin{multline*}
P_i(m) = \frac{w_i}{\sqrt{2 \pi} \sigma(\genjetpt)} \\ \times\exp \left[ -\frac{1}{2} \left( \frac{m-s(\genjetpt)\,s\mWVpart}{\sigma(\genjetpt)} \right )^2 \right].
\end{multline*}
}
{
\begin{equation*}
\begin{aligned}
P_i(m) = \frac{w_i}{\sqrt{2 \pi} \sigma(\genjetpt)} \exp \left[ -\frac{1}{2} \left( \frac{m-s(\genjetpt)\,\mWVpart}{\sigma(\genjetpt)} \right )^2 \right].
\end{aligned}
\end{equation*}
}
\end{linenomath}
Here, $m$ runs over the allowed values of \mWV, 
$w_i$ is the event weight of the simulation, corrected for differences in the \mWV spectrum of \nonRes observed between data and simulation in the control region, 
and $s(\genjetpt)$ and $\sigma(\genjetpt)$ are the scale and resolution parameters from the detector response model. 
This modified kernel procedure, seeded by the resolution instead of the event density, is  robust in describing rapidly changing shapes, however it has limited performance at very low statistics. 
Therefore, since simulated samples do not have enough events to provide regular shapes up to the largest values of \mWV, these high-mass tails have to be smoothed.
This smoothing is applied in the region where the event yield is dominated by Poisson statistics, namely for events with \mWV larger than a threshold that varies between $1.1$ and $1.6\TeV$ depending on the background class and category.
The high-\mWV distribution is fitted to a power-law function, which is used as the shape in this region.
To improve the robustness of the templates, the electron and muon channels are here merged for the purpose of constructing the template, and in the case of \nonRes, the VBF-tagged, \bbbar, and \nobbbar regions are merged as well.
The residual differences between the 24 categories are later corrected for by the final 2D fit, deploying uncorrelated shape uncertainties that are  described in Section~\ref{sec:BkgdSystUnc}.

The \mjet pdf is built separately for each of the 24 search categories, in order to account for the specific kinematic configuration of each category.
For the nonresonant \nonRes background, the \mjet pdf is obtained by smoothing the 1D histogram of selected simulated events in each category using cubic spline interpolation between bins. A kernel method, similar to the one followed for \mWV has limited performance in this variable since the resolution model is complex for the soft-drop mass variable. 
In the case of the \res background, the \mjet distribution features a peak around the \PW boson mass, which is dominated by top quark jets where only the \Wtoqqbarpr products were reconstructed inside the large-radius jet, and a peak around the top quark mass, where the \PW boson and the \PQb quark decays are merged.
To define the nominal background shapes, an analytical function is fitted to the distribution of simulated events, where the two-peak structure is modeled as the sum of two dCB functions and an exponential function that is used only in the LP categories.
The inclusion of the top peak in the search region improves the fit precision, because constraining the relative fraction of the two peaks helps capture the convoluted effects of the top quark \pt spectrum and jet grooming. 
The \mjet shape of the \res background in the region of the \PW mass peak is found to differ from that of the \XtoWW signals even if, in both cases, a \PW boson is present. This difference is attributed to the fact that the background consists of high-momentum \ttbar events, in which a part of the \PQb jet from the $\text{t}\to\PW\PQb$ decay overlaps with the \PV jet from the \Wtoqqbarpr decay, while in the case of the signal, the \PV jet is isolated.

\section{Systematic uncertainties}
\label{sec:SystUnc}

Systematic uncertainties affecting the normalization and shape of the signal and backgrounds are modeled by nuisance parameters, each of which is profiled in the likelihood maximization.
All sources of systematic uncertainties are listed in Table~\ref{tab:systematics}.
When specified, the magnitude of the uncertainty is the width of the function used to constrain the nuisance parameter, which is a log-normal distribution for uncertainties related to normalization, and a Gaussian distribution for parameters that control shape uncertainties. The following sections describe these uncertainties in more detail.

\begin{table*}[htbp]
  \centering
  \topcaption{Summary of the systematic uncertainties considered in the 2D fit, the quantities they affect, and their magnitude, when applicable. When ranges are given, the magnitude of the uncertainty depends on the signal model or mass. The three parts of the table concern shape uncertainties only affecting backgrounds, shape uncertainties in the scales and resolutions, and normalization uncertainties.}
  \cmsTable {
  \begin{scotch}{llc}
    {Source}                          & {Relevant quantity}         & {Magnitude}   \\
    \hline
    \multicolumn{3}{c}{\textit{Shape uncertainties only affecting backgrounds}} \\
    Jet \pt spectrum                      & \nonRes and \res \mWV shape     &             \\ 
    Correlation between jet mass and \pt  & \nonRes \mWV and \mjet shape    &             \\ 
    Jet mass scale                        & \nonRes \mjet shape             &             \\ 
    Hadronization modeling                & \nonRes \mjet shape             &             \\ 
    High-\mWV tail                        & \res \mWV shape                 &             \\ 
    \PW boson and top quark mass peak ratio & \res \mjet shape              &             \\[\cmsTabSkip]
    \multicolumn{3}{c}{\textit{Shape uncertainties in scale and resolution}} \\
    Jet mass scale                        & Signal and \res \mjet mean      & 1\%         \\
    Jet mass resolution                   & Signal and \res \mjet width     & 8\%         \\
    Jet energy scale                      & Signal \mWV mean                & 2\%         \\
    Jet energy resolution                 & Signal \mWV width               & 5\%         \\
    \ptmiss scale                         & Signal \mWV mean                & 2\%         \\
    \ptmiss resolution                    & Signal \mWV width               & 1\%         \\
    Lepton energy scale                   & Signal \mWV mean                & 0.5\% (\Pe), 0.3\% ($\mu$)          \\[\cmsTabSkip]
    \multicolumn{3}{c}{\textit{Normalization uncertainties}} \\
    \nonRes normalization                 & \nonRes yield                   & 25\%        \\
    \res normalization                    & \res yield                      & 25\%        \\
    Lepton selection efficiency           & \nonRes, \res and signal yield  & 5\%         \\
    \PV tagging                           & Signal yield                    & 4\% (HP), 4\% (LP)                \\ 
    \pt-dependence of \PV tagging         & Signal yield                    & 1.7--19\% (HP), 1.2--14\% (LP)    \\
    \DoubleB tagging                      & Signal yield                    & 6--9\% (\bbbar), 0.4--2\% (\nobbbar)      \\
    $\Dy$-based categorization            & Signal yield                    & 2--6\% (LDy), 1.5--5.5\% (HDy)    \\
    Integrated luminosity                 & Signal yield                    & 1.6\%       \\
    Pileup reweighting                    & Signal yield                    & 1.5\%       \\
    b tagging veto                        & Signal yield                    & 2\%         \\
    PDFs                                  & Signal yield                    & 0.1--2\%    \\
  \end{scotch}
 }
  \label{tab:systematics}
\end{table*}

\subsection{Systematic uncertainties in the background estimation}
\label{sec:BkgdSystUnc}

The 2D fit is designed to predict correctly the normalization and shapes of background contributions directly from the data by introducing nuisance parameters that vary the shapes and the yields of each contribution during the likelihood minimization process. 
To implement nuisance parameters that affect the shapes, the template-building procedure described in Section~\ref{sec:Templates} is repeated with additional event weights or modified shape parameters.
For each parameter, this produces two alternative 2D templates that represent an upward and a downward shift, between which the 2D fit performs an interpolation based on the value of the nuisance parameter.
The magnitude of the shape variations are chosen to cover the differences between data and simulation observed in the control region.

First, both classes of backgrounds are assigned shape uncertainties that modify the conditional \mWV factor in the 2D likelihood.
These have to account not only for differences between data and simulation but also for the use of common conditional likelihoods in different categories. The main difference of the shapes between the data and the templates arises from differences in the \pt spectrum  after categorization and jet substructure requirements are applied. Therefore, we define alternative shapes by performing a linear reweighting of the jet \pt spectrum in each category.  
This variation is motivated by the imperfect modeling of the parton distribution functions and initial-state radiation, and is conservative given that the fit has the statistical power to constrain these uncertainties in regions where no signal is expected for a given model.
The \nonRes background has another shape variation related to the correlation between the jet mass and the jet \pt, which simultaneously modifies both dimensions of the conditional likelihood, while the \res background is assigned an uncertainty that modifies the power-law function used to populate the high-\mWV tails.
Since each category involves jets with different \pt spectra as a result of the selection requirements imposed on different particles in the event, these sets of nuisance parameters are left uncorrelated between categories.

Uncertainties in the \mjet shape of the \nonRes background mostly arise from hadronization-related effects and their interplay with the soft-drop algorithm.
Two \mjet shape variations are defined.
The first one is chosen to be a simple shift of the \mjet scale. 
The second one is motivated by the study of the scaling variable $\log(\mjet^2/\pt)$, which reveals a difference in hadronization behavior between data and simulation.
The discrepancy in the distribution of this variable is measured in the region of $\mjet<50 \GeV$ that is dominated by \nonRes events. Consequently, an uncertainty is introduced that generates alternative shapes after reweighting the simulation to match the data.
Since the jet substructure variables are very sensitive to the apparent difference of the jet \pt spectrum in different categories, these uncertainties are treated as uncorrelated.

The \mjet shape of the \res background is affected by uncertainties in the scale and resolution of the soft-drop jet mass, which are estimated by the top-enriched sample and are encoded in nuisance parameters that modify the width and the peak of the fitted dCB functions, respectively.
Since the scale and resolution effects of the softdrop mass are different for two-prong and three-prong objects, these uncertainties  are left uncorrelated between the \PW boson and top quark mass peaks and between the HP and LP categories but are considered fully correlated across other categorization criteria.

Additionally, an uncertainty of 13\% in the ratio of the \PW boson mass peak normalization to the sum of the \PW boson and top quark mass peaks, derived by fitting the \mjet spectrum in the top-enriched control region, is introduced with one parameter per category, effectively measuring the \pt spectrum of the top quark. The shape uncertainties introduced are constrained in the fit by the shapes and relative normalizations of the \PW and top peaks in data.

Both the \nonRes and the \res backgrounds are assigned a large normalization uncertainty of 25\% based on agreement between data and simulation in the low \mjet sideband and the top-enriched control region respectively. 
While the cross section measurements of these processes at the LHC are known to better precision, the effects of jet substructure, the requirement of jet mass windows, and the categorization, all introduce larger discrepancies. 
The motivation behind the 2D-fit signal extraction procedure is to constrain those differences from the data in each category without being sensitive to the initial value of the uncertainty.
The corresponding parameters are fully correlated between lepton channels and uncorrelated across other categorization criteria and between the two background classes.
Two other uncorrelated 5\% background normalization parameters are assigned to the electron and muon channels in order to account for lepton triggering, reconstruction, and isolation efficiencies based on measurements from the data.

\subsection{Systematic uncertainties in the signal prediction}
\label{sec:SignalSystUnc}

The two-dimensional signal shapes in the (\mjet, \mWV) plane are affected by several uncertainties.
Relative scale factors on the mean and width of the dCB function modeling the \mWV shape encode uncertainties in the scale and resolution of the jet energy and the \ptmiss, and electron and muon energy scales.

In the \mjet dimension, two parameters account for the impact of grooming on the scale and resolution of the soft-drop jet mass, and are fully correlated with the analogous shape uncertainties of the \PW mass peak of the \res background. 

The dominant uncertainties in the signal normalizations arise from uncertainties in the efficiency of the \PV tagging, \doubleB tagging, and $\Dy$-based selections.
The corresponding nuisance parameters are anticorrelated between HP and LP, between \bbbar and \nobbbar, and between LDy and HDy categories, respectively, therefore inducing a migration of events between the categories in which they apply.
Two nuisance parameters are associated with the \nsubjDDT-based categorization: one for its efficiency and another for its dependence on the jet \pt. The values of these nuisance parameters correspond to the statistical and systematic
  uncertainties of the measurement of \PV tagging in data for different ranges of the jet \pt. These measurements are performed in the disjoint region enriched in top quark events with hadronic \PW boson decays, by splitting the sample into events that pass or fail \PV tagging and then simultaneously fitting the \PW jet mass distributions in data and simulation.

The uncertainty caused by the \Dy requirement is evaluated by studying the distributions of data and simulated events in the top quark-enriched control region.

Other uncertainties that apply to the normalization of signal events are associated with the integrated luminosity~\cite{CMS:2021xjt,CMS-PAS-LUM-17-004,CMS-PAS-LUM-18-002}, the pileup reweighting, the efficiency of the b tagging veto on standard jets, and the lepton triggering, reconstruction, and isolation. 
Finally, uncertainties in the signal yield due to the choice of PDFs, and the factorization and renormalization scales are also taken into account: the scale uncertainties are evaluated following the proposal in Refs.~\cite{Cacciari:2003fi,Catani:2003zt}, while the PDF uncertainties are evaluated using the NNPDF 3.0~\cite{Ball:2014uwa} PDF set.
The resulting uncertainties in acceptance are found to be negligible for the scale variation and range from 0.1 to 2\% for the PDF evaluation.
On the other hand, the uncertainties in the signal cross section due to PDFs and scales are not taken into account in the statistical interpretation but are instead considered as uncertainties in the theoretical cross section. 
\begin{figure*}[!tb]
  \centering
  \includegraphics[width=0.32\textwidth]{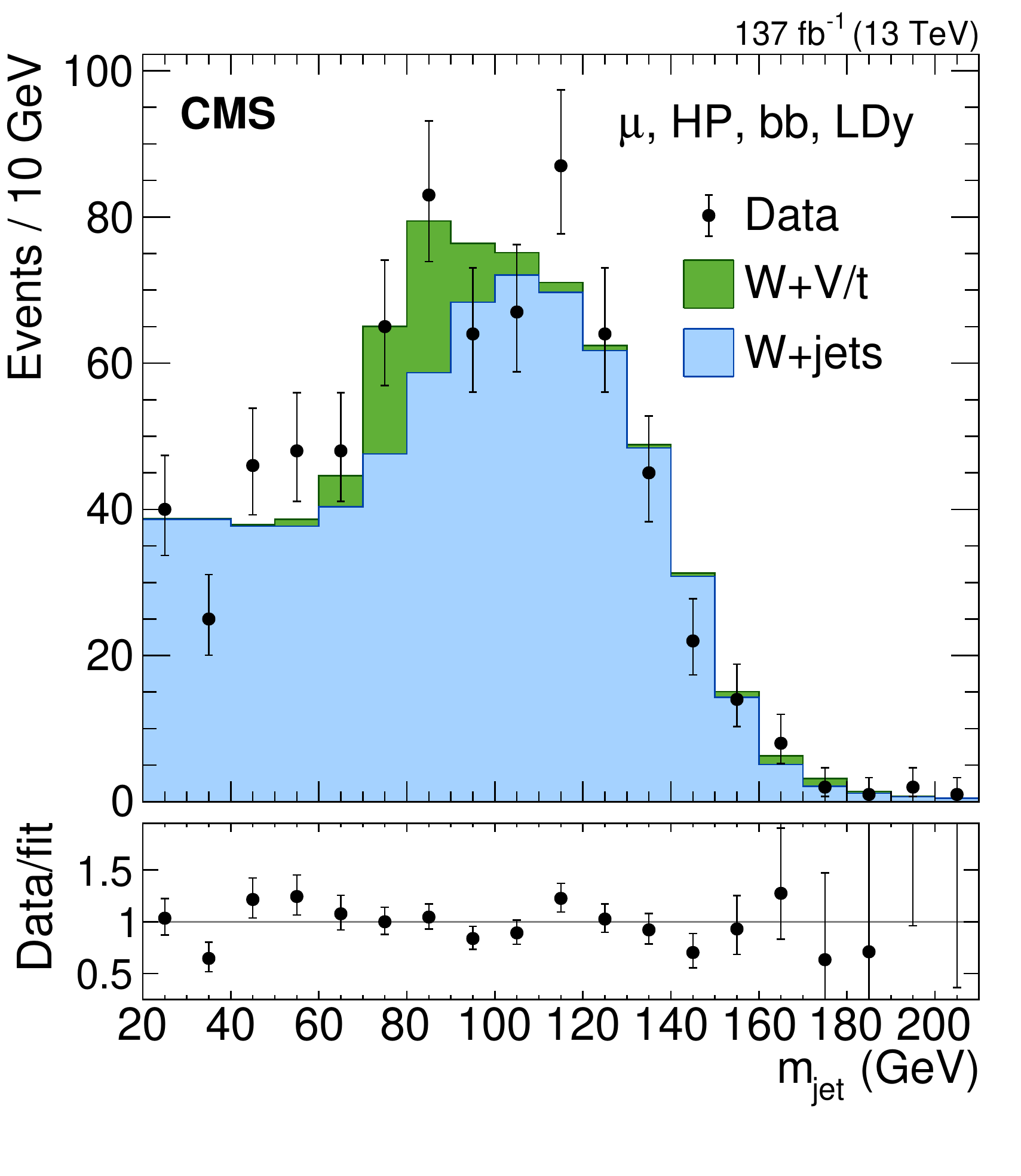}
  \includegraphics[width=0.32\textwidth]{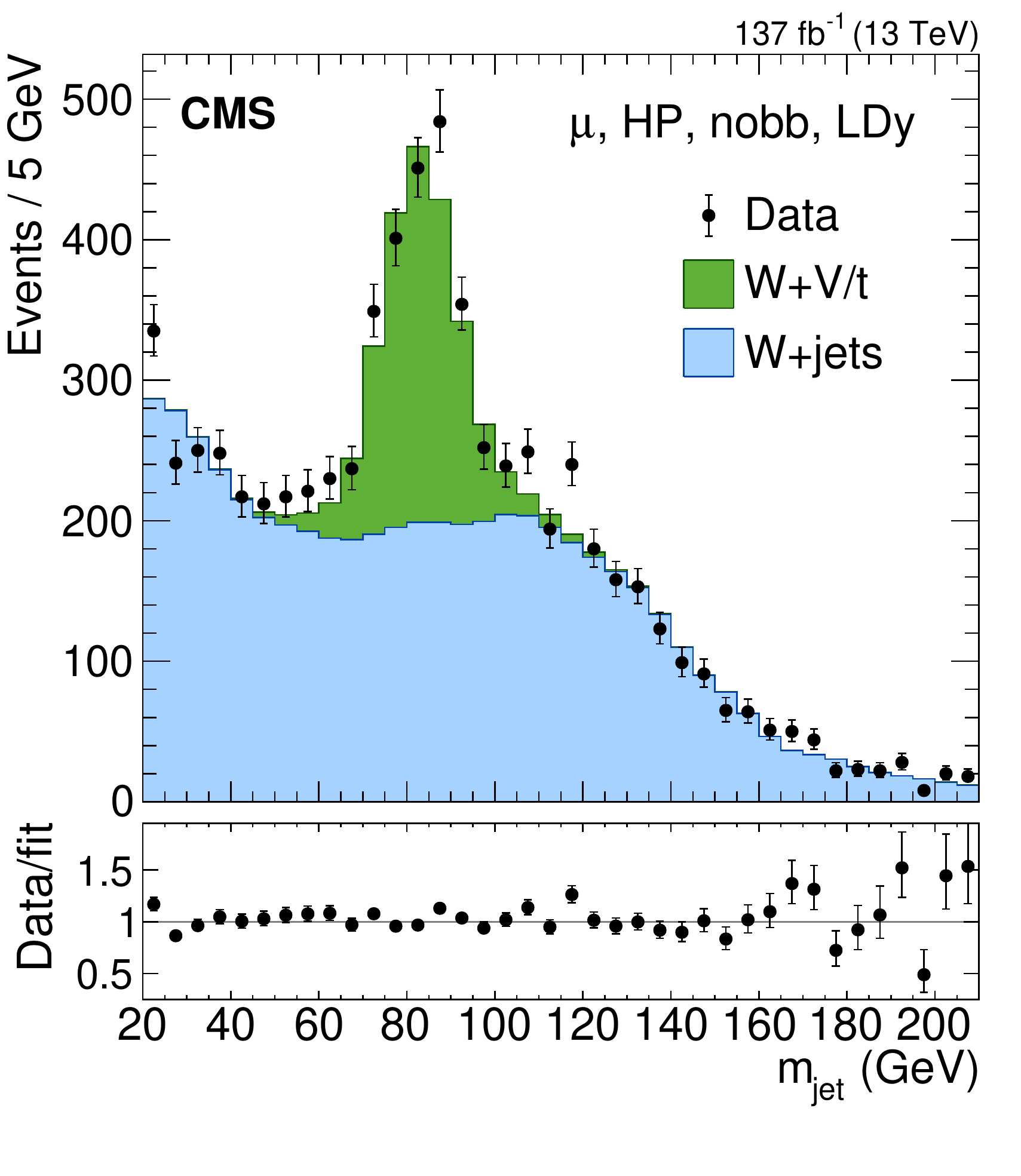}
  \includegraphics[width=0.32\textwidth]{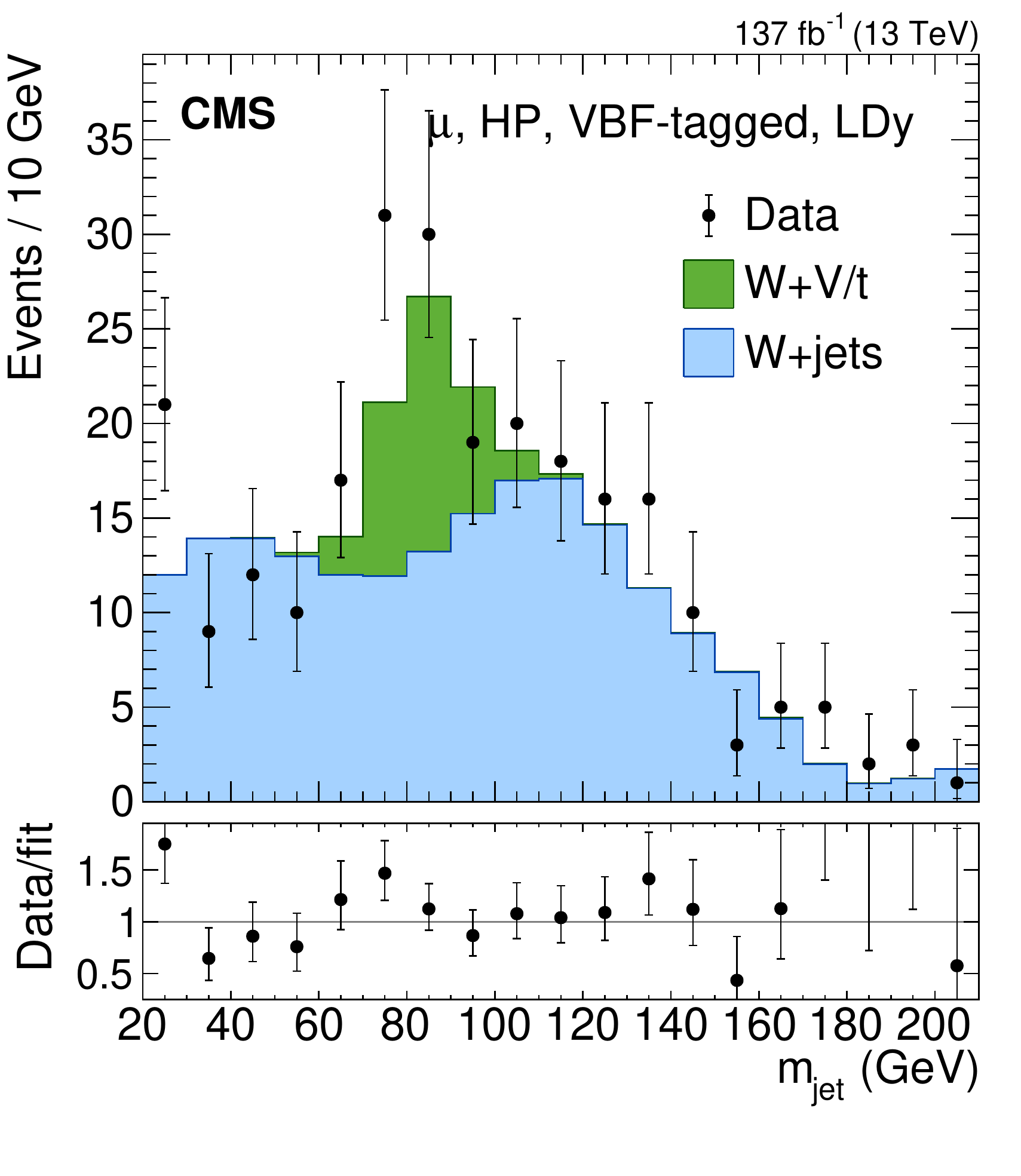} 
  \includegraphics[width=0.32\textwidth]{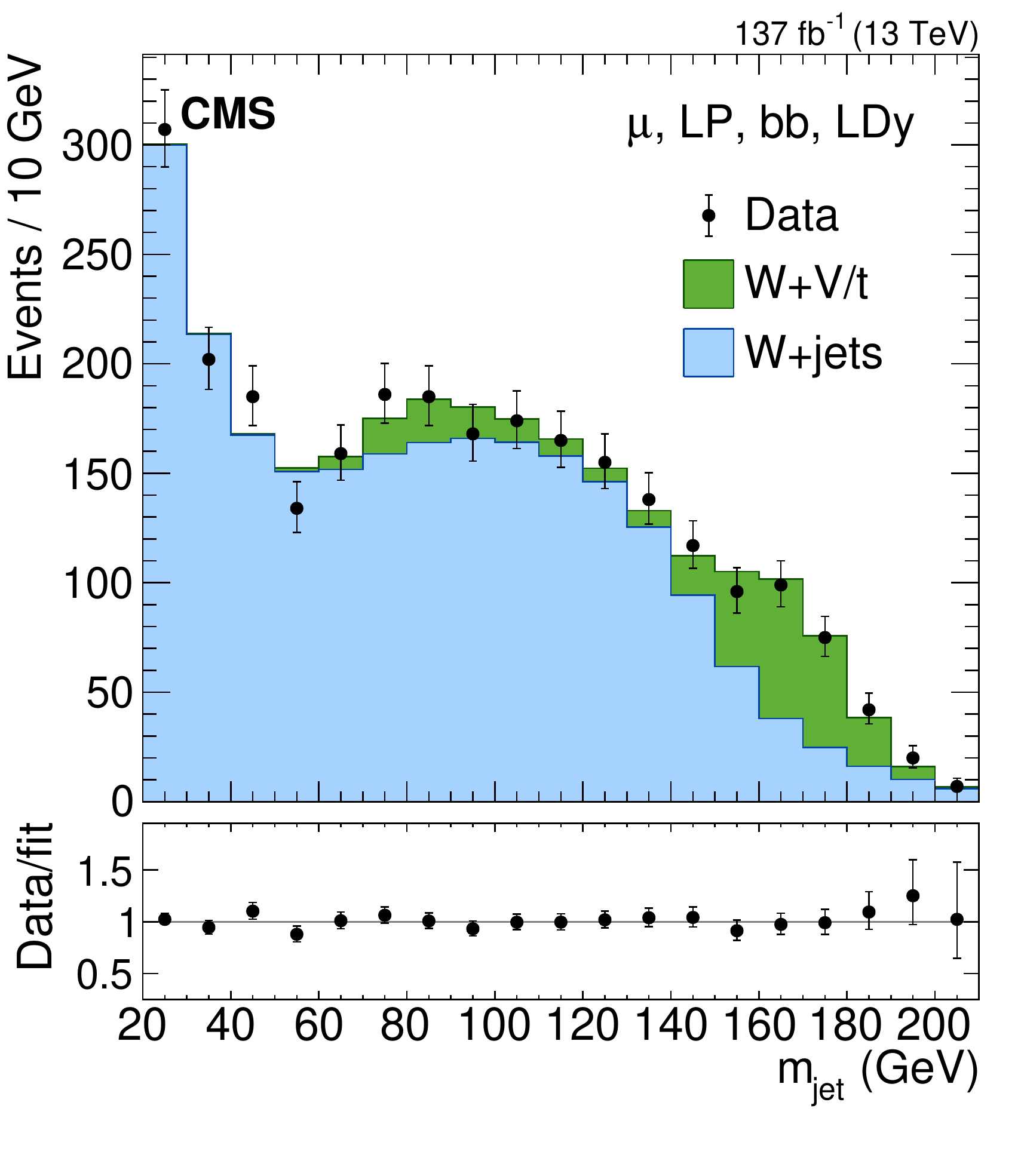}
  \includegraphics[width=0.32\textwidth]{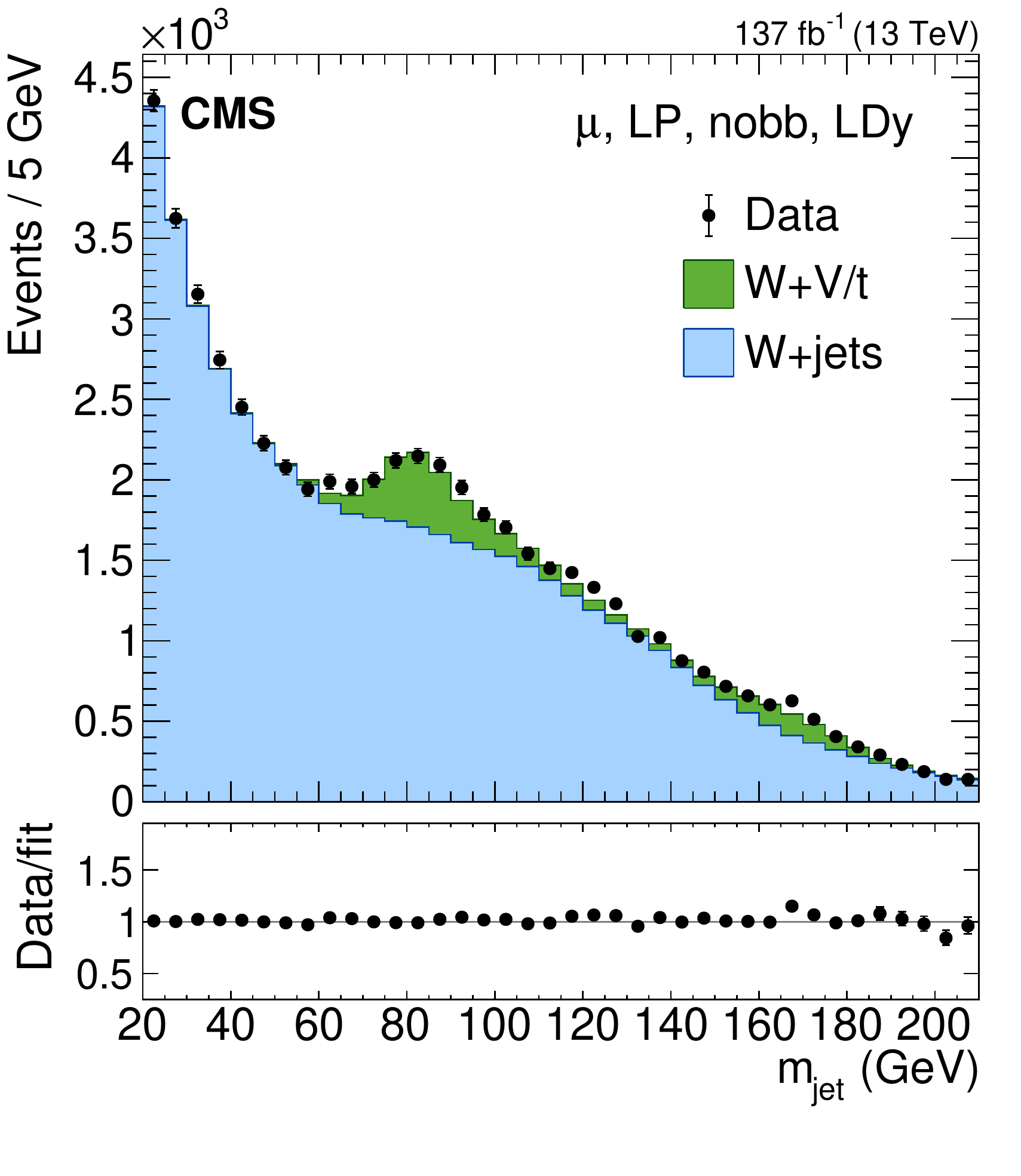}
  \includegraphics[width=0.32\textwidth]{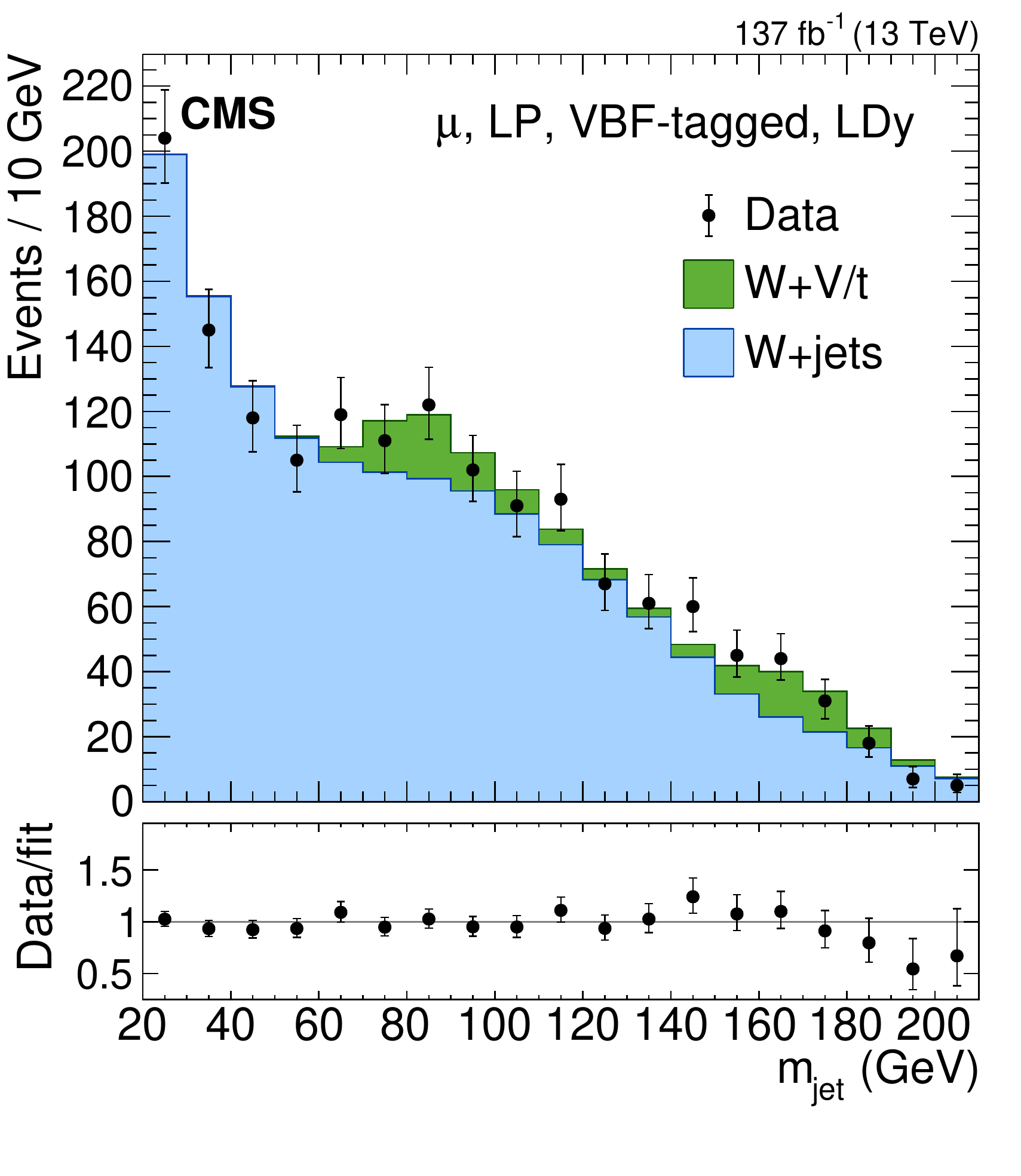} 
  \caption{Comparison between the fit result and data distributions of \mjet in six representative  muon-LDy categories. The distributions in the remaining 18 categories show very similar levels of agreement. 
    The statistical uncertainties of the data are shown as vertical bars.
    The lower panels show the ratio of the data to the fit result.
  }    
  \label{fig:postfit-mjet}
\end{figure*}

\section{Results}
\label{sec:Results}

\begin{figure*}[!htb]
  \centering
  \includegraphics[width=0.32\textwidth]{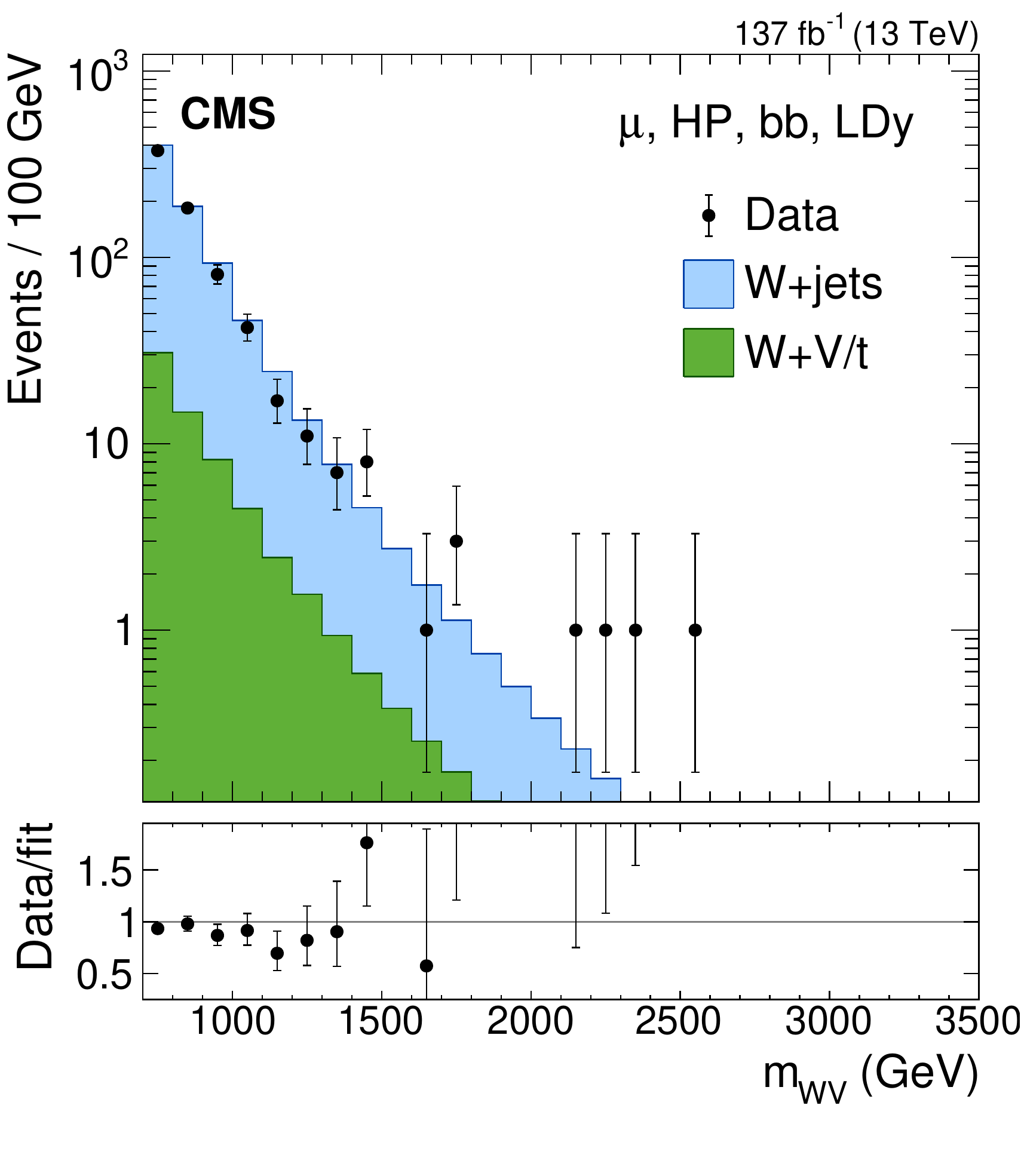}
  \includegraphics[width=0.32\textwidth]{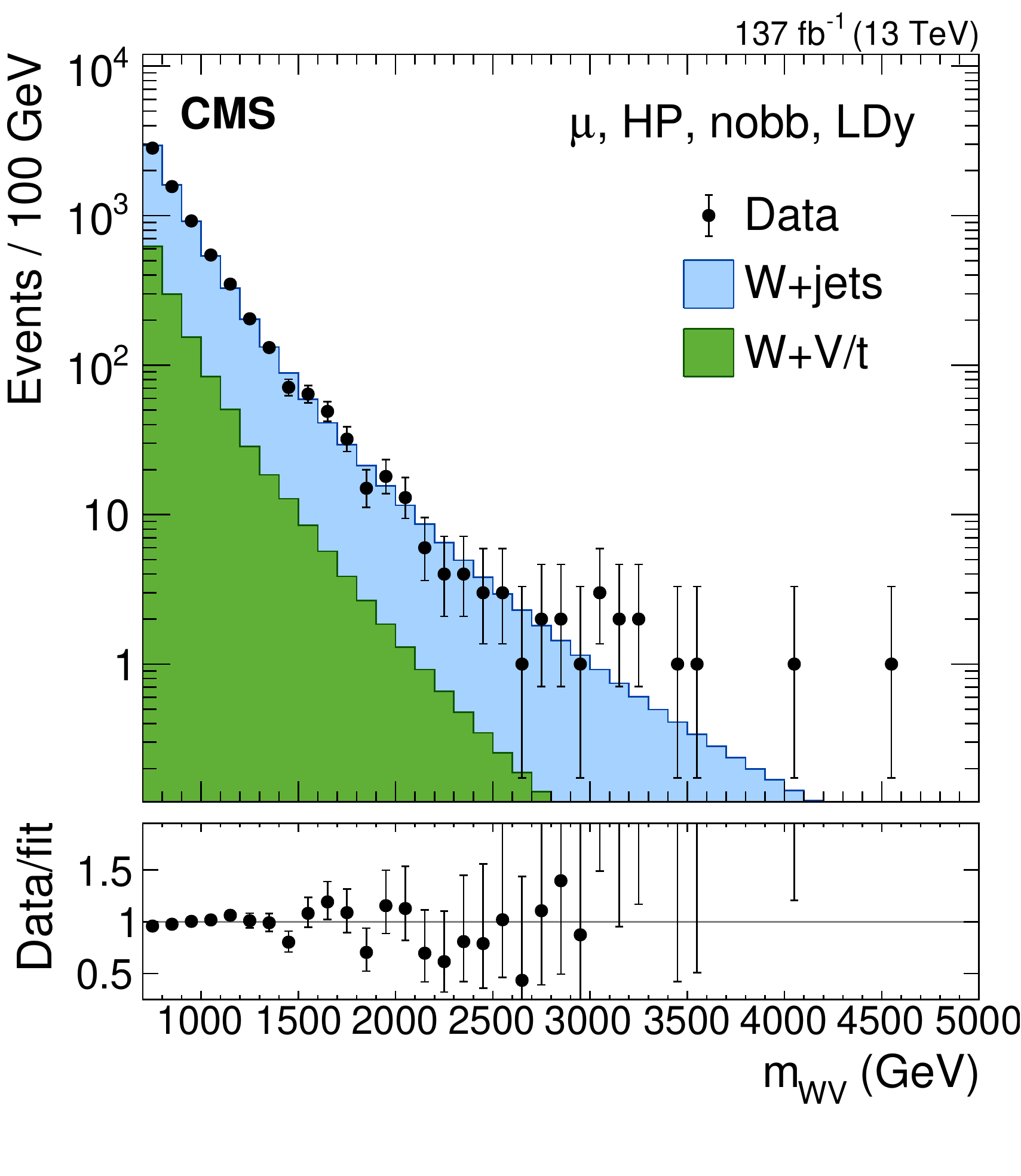}
  \includegraphics[width=0.32\textwidth]{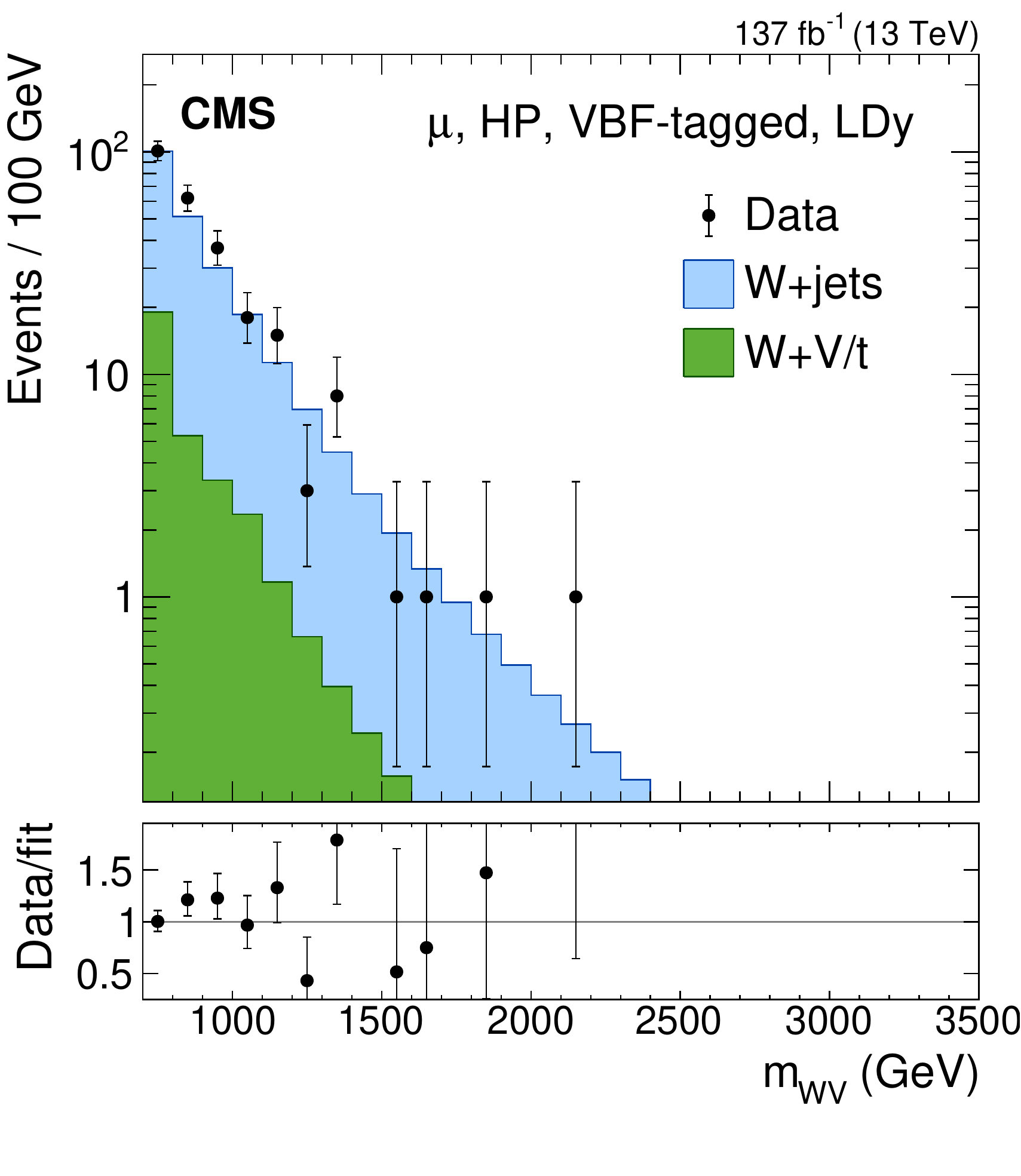} 
  \includegraphics[width=0.32\textwidth]{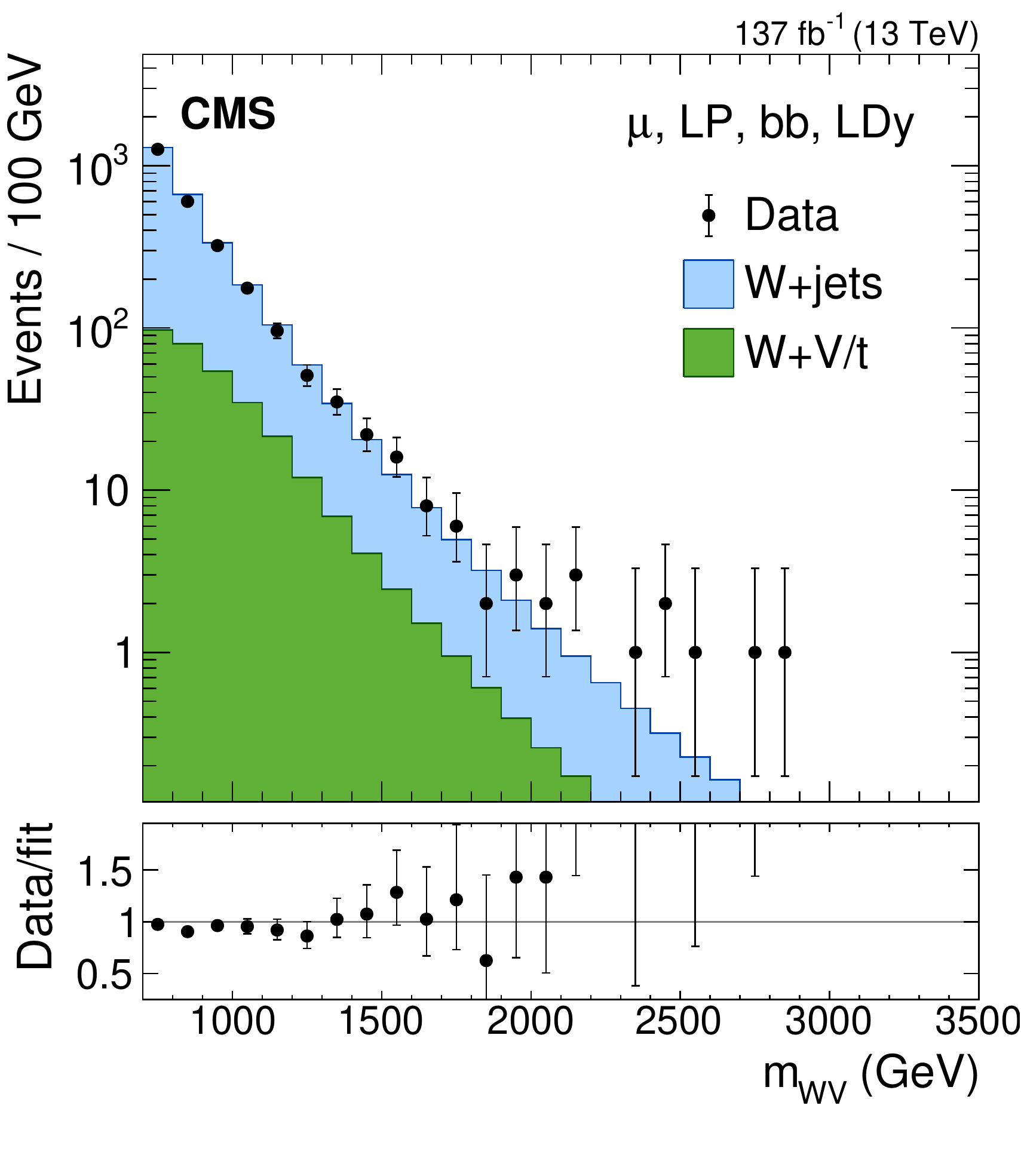}
  \includegraphics[width=0.32\textwidth]{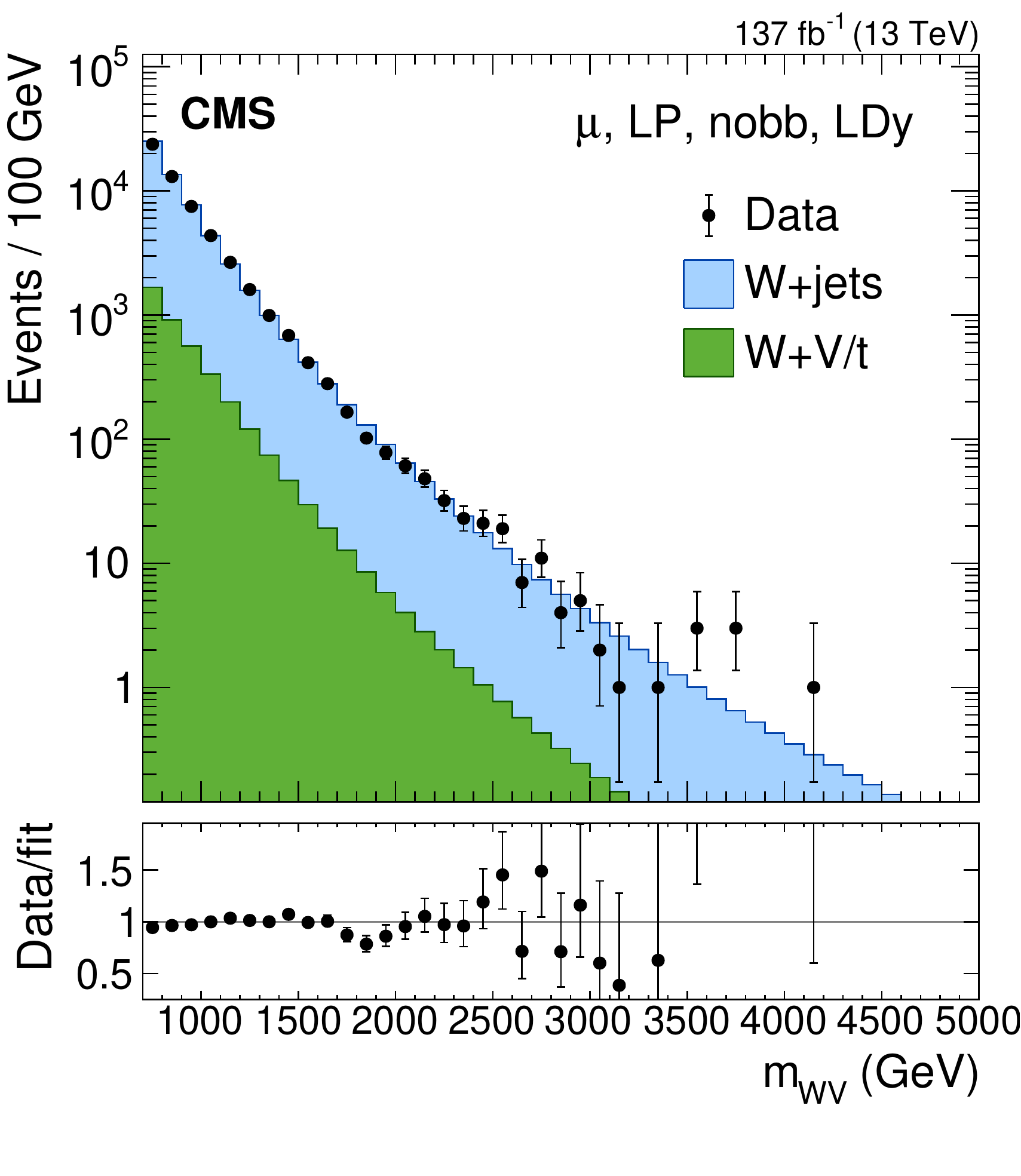}
  \includegraphics[width=0.32\textwidth]{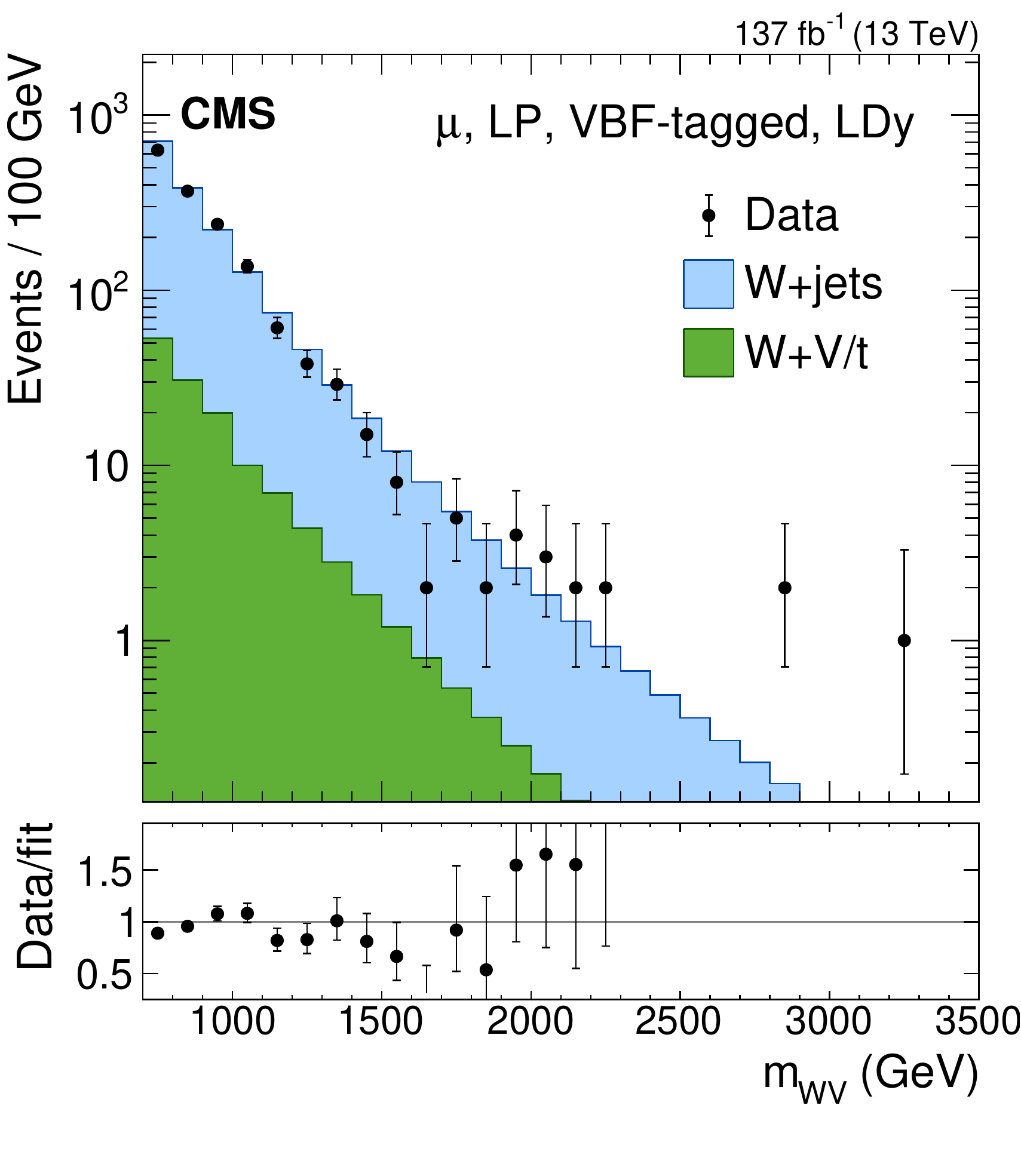} 
  \caption{Comparison between the fit result and data distributions of \mWV in  six representative muon-LDy categories. The distributions in the remaining 18 categories show very similar levels of agreement. The statistical uncertainties of the data are shown as vertical bars. The lower panels show the ratio of the data to the fit result.
  }    
  \label{fig:postfit-mWV}
\end{figure*}

The 2D maximum likelihood fit is performed simultaneously in all 24 search categories.
To assess the fit quality, the fit is first performed without signal contributions.  
The results of the background-only fit are illustrated for six representative categories in Figs.~\ref{fig:postfit-mjet} and~\ref{fig:postfit-mWV}, where projections of the 2D post-fit distributions are shown in the \mjet and \mWV dimensions, respectively. The distributions for the remaining 18 categories show very similar levels of agreement. The jet mass distributions demonstrate good modeling of both the resonant peaks and the continuum for all categories. 
In the LP categories, the \res background has significant contributions from the \PW boson peak and top quark peak, while only the \PW peak is visible in the HP categories.

The post-fit pull distribution of the nuisance parameters is consistent with a Gaussian distribution centered around zero with a standard deviation of unity, while the best-fit values of most nuisance parameters are found to lie within the $\pm1\sigma$ range initially associated with each uncertainty.
The quality of the fit is also assessed with a goodness-of-fit estimator that uses the saturated model~\cite{Baker:1983tu}, and the observed value of the estimator falls well within the central 68\% interval defined from pseudo-experiments.

Then, the fit is repeated for each benchmark model, including a signal contribution, and extracting the signal production cross section . No significant excess is observed over the estimated background. The largest deviation from the background hypothesis is observed for a VBF-produced charged spin-1 resonance decaying to \WZ with mass around $1\TeV$, with a local significance of $3.0$ standard deviations.

The results are interpreted in terms of exclusion limits at 95\% confidence level (\CL). While the interpretation is performed for a well-defined set of benchmark signal models, the results are generally relevant for narrow resonances of a given spin and production mechanism. The limits are evaluated using the asymptotic approximation~\cite{Cowan:2010js} of the \CLs\ method~\cite{Junk:1999kv,Read:2002hq}.

\begin{figure*}[!htb]
 \centering
 \includegraphics[width=0.4\textwidth]{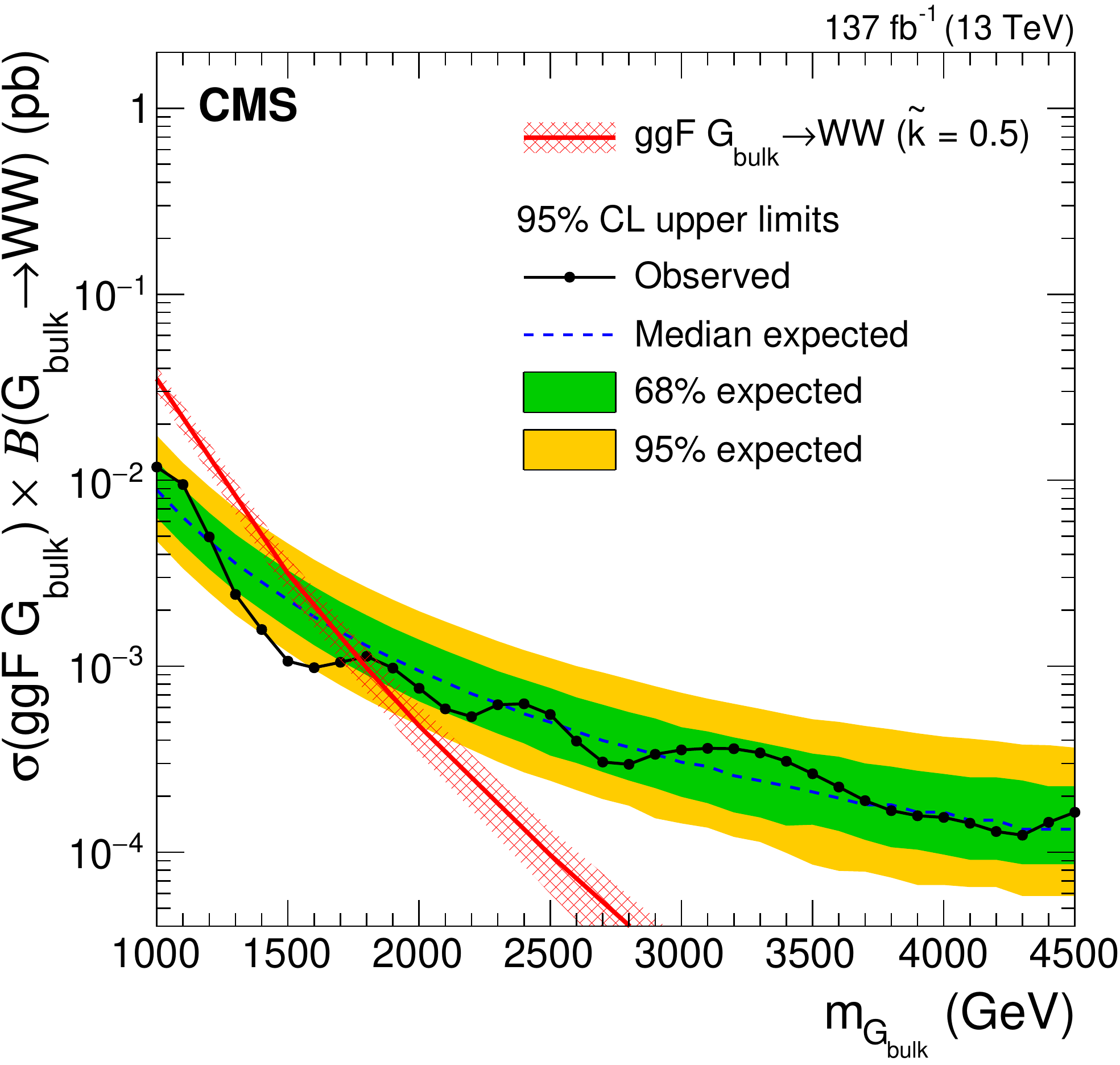}
 \includegraphics[width=0.4\textwidth]{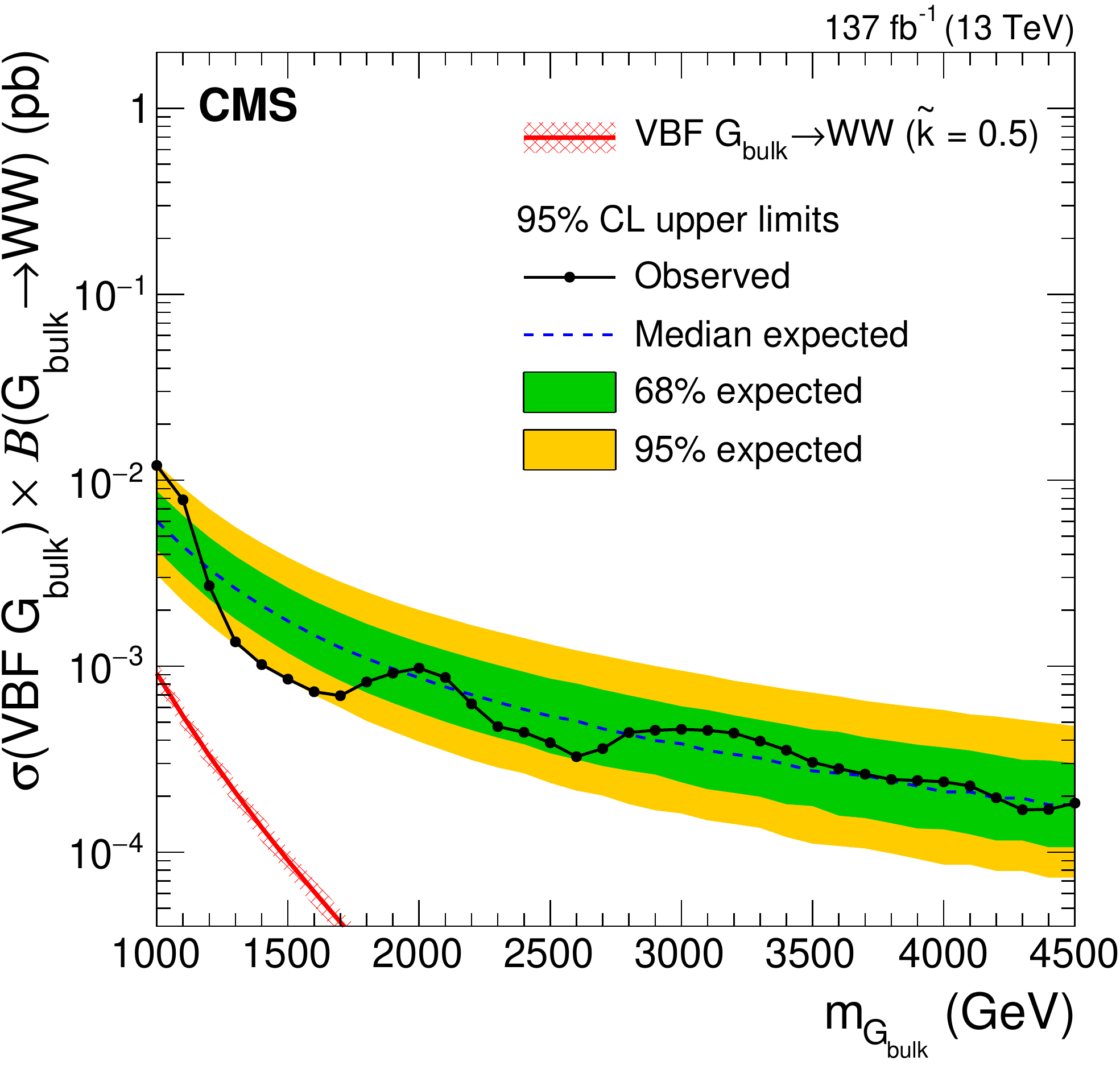}
 \caption{Exclusion limits on the product of the production cross section and the branching fraction for a new spin-2 resonance produced via gluon-gluon fusion (left) or vector boson fusion (right) and decaying to \WW, as functions of the resonance mass hypothesis, compared with the predicted cross sections for a spin-2 bulk graviton with $\ktilde=0.5$. Signal cross section uncertainties are shown as red cross-hatched bands.}
 \label{fig:exclusion_limits_spin2}
\end{figure*}

\begin{figure*}[!htb]
 \centering
 \includegraphics[width=0.4\textwidth]{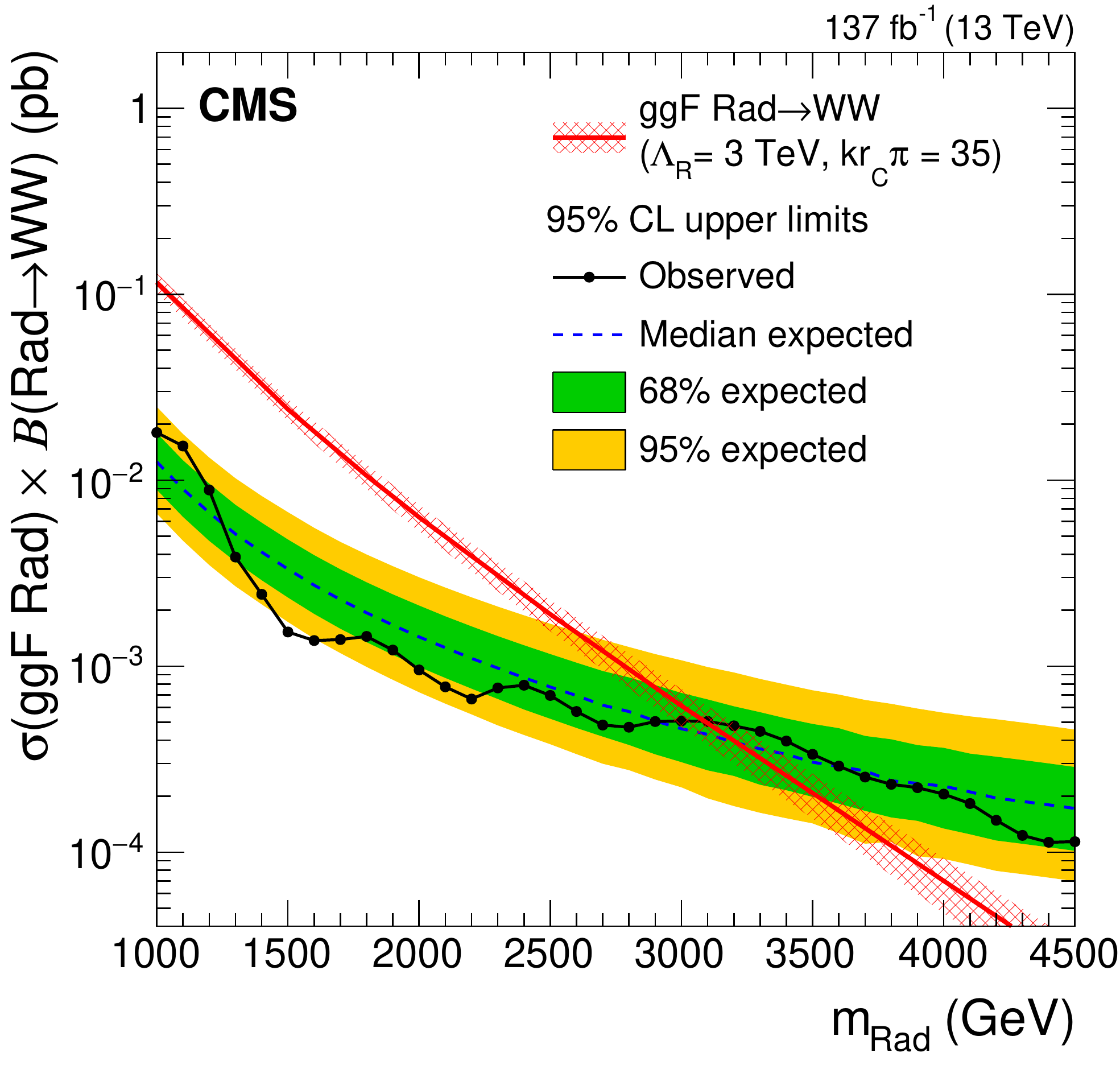}
 \includegraphics[width=0.4\textwidth]{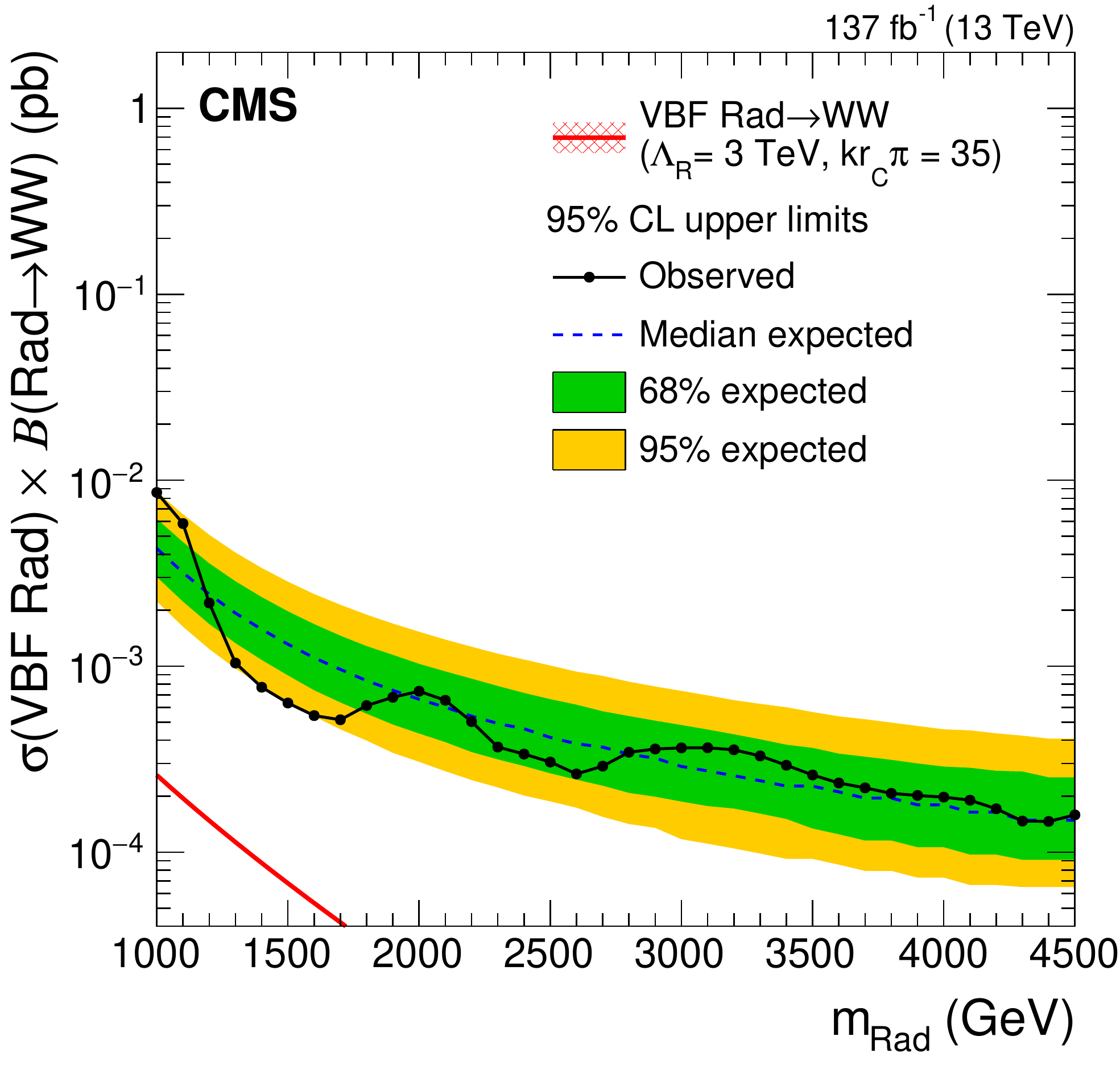}
 \caption{Exclusion limits on the product of the production cross section and the branching fraction for a new spin-0 resonance produced via gluon-gluon fusion (left) or vector boson fusion (right) and decaying to \WW, as functions of the resonance mass hypothesis, compared with the predicted cross sections for a spin-0 bulk radion with $\LambdaR=3\TeV$ and $k r_\text{c} \pi = 35$. Signal cross section uncertainties are shown as red cross-hatched bands.}
 \label{fig:exclusion_limits_spin0}
\end{figure*}

\begin{figure*}[!htb]
 \centering
 \includegraphics[width=0.4\textwidth]{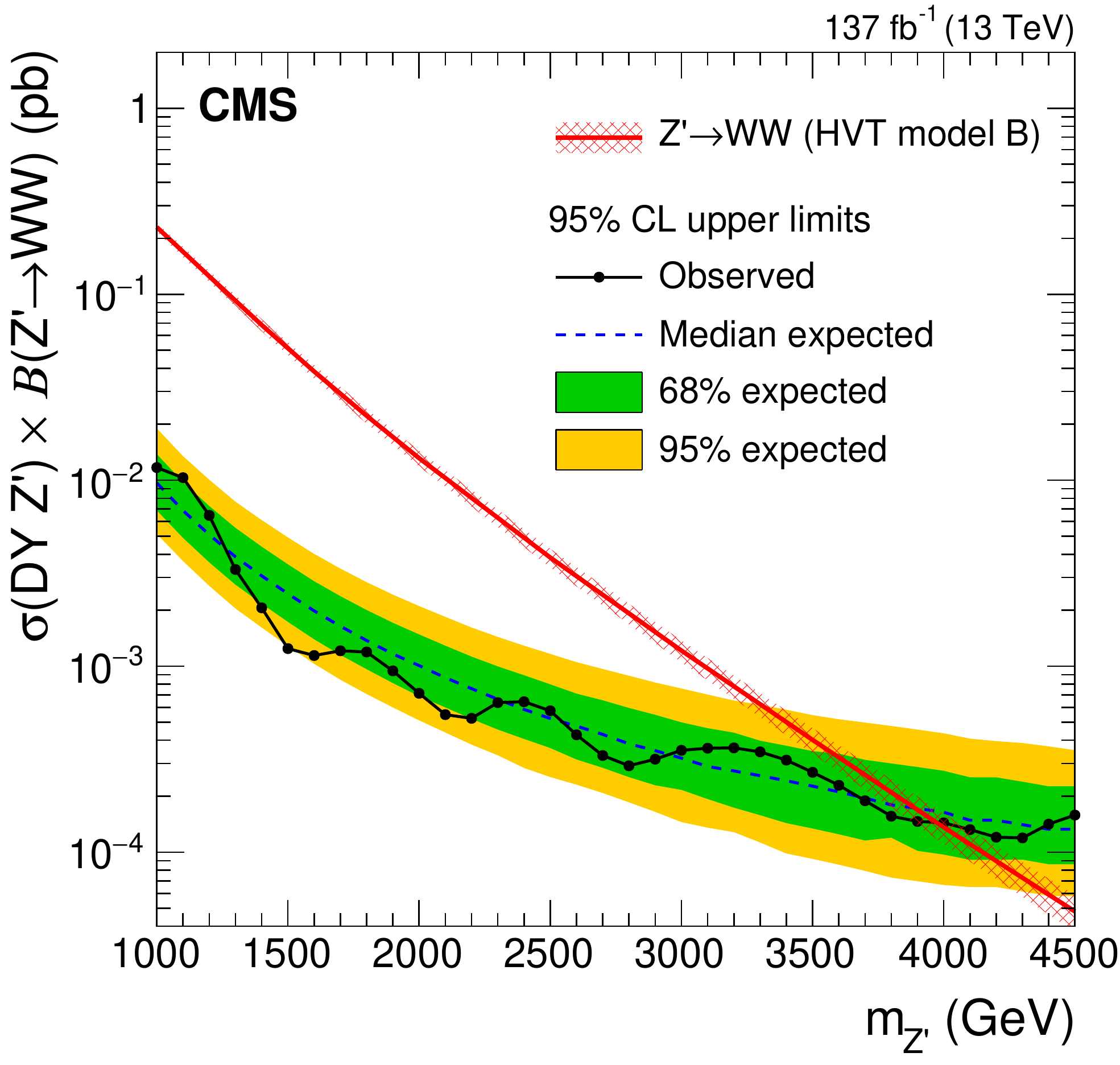}
 \includegraphics[width=0.4\textwidth]{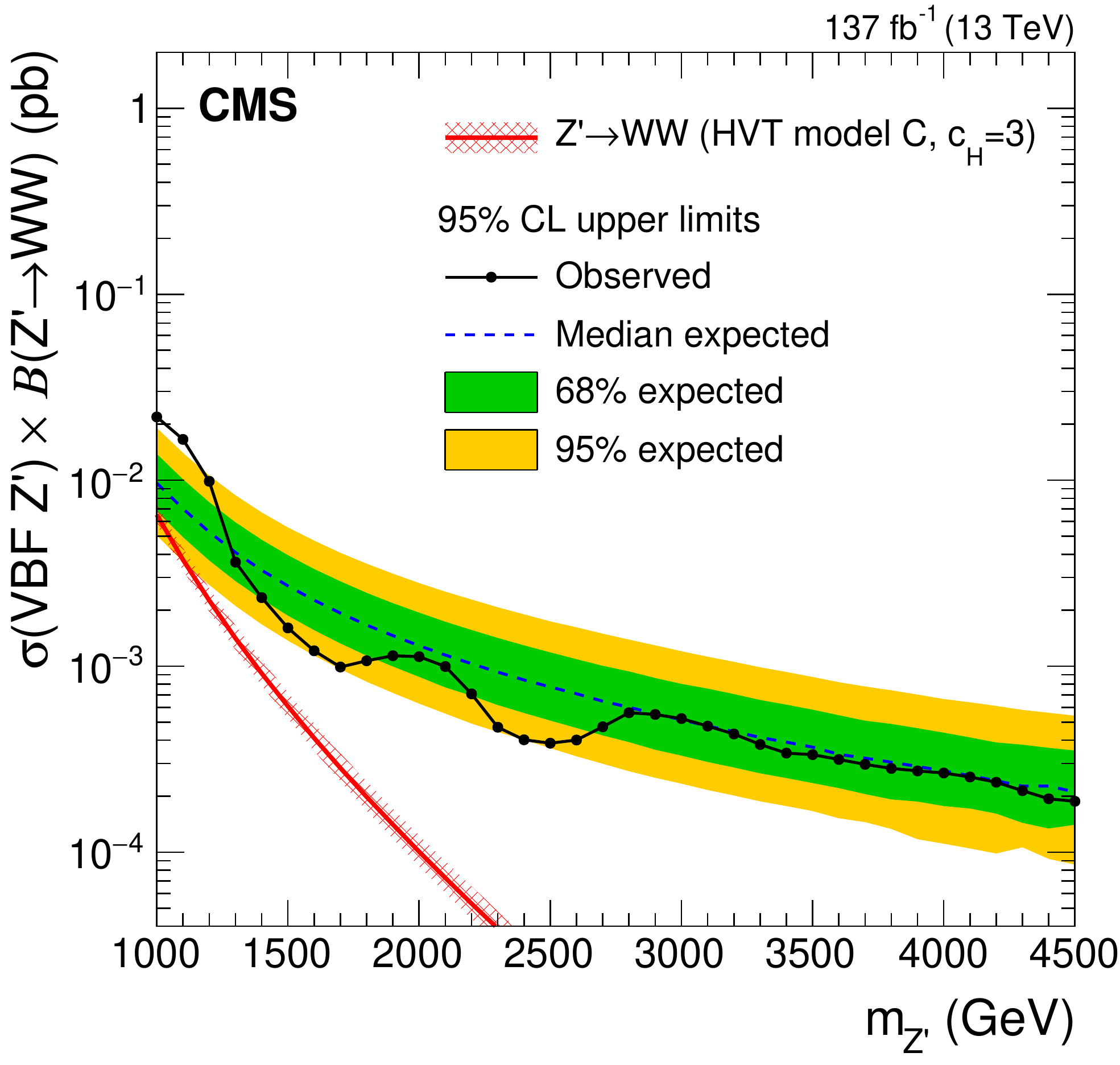}
 \includegraphics[width=0.4\textwidth]{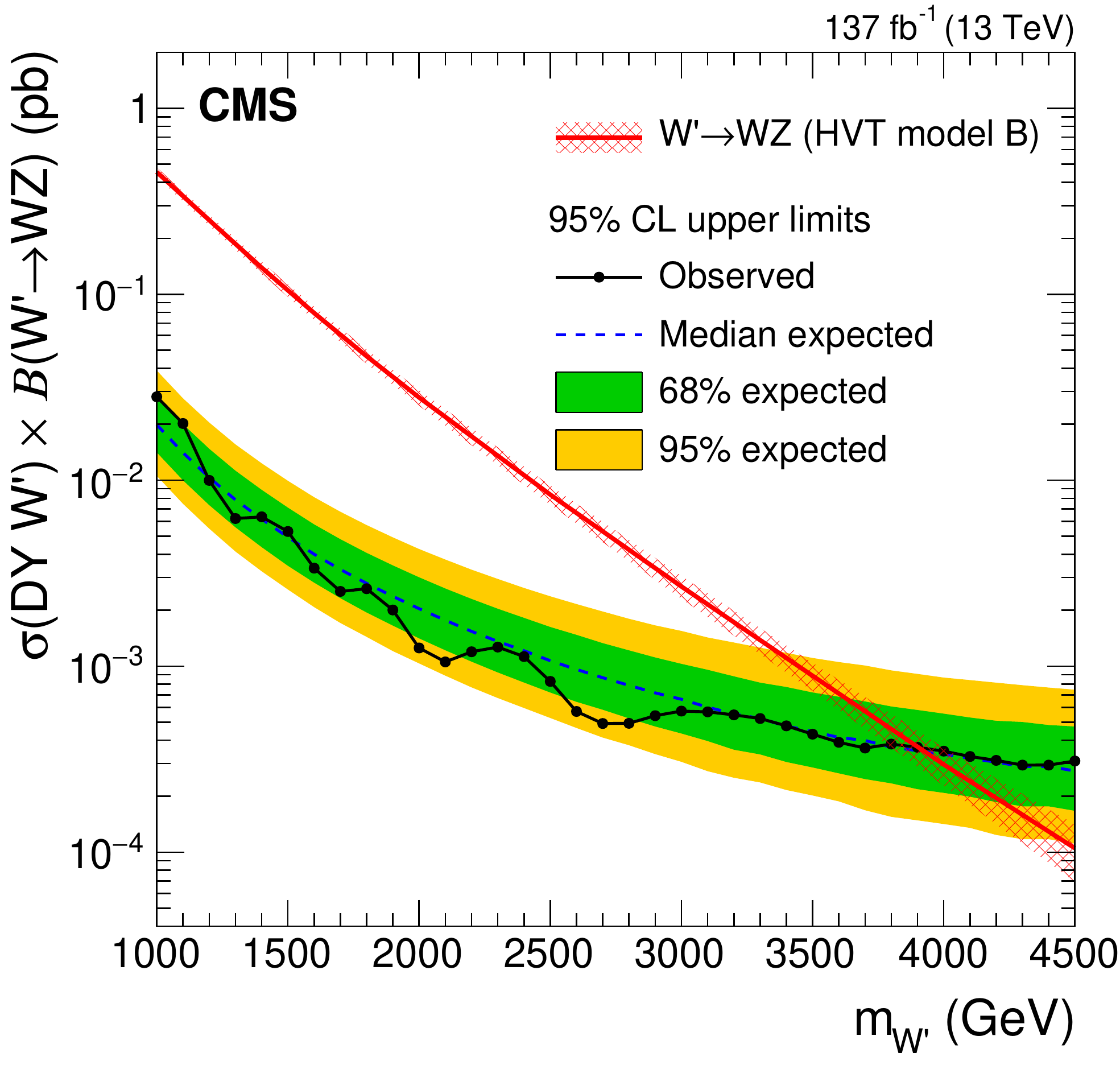}
 \includegraphics[width=0.4\textwidth]{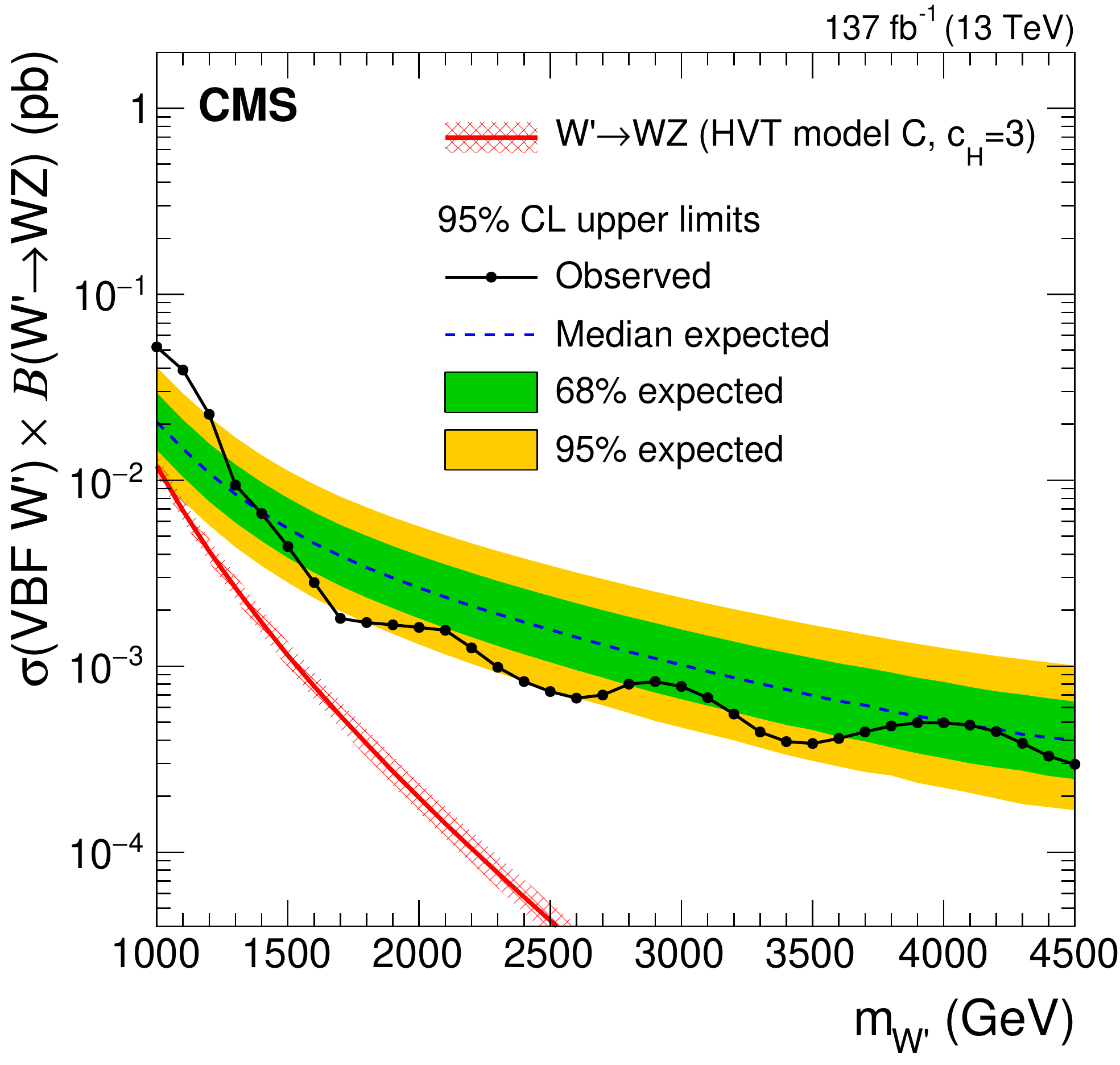}
 \includegraphics[width=0.4\textwidth]{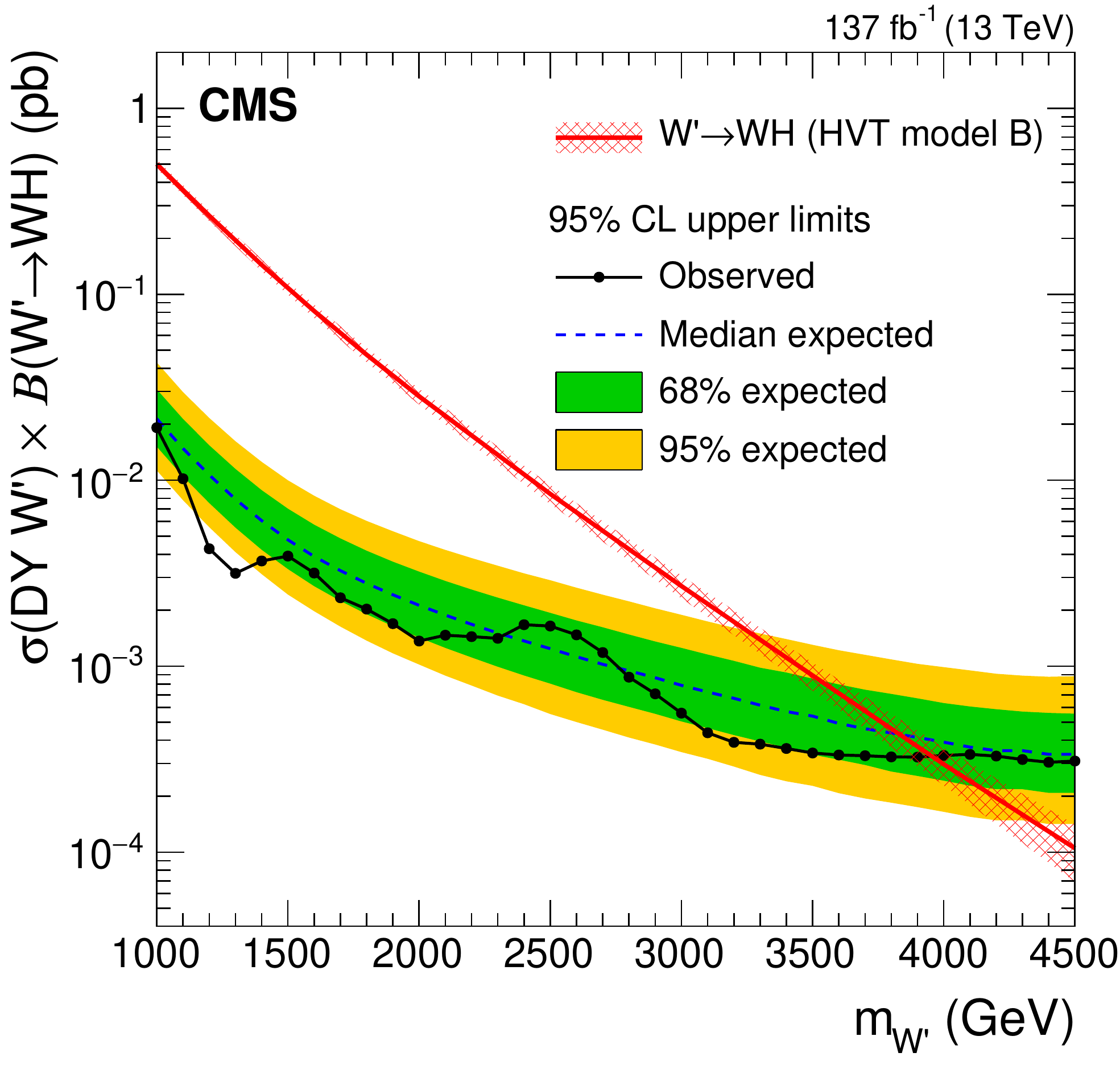}
 \caption{Exclusion limits on the product of the production cross section and the branching fraction for a new neutral spin-1 resonance produced via \qqbar annihilation (upper left) or vector boson fusion (upper right) and decaying to \WW, for a new charged spin-1 resonance produced via \qqbar annihilation (center left) or vector boson fusion (center right) and decaying to \WZ, and for a new charged spin-1 resonance produced via \qqbar annihilation and decaying to \WH (lower), as functions of the resonance mass hypothesis, compared with the predicted cross sections for a \PWpr or \PZpr from HVT model B (for DY) or HVT model C with $\cH=3$ (for VBF). Signal cross section uncertainties are shown as red cross-hatched bands.}
 \label{fig:exclusion_limits_spin1}
\end{figure*}

Figures~\ref{fig:exclusion_limits_spin2},~\ref{fig:exclusion_limits_spin0}, and~\ref{fig:exclusion_limits_spin1} show the upper exclusion limits on the product of the resonance production cross section and the branching fraction to pairs of bosons, as functions of the resonance mass, for spin-2, spin-0, and spin-1 signal models, respectively.
For ggF-produced \GbuToWW resonances and for \WprToWZ resonances from HVT model B, the median expected limits are more stringent than those presented in Ref.~\cite{Sirunyan:2018iff} by a factor of 4 to 5, benefiting from both the larger data sample, the improved analysis techniques, and the new event categories based on \nsubjDDT and \Dy.

By comparing the observed limits to the expected cross sections from theoretical calculations, mass exclusion limits can be set for resonances produced via ggF and DY.
For spin-0 resonances decaying to \WW,
ggF-produced bulk radions with masses below \RadToWWexcl
are excluded at 95\% \CLnp.
For the spin-1 resonances of HVT model B,
DY-produced \ZprToWW resonances lighter than \ZprToWWexcl, 
\WprToWZ resonances lighter than \WprToWZexcl, 
and \WprToWH resonances lighter than \WprToWHexcl  
are excluded at 95\% \CLnp.
For spin-2 resonances decaying to \WW,
ggF-produced bulk gravitons with masses below \GbuToWWexcl are excluded at 95\% \CLnp.
For resonances produced only via VBF, the present data do not yet have sensitivity to exclude resonance masses for the benchmark scenarios and mass range under study.

\clearpage

\section{Summary}
\label{sec:Summary}

A search for new narrow heavy resonances with mass larger than 1\TeV and decaying to \WW, \WZ, or \WH boson pairs is performed using proton-proton collision events at $\sqrt{s}=13\TeV$ containing one high-\pt electron or muon, large missing transverse momentum, and a massive large-radius jet. 
The data were collected with the CMS detector at the LHC in 2016--2018 and correspond to an integrated luminosity of \LUMIFULL.
The signal extraction strategy is structured around a two-dimensional maximum-likelihood fit to the distributions of the diboson reconstructed mass and the soft-drop jet mass.
The sensitivity to different final states and production mechanisms is enhanced by the use of event categories that exploit the mass-decorrelated $N$-subjettiness ratio, the \doubleB tagger, the presence of a pair of forward jets compatible with vector boson fusion production, and the difference in rapidity between the reconstructed bosons. 
No significant excess is found, and the results are interpreted in terms of upper limits on the production cross section of new narrow resonances in several benchmark models. Spin-2 ggF-produced bulk gravitons with masses below \GbuToWWexcl and decaying to \WW are excluded at 95\% \CLnp. Spin-1 DY-produced \ZprToWW resonances lighter than \ZprToWWexcl, \WprToWZ resonances lighter than \WprToWZexcl, and \WprToWH resonances lighter than \WprToWHexcl in the context of HVT model B are excluded at 95\% \CLnp. Spin-0 ggF-produced bulk radions with masses below \RadToWWexcl,  decaying to \WW, are excluded at 95\% \CLnp. Finally, for particles produced exclusively by vector boson fusion, the present data do not yet have sensitivity to exclude the benchmark scenarios under study. The reported limits, also provided in tabulated form in the HEPData record~\cite{hepdata} for this analysis, are generally relevant for any narrow heavy resonance with a given spin produced by gluon fusion, \qqbar annihilation, or vector boson fusion. The excluded cross section values set the most stringent experimental bounds to date.

\begin{acknowledgments}
  We congratulate our colleagues in the CERN accelerator departments for the excellent performance of the LHC and thank the technical and administrative staffs at CERN and at other CMS institutes for their contributions to the success of the CMS effort. In addition, we gratefully acknowledge the computing centers and personnel of the Worldwide LHC Computing Grid and other centers for delivering so effectively the computing infrastructure essential to our analyses. Finally, we acknowledge the enduring support for the construction and operation of the LHC, the CMS detector, and the supporting computing infrastructure provided by the following funding agencies: BMBWF and FWF (Austria); FNRS and FWO (Belgium); CNPq, CAPES, FAPERJ, FAPERGS, and FAPESP (Brazil); MES and BNSF (Bulgaria); CERN; CAS, MoST, and NSFC (China); MINCIENCIAS (Colombia); MSES and CSF (Croatia); RIF (Cyprus); SENESCYT (Ecuador); MoER, ERC PUT and ERDF (Estonia); Academy of Finland, MEC, and HIP (Finland); CEA and CNRS/IN2P3 (France); BMBF, DFG, and HGF (Germany); GSRI (Greece); NKFIA (Hungary); DAE and DST (India); IPM (Iran); SFI (Ireland); INFN (Italy); MSIP and NRF (Republic of Korea); MES (Latvia); LAS (Lithuania); MOE and UM (Malaysia); BUAP, CINVESTAV, CONACYT, LNS, SEP, and UASLP-FAI (Mexico); MOS (Montenegro); MBIE (New Zealand); PAEC (Pakistan); MSHE and NSC (Poland); FCT (Portugal); JINR (Dubna); MON, RosAtom, RAS, RFBR, and NRC KI (Russia); MESTD (Serbia); SEIDI, CPAN, PCTI, and FEDER (Spain); MOSTR (Sri Lanka); Swiss Funding Agencies (Switzerland); MST (Taipei); ThEPCenter, IPST, STAR, and NSTDA (Thailand); TUBITAK and TAEK (Turkey); NASU (Ukraine); STFC (United Kingdom); DOE and NSF (USA).
  
  \hyphenation{Rachada-pisek} Individuals have received support from the Marie-Curie program and the European Research Council and Horizon 2020 Grant, contract Nos.\ 675440, 724704, 752730, 758316, 765710, 824093, 884104, and COST Action CA16108 (European Union); the Leventis Foundation; the Alfred P.\ Sloan Foundation; the Alexander von Humboldt Foundation; the Belgian Federal Science Policy Office; the Fonds pour la Formation \`a la Recherche dans l'Industrie et dans l'Agriculture (FRIA-Belgium); the Agentschap voor Innovatie door Wetenschap en Technologie (IWT-Belgium); the F.R.S.-FNRS and FWO (Belgium) under the ``Excellence of Science -- EOS" -- be.h project n.\ 30820817; the Beijing Municipal Science \& Technology Commission, No. Z191100007219010; the Ministry of Education, Youth and Sports (MEYS) of the Czech Republic; the Deutsche Forschungsgemeinschaft (DFG), under Germany's Excellence Strategy -- EXC 2121 ``Quantum Universe" -- 390833306, and under project number 400140256 - GRK2497; the Lend\"ulet (``Momentum") Program and the J\'anos Bolyai Research Scholarship of the Hungarian Academy of Sciences, the New National Excellence Program \'UNKP, the NKFIA research grants 123842, 123959, 124845, 124850, 125105, 128713, 128786, and 129058 (Hungary); the Council of Science and Industrial Research, India; the Latvian Council of Science; the Ministry of Science and Higher Education and the National Science Center, contracts Opus 2014/15/B/ST2/03998 and 2015/19/B/ST2/02861 (Poland); the Funda\c{c}\~ao para a Ci\^encia e a Tecnologia, grant CEECIND/01334/2018 (Portugal); the National Priorities Research Program by Qatar National Research Fund; the Ministry of Science and Higher Education, project no. 14.W03.31.0026 (Russia); the Programa Estatal de Fomento de la Investigaci{\'o}n Cient{\'i}fica y T{\'e}cnica de Excelencia Mar\'{\i}a de Maeztu, grant MDM-2015-0509 and the Programa Severo Ochoa del Principado de Asturias; the Stavros Niarchos Foundation (Greece); the Rachadapisek Sompot Fund for Postdoctoral Fellowship, Chulalongkorn University and the Chulalongkorn Academic into Its 2nd Century Project Advancement Project (Thailand); the Kavli Foundation; the Nvidia Corporation; the SuperMicro Corporation; the Welch Foundation, contract C-1845; and the Weston Havens Foundation (USA).

\end{acknowledgments}

\bibliography{auto_generated}

\cleardoublepage \appendix\section{The CMS Collaboration \label{app:collab}}\begin{sloppypar}\hyphenpenalty=5000\widowpenalty=500\clubpenalty=5000\input{B2G-19-002-authorlist.tex}\end{sloppypar}
\end{document}

%% file: B2G-19-002-authorlist.tex
\vskip\cmsinstskip
\textbf{Yerevan Physics Institute, Yerevan, Armenia}\\*[0pt]
A.~Tumasyan
\vskip\cmsinstskip
\textbf{Institut f\"{u}r Hochenergiephysik, Vienna, Austria}\\*[0pt]
W.~Adam, J.W.~Andrejkovic, T.~Bergauer, S.~Chatterjee, M.~Dragicevic, A.~Escalante~Del~Valle, R.~Fr\"{u}hwirth\cmsAuthorMark{1}, M.~Jeitler\cmsAuthorMark{1}, N.~Krammer, L.~Lechner, D.~Liko, I.~Mikulec, P.~Paulitsch, F.M.~Pitters, J.~Schieck\cmsAuthorMark{1}, R.~Sch\"{o}fbeck, M.~Spanring, S.~Templ, W.~Waltenberger, C.-E.~Wulz\cmsAuthorMark{1}
% ...
\vskip\cmsinstskip
% ...
1:  Also at TU Wien, Wien, Austria\\
% ...


%% file: B2G-19-002_temp.bbl
\providecommand{\href}[2]{#2}\begingroup\raggedright\begin{thebibliography}{10}%
\makeatletter
\providecommand{\hrefCMSnoop }[0]{\@secondoftwo}%
\makeatother
\providecommand{\doi}{\texttt{doi:}\begingroup \urlstyle{tt}\Url}

\bibitem{Glashow:1961tr}
\hrefCMSnoop {}{S.~L. Glashow, ``Partial symmetries of weak interactions'',}
  \textit{ Nucl. Phys.} \textbf{ 22} (1961) 579,
\href{http://dx.doi.org/10.1016/0029-5582(61)90469-2}{\doi{10.1016/0029-5582(61)90469-2}}.

\bibitem{Salam:1964ry}
\hrefCMSnoop {}{A.~Salam and J.~C. Ward, ``Electromagnetic and weak
  interactions'',} \textit{ Phys. Lett.} \textbf{ 13} (1964) 168,
\href{http://dx.doi.org/10.1016/0031-9163(64)90711-5}{\doi{10.1016/0031-9163(64)90711-5}}.

\bibitem{Weinberg:1967tq}
\hrefCMSnoop {}{S.~Weinberg, ``A model of leptons'',} \textit{ Phys. Rev.
  Lett.} \textbf{ 19} (1967) 1264,
\href{http://dx.doi.org/10.1103/PhysRevLett.19.1264}{\doi{10.1103/PhysRevLett.19.1264}}.

\bibitem{Aad:2012tfa}
\hrefCMSnoop {}{{ATLAS Collaboration}, ``Observation of a new particle in the
  search for the standard model {Higgs} boson with the {ATLAS} detector at the
  {LHC}'',} \textit{ Phys. Lett. B} \textbf{ 716} (2012) 1,
  \href{http://dx.doi.org/10.1016/j.physletb.2012.08.020}{\doi{10.1016/j.physletb.2012.08.020}},
\href{http://www.arXiv.org/abs/1207.7214}{\texttt{arXiv:1207.7214}}.

\bibitem{Chatrchyan:2012xdj}
\hrefCMSnoop {}{{CMS Collaboration}, ``Observation of a new boson at a mass of
  125\gev with the {CMS} experiment at the {LHC}'',} \textit{ Phys. Lett. B}
  \textbf{ 716} (2012) 30,
  \href{http://dx.doi.org/10.1016/j.physletb.2012.08.021}{\doi{10.1016/j.physletb.2012.08.021}},
\href{http://www.arXiv.org/abs/1207.7235}{\texttt{arXiv:1207.7235}}.

\bibitem{Chatrchyan:2013lba}
\hrefCMSnoop {}{{CMS Collaboration}, ``Observation of a new boson with mass
  near 125\gev in pp collisions at $\sqrt{s} = 7$ and 8 {TeV}'',} \textit{
  JHEP} \textbf{ 06} (2013) 081,
  \href{http://dx.doi.org/10.1007/JHEP06(2013)081}{\doi{10.1007/JHEP06(2013)081}},
\href{http://www.arXiv.org/abs/1303.4571}{\texttt{arXiv:1303.4571}}.

\bibitem{Randall:1999ee}
\hrefCMSnoop {}{L.~Randall and R.~Sundrum, ``A large mass hierarchy from a
  small extra dimension'',} \textit{ Phys. Rev. Lett.} \textbf{ 83} (1999)
  3370,
  \href{http://dx.doi.org/10.1103/PhysRevLett.83.3370}{\doi{10.1103/PhysRevLett.83.3370}},
\href{http://www.arXiv.org/abs/hep-ph/9905221}{\texttt{arXiv:hep-ph/9905221}}.

\bibitem{Randall:1999vf}
\hrefCMSnoop {}{L.~Randall and R.~Sundrum, ``An alternative to
  compactification'',} \textit{ Phys. Rev. Lett.} \textbf{ 83} (1999) 4690,
  \href{http://dx.doi.org/10.1103/PhysRevLett.83.4690}{\doi{10.1103/PhysRevLett.83.4690}},
\href{http://www.arXiv.org/abs/hep-th/9906064}{\texttt{arXiv:hep-th/9906064}}.

\bibitem{Pappadopulo:2014qza}
\hrefCMSnoop {}{D.~Pappadopulo, A.~Thamm, R.~Torre, and A.~Wulzer, ``Heavy
  vector triplets: Bridging theory and data'',} \textit{ JHEP} \textbf{ 09}
  (2014) 060,
  \href{http://dx.doi.org/10.1007/JHEP09(2014)060}{\doi{10.1007/JHEP09(2014)060}},
\href{http://www.arXiv.org/abs/1402.4431}{\texttt{arXiv:1402.4431}}.

\bibitem{Bellazzini:2014yua}
\hrefCMSnoop {}{B.~Bellazzini, C.~Cs{\'a}ki, and J.~Serra, ``Composite
  {Higgses}'',} \textit{ Eur. Phys. J. C} \textbf{ 74} (2014) 2766,
  \href{http://dx.doi.org/10.1140/epjc/s10052-014-2766-x}{\doi{10.1140/epjc/s10052-014-2766-x}},
\href{http://www.arXiv.org/abs/1401.2457}{\texttt{arXiv:1401.2457}}.

\bibitem{Contino:2011np}
\hrefCMSnoop {}{R.~Contino, D.~Marzocca, D.~Pappadopulo, and R.~Rattazzi, ``On
  the effect of resonances in composite {Higgs} phenomenology'',} \textit{
  JHEP} \textbf{ 10} (2011) 081,
  \href{http://dx.doi.org/10.1007/JHEP10(2011)081}{\doi{10.1007/JHEP10(2011)081}},
\href{http://www.arXiv.org/abs/1109.1570}{\texttt{arXiv:1109.1570}}.

\bibitem{Marzocca:2012zn}
\hrefCMSnoop {}{D.~Marzocca, M.~Serone, and J.~Shu, ``General composite {Higgs}
  models'',} \textit{ JHEP} \textbf{ 08} (2012) 013,
  \href{http://dx.doi.org/10.1007/JHEP08(2012)013}{\doi{10.1007/JHEP08(2012)013}},
\href{http://www.arXiv.org/abs/1205.0770}{\texttt{arXiv:1205.0770}}.

\bibitem{Greco:2014aza}
\hrefCMSnoop {}{D.~Greco and D.~Liu, ``Hunting composite vector resonances at
  the {LHC}: naturalness facing data'',} \textit{ JHEP} \textbf{ 12} (2014)
  126,
  \href{http://dx.doi.org/10.1007/JHEP12(2014)126}{\doi{10.1007/JHEP12(2014)126}},
\href{http://www.arXiv.org/abs/1410.2883}{\texttt{arXiv:1410.2883}}.

\bibitem{Lane:2016kvg}
\hrefCMSnoop {}{K.~Lane and L.~Pritchett, ``The light composite {Higgs} boson
  in strong extended technicolor'',} \textit{ JHEP} \textbf{ 06} (2017) 140,
  \href{http://dx.doi.org/10.1007/JHEP06(2017)140}{\doi{10.1007/JHEP06(2017)140}},
\href{http://www.arXiv.org/abs/1604.07085}{\texttt{arXiv:1604.07085}}.

\bibitem{Schmaltz:2005ky}
\hrefCMSnoop {}{M.~Schmaltz and D.~Tucker-Smith, ``{Little Higgs review}'',}
  \textit{ Ann. Rev. Nucl. Part. Sci.} \textbf{ 55} (2005) 229,
  \href{http://dx.doi.org/10.1146/annurev.nucl.55.090704.151502}{\doi{10.1146/annurev.nucl.55.090704.151502}},
\href{http://www.arXiv.org/abs/hep-ph/0502182}{\texttt{arXiv:hep-ph/0502182}}.

\bibitem{ArkaniHamed:2002qy}
\hrefCMSnoop {}{N.~Arkani-Hamed, A.~G. Cohen, E.~Katz, and A.~E. Nelson, ``The
  littlest {Higgs}'',} \textit{ JHEP} \textbf{ 07} (2002) 034,
  \href{http://dx.doi.org/10.1088/1126-6708/2002/07/034}{\doi{10.1088/1126-6708/2002/07/034}},
\href{http://www.arXiv.org/abs/hep-ph/0206021}{\texttt{arXiv:hep-ph/0206021}}.

\bibitem{Aad:2015ufa}
\hrefCMSnoop {}{{ATLAS Collaboration}, ``Search for production of {\WW}/{\WZ}
  resonances decaying to a lepton, neutrino and jets in pp collisions at
  $\sqrt{s}=8{\TeV}$ with the {ATLAS} detector'',} \textit{ Eur. Phys. J. C}
  \textbf{ 75} (2015) 209,
  \href{http://dx.doi.org/10.1140/epjc/s10052-015-3425-6}{\doi{10.1140/epjc/s10052-015-3425-6}},
  \href{http://www.arXiv.org/abs/1503.04677}{\texttt{arXiv:1503.04677}}.
[Erratum: \DOI{10.1140/epjc/s10052-015-3593-4}].

\bibitem{Aaboud:2016okv}
\hrefCMSnoop {}{{ATLAS Collaboration}, ``Searches for heavy diboson resonances
  in pp collisions at $\sqrt{s}=13{\TeV}$ with the {ATLAS} detector'',}
  \textit{ JHEP} \textbf{ 09} (2016) 173,
  \href{http://dx.doi.org/10.1007/JHEP09(2016)173}{\doi{10.1007/JHEP09(2016)173}},
\href{http://www.arXiv.org/abs/1606.04833}{\texttt{arXiv:1606.04833}}.

\bibitem{Aaboud:2017fgj}
\hrefCMSnoop {}{{ATLAS Collaboration}, ``Search for {\WW}/{\WZ} resonance
  production in $\ell{\Pgn}{\cPq}{\cPq}$ final states in pp collisions at
  $\sqrt{s} = 13{\TeV}$ with the {ATLAS} detector'',} \textit{ JHEP} \textbf{
  03} (2018) 042,
  \href{http://dx.doi.org/10.1007/JHEP03(2018)042}{\doi{10.1007/JHEP03(2018)042}},
\href{http://www.arXiv.org/abs/1710.07235}{\texttt{arXiv:1710.07235}}.

\bibitem{Khachatryan:2014gha}
\hrefCMSnoop {}{{CMS Collaboration}, ``Search for massive resonances decaying
  into pairs of boosted bosons in semi-leptonic final states at $\sqrt{s} =
  8{\TeV}$'',} \textit{ JHEP} \textbf{ 08} (2014) 174,
  \href{http://dx.doi.org/10.1007/JHEP08(2014)174}{\doi{10.1007/JHEP08(2014)174}},
\href{http://www.arXiv.org/abs/1405.3447}{\texttt{arXiv:1405.3447}}.

\bibitem{Sirunyan:2016cao}
\hrefCMSnoop {}{{CMS Collaboration}, ``Search for massive resonances decaying
  into {\WW}, {\WZ} or {\PZ}{\PZ} bosons in proton-proton collisions at
  $\sqrt{s} = 13{\TeV}$'',} \textit{ JHEP} \textbf{ 03} (2017) 162,
  \href{http://dx.doi.org/10.1007/JHEP03(2017)162}{\doi{10.1007/JHEP03(2017)162}},
\href{http://www.arXiv.org/abs/1612.09159}{\texttt{arXiv:1612.09159}}.

\bibitem{Sirunyan:2018iff}
\hrefCMSnoop {}{{CMS Collaboration}, ``{Search for a heavy resonance decaying
  to a pair of vector bosons in the lepton plus merged jet final state at $
  \sqrt{s}=13 $ TeV}'',} \textit{ JHEP} \textbf{ 05} (2018) 088,
  \href{http://dx.doi.org/10.1007/JHEP05(2018)088}{\doi{10.1007/JHEP05(2018)088}},
\href{http://www.arXiv.org/abs/1802.09407}{\texttt{arXiv:1802.09407}}.

\bibitem{ATLAS:2020fry}
\hrefCMSnoop {}{{ATLAS Collaboration}, ``{Search for heavy diboson resonances
  in semileptonic final states in pp collisions at $\sqrt{s}=13$ TeV with the
  ATLAS detector}'',} \textit{ Eur. Phys. J. C} \textbf{ 80} (2020), no.~12,
  1165,
  \href{http://dx.doi.org/10.1140/epjc/s10052-020-08554-y}{\doi{10.1140/epjc/s10052-020-08554-y}},
  \href{http://www.arXiv.org/abs/2004.14636}{\texttt{arXiv:2004.14636}}.

\bibitem{Aad:2015yza}
\hrefCMSnoop {}{{ATLAS Collaboration}, ``Search for a new resonance decaying to
  a {\PW} or {\PZ} boson and a {Higgs} boson in the $\ell\ell / \ell {\Pgn} /
  {\Pgn\Pgn} + \bbbar{}$ final states with the {ATLAS} detector'',} \textit{
  Eur. Phys. J. C} \textbf{ 75} (2015) 263,
  \href{http://dx.doi.org/10.1140/epjc/s10052-015-3474-x}{\doi{10.1140/epjc/s10052-015-3474-x}},
\href{http://www.arXiv.org/abs/1503.08089}{\texttt{arXiv:1503.08089}}.

\bibitem{Aaboud:2016lwx}
\hrefCMSnoop {}{{ATLAS Collaboration}, ``Search for new resonances decaying to
  a {\PW} or {\PZ} boson and a {Higgs} boson in the $\ell^+ \ell^- \bbbar{}$,
  $\ell {\Pgn}\bbbar{}$, and ${\Pgn\Pagn} \bbbar{}$ channels with pp collisions
  at $\sqrt s = 13{\TeV}$ with the {ATLAS} detector'',} \textit{ Phys. Lett. B}
  \textbf{ 765} (2017) 32,
  \href{http://dx.doi.org/10.1016/j.physletb.2016.11.045}{\doi{10.1016/j.physletb.2016.11.045}},
\href{http://www.arXiv.org/abs/1607.05621}{\texttt{arXiv:1607.05621}}.

\bibitem{Aaboud:2017cxo}
\hrefCMSnoop {}{{ATLAS Collaboration}, ``{Search for heavy resonances decaying
  into a $W$ or $Z$ boson and a Higgs boson in final states with leptons and
  $b$-jets in 36 fb$^{-1}$ of $\sqrt s = 13$ TeV $pp$ collisions with the
  {ATLAS} detector}'',} \textit{ JHEP} \textbf{ 03} (2018) 174,
  \href{http://dx.doi.org/10.1007/JHEP03(2018)174}{\doi{10.1007/JHEP03(2018)174}},
\href{http://www.arXiv.org/abs/1712.06518}{\texttt{arXiv:1712.06518}}.

\bibitem{Khachatryan:2016yji}
\hrefCMSnoop {}{{CMS Collaboration}, ``Search for massive {\PW}{\PH} resonances
  decaying into the $\ell${\Pgn}\bbbar{} final state at $\sqrt{s}=8{\TeV}$'',}
  \textit{ Eur. Phys. J. C} \textbf{ 76} (2016) 237,
  \href{http://dx.doi.org/10.1140/epjc/s10052-016-4067-z}{\doi{10.1140/epjc/s10052-016-4067-z}},
\href{http://www.arXiv.org/abs/1601.06431}{\texttt{arXiv:1601.06431}}.

\bibitem{Khachatryan:2016cfx}
\hrefCMSnoop {}{{CMS Collaboration}, ``Search for heavy resonances decaying
  into a vector boson and a {Higgs} boson in final states with charged leptons,
  neutrinos, and {\cPqb} quarks'',} \textit{ Phys. Lett. B} \textbf{ 768}
  (2017) 137,
  \href{http://dx.doi.org/10.1016/j.physletb.2017.02.040}{\doi{10.1016/j.physletb.2017.02.040}},
\href{http://www.arXiv.org/abs/1610.08066}{\texttt{arXiv:1610.08066}}.

\bibitem{Sirunyan:2018qob}
\hrefCMSnoop {}{{CMS Collaboration}, ``{Search for heavy resonances decaying
  into a vector boson and a Higgs boson in final states with charged leptons,
  neutrinos and b quarks at $ \sqrt{s}=13 $ TeV}'',} \textit{ JHEP} \textbf{
  11} (2018) 172,
  \href{http://dx.doi.org/10.1007/JHEP11(2018)172}{\doi{10.1007/JHEP11(2018)172}},
\href{http://www.arXiv.org/abs/1807.02826}{\texttt{arXiv:1807.02826}}.

\bibitem{Chatrchyan:2008zzk}
\hrefCMSnoop {}{{CMS Collaboration}, ``The {CMS} experiment at the {CERN}
  {LHC}'',} \textit{ JINST} \textbf{ 3} (2008) S08004,
  \href{http://dx.doi.org/10.1088/1748-0221/3/08/S08004}{\doi{10.1088/1748-0221/3/08/S08004}}.

\bibitem{CMS-PRF-14-001}
\hrefCMSnoop {}{{CMS Collaboration}, ``Particle-flow reconstruction and global
  event description with the {CMS} detector'',} \textit{ JINST} \textbf{ 12}
  (2017) P10003,
  \href{http://dx.doi.org/10.1088/1748-0221/12/10/P10003}{\doi{10.1088/1748-0221/12/10/P10003}},
\href{http://www.arXiv.org/abs/1706.04965}{\texttt{arXiv:1706.04965}}.

\bibitem{Sirunyan:2019kia}
\hrefCMSnoop {}{{CMS Collaboration}, ``Performance of missing transverse
  momentum reconstruction in proton-proton collisions at $\sqrt{s} = 13$\,{TeV}
  using the {CMS} detector'',} \textit{ JINST} \textbf{ 14} (2019) P07004,
  \href{http://dx.doi.org/10.1088/1748-0221/14/07/P07004}{\doi{10.1088/1748-0221/14/07/P07004}},
\href{http://www.arXiv.org/abs/1903.06078}{\texttt{arXiv:1903.06078}}.

\bibitem{Cacciari:2008gp}
\hrefCMSnoop {}{M.~Cacciari, G.~P. Salam, and G.~Soyez, ``{The anti-\kt jet
  clustering algorithm}'',} \textit{ JHEP} \textbf{ 04} (2008) 063,
  \href{http://dx.doi.org/10.1088/1126-6708/2008/04/063}{\doi{10.1088/1126-6708/2008/04/063}},
  \href{http://www.arXiv.org/abs/0802.1189}{\texttt{arXiv:0802.1189}}.

\bibitem{Cacciari:2011ma}
\hrefCMSnoop {}{M.~Cacciari, G.~P. Salam, and G.~Soyez, ``{FastJet user
  manual}'',} \textit{ Eur. Phys. J. C} \textbf{ 72} (2012) 1896,
  \href{http://dx.doi.org/10.1140/epjc/s10052-012-1896-2}{\doi{10.1140/epjc/s10052-012-1896-2}},
\href{http://www.arXiv.org/abs/1111.6097}{\texttt{arXiv:1111.6097}}.

\bibitem{Khachatryan:2016kdb}
\hrefCMSnoop {}{{CMS Collaboration}, ``{Jet energy scale and resolution in the
  CMS experiment in pp collisions at 8 TeV}'',} \textit{ JINST} \textbf{ 12}
  (2017) P02014,
  \href{http://dx.doi.org/10.1088/1748-0221/12/02/P02014}{\doi{10.1088/1748-0221/12/02/P02014}},
\href{http://www.arXiv.org/abs/1607.03663}{\texttt{arXiv:1607.03663}}.

\bibitem{Sirunyan:2020foa}
\hrefCMSnoop {}{{CMS Collaboration}, ``{Pileup mitigation at CMS in 13 TeV
  data}'',} \textit{ JINST} \textbf{ 15} (2020) P09018,
  \href{http://dx.doi.org/10.1088/1748-0221/15/09/p09018}{\doi{10.1088/1748-0221/15/09/p09018}},
  \href{http://www.arXiv.org/abs/2003.00503}{\texttt{arXiv:2003.00503}}.

\bibitem{Bertolini:2014bba}
\hrefCMSnoop {}{D.~Bertolini, P.~Harris, M.~Low, and N.~Tran, ``{Pileup per
  particle identification}'',} \textit{ JHEP} \textbf{ 10} (2014) 059,
  \href{http://dx.doi.org/10.1007/JHEP10(2014)059}{\doi{10.1007/JHEP10(2014)059}},
\href{http://www.arXiv.org/abs/1407.6013}{\texttt{arXiv:1407.6013}}.

\bibitem{Sirunyan:2020zal}
\hrefCMSnoop {}{{CMS Collaboration}, ``{Performance of the CMS Level-1 trigger
  in proton-proton collisions at $\sqrt{s} = 13$\,TeV}'',} \textit{ JINST}
  \textbf{ 15} (2020) P10017,
  \href{http://dx.doi.org/10.1088/1748-0221/15/10/P10017}{\doi{10.1088/1748-0221/15/10/P10017}},
  \href{http://www.arXiv.org/abs/2006.10165}{\texttt{arXiv:2006.10165}}.

\bibitem{Khachatryan:2016bia}
\hrefCMSnoop {}{{CMS Collaboration}, ``{The CMS trigger system}'',} \textit{
  JINST} \textbf{ 12} (2017) P01020,
  \href{http://dx.doi.org/10.1088/1748-0221/12/01/P01020}{\doi{10.1088/1748-0221/12/01/P01020}},
\href{http://www.arXiv.org/abs/1609.02366}{\texttt{arXiv:1609.02366}}.

\bibitem{Goldberger:1999uk}
\hrefCMSnoop {}{W.~D. Goldberger and M.~B. Wise, ``Modulus stabilization with
  bulk fields'',} \textit{ Phys. Rev. Lett.} \textbf{ 83} (1999) 4922,
  \href{http://dx.doi.org/10.1103/PhysRevLett.83.4922}{\doi{10.1103/PhysRevLett.83.4922}},
\href{http://www.arXiv.org/abs/hep-ph/9907447}{\texttt{arXiv:hep-ph/9907447}}.

\bibitem{Csaki:1999mp}
\hrefCMSnoop {}{C.~Csaki, M.~Graesser, L.~Randall, and J.~Terning, ``{Cosmology
  of brane models with radion stabilization}'',} \textit{ Phys. Rev. D}
  \textbf{ 62} (2000) 045015,
  \href{http://dx.doi.org/10.1103/PhysRevD.62.045015}{\doi{10.1103/PhysRevD.62.045015}},
\href{http://www.arXiv.org/abs/hep-ph/9911406}{\texttt{arXiv:hep-ph/9911406}}.

\bibitem{Csaki:2000zn}
\hrefCMSnoop {}{C.~Csaki, M.~L. Graesser, and G.~D. Kribs, ``{Radion dynamics
  and electroweak physics}'',} \textit{ Phys. Rev. D} \textbf{ 63} (2001)
  065002,
  \href{http://dx.doi.org/10.1103/PhysRevD.63.065002}{\doi{10.1103/PhysRevD.63.065002}},
\href{http://www.arXiv.org/abs/hep-th/0008151}{\texttt{arXiv:hep-th/0008151}}.

\bibitem{Agashe:2007zd}
\hrefCMSnoop {}{K.~Agashe, H.~Davoudiasl, G.~Perez, and A.~Soni, ``Warped
  gravitons at the {LHC} and beyond'',} \textit{ Phys. Rev. D} \textbf{ 76}
  (2007) 036006,
  \href{http://dx.doi.org/10.1103/PhysRevD.76.036006}{\doi{10.1103/PhysRevD.76.036006}},
\href{http://www.arXiv.org/abs/hep-ph/0701186}{\texttt{arXiv:hep-ph/0701186}}.

\bibitem{Fitzpatrick:2007qr}
\hrefCMSnoop {}{A.~L. Fitzpatrick, J.~Kaplan, L.~Randall, and L.-T. Wang,
  ``Searching for the {Kaluza}-{Klein} graviton in bulk {RS} models'',}
  \textit{ JHEP} \textbf{ 09} (2007) 013,
  \href{http://dx.doi.org/10.1088/1126-6708/2007/09/013}{\doi{10.1088/1126-6708/2007/09/013}},
\href{http://www.arXiv.org/abs/hep-ph/0701150}{\texttt{arXiv:hep-ph/0701150}}.

\bibitem{Antipin:2007pi}
\hrefCMSnoop {}{O.~Antipin, D.~Atwood, and A.~Soni, ``Search for {RS} gravitons
  via $\mathrm{W}_\mathrm{L}\mathrm{W}_\mathrm{L}$ decays'',} \textit{ Phys.
  Lett. B} \textbf{ 666} (2008) 155,
  \href{http://dx.doi.org/10.1016/j.physletb.2008.07.009}{\doi{10.1016/j.physletb.2008.07.009}},
\href{http://www.arXiv.org/abs/0711.3175}{\texttt{arXiv:0711.3175}}.

\bibitem{Oliveira:2014kla}
\hrefCMSnoop {}{A.~Oliveira, ``Gravity particles from warped extra dimensions,
  predictions for {LHC}'',} 2014.
\href{http://www.arXiv.org/abs/1404.0102}{\texttt{arXiv:1404.0102}}.

\bibitem{Alwall:2014hca}
J.~Alwall\hrefCMSnoop {}{ {et~al.}, ``The automated computation of tree-level
  and next-to-leading order differential cross sections, and their matching to
  parton shower simulations'',} \textit{ JHEP} \textbf{ 07} (2014) 079,
  \href{http://dx.doi.org/10.1007/JHEP07(2014)079}{\doi{10.1007/JHEP07(2014)079}},
\href{http://www.arXiv.org/abs/1405.0301}{\texttt{arXiv:1405.0301}}.

\bibitem{Nason:2004rx}
\hrefCMSnoop {}{P.~Nason, ``A new method for combining {NLO} {QCD} with shower
  {Monte} {Carlo} algorithms'',} \textit{ JHEP} \textbf{ 11} (2004) 040,
  \href{http://dx.doi.org/10.1088/1126-6708/2004/11/040}{\doi{10.1088/1126-6708/2004/11/040}},
\href{http://www.arXiv.org/abs/hep-ph/0409146}{\texttt{arXiv:hep-ph/0409146}}.

\bibitem{Frixione:2007vw}
\hrefCMSnoop {}{S.~Frixione, P.~Nason, and C.~Oleari, ``Matching {NLO} {QCD}
  computations with parton shower simulations: the {POWHEG} method'',} \textit{
  JHEP} \textbf{ 11} (2007) 070,
  \href{http://dx.doi.org/10.1088/1126-6708/2007/11/070}{\doi{10.1088/1126-6708/2007/11/070}},
\href{http://www.arXiv.org/abs/0709.2092}{\texttt{arXiv:0709.2092}}.

\bibitem{Alioli:2010xd}
\hrefCMSnoop {}{S.~Alioli, P.~Nason, C.~Oleari, and E.~Re, ``A general
  framework for implementing {NLO} calculations in shower {Monte} {Carlo}
  programs: the {POWHEG} {BOX}'',} \textit{ JHEP} \textbf{ 06} (2010) 043,
  \href{http://dx.doi.org/10.1007/JHEP06(2010)043}{\doi{10.1007/JHEP06(2010)043}},
\href{http://www.arXiv.org/abs/1002.2581}{\texttt{arXiv:1002.2581}}.

\bibitem{Alioli:2011as}
\hrefCMSnoop {}{S.~Alioli, S.-O. Moch, and P.~Uwer, ``Hadronic top-quark
  pair-production with one jet and parton showering'',} \textit{ JHEP} \textbf{
  01} (2012) 137,
  \href{http://dx.doi.org/10.1007/JHEP01(2012)137}{\doi{10.1007/JHEP01(2012)137}},
\href{http://www.arXiv.org/abs/1110.5251}{\texttt{arXiv:1110.5251}}.

\bibitem{Alioli:2009je}
\hrefCMSnoop {}{S.~Alioli, P.~Nason, C.~Oleari, and E.~Re, ``{NLO} single-top
  production matched with shower in {POWHEG}: $s$- and $t$-channel
  contributions'',} \textit{ JHEP} \textbf{ 09} (2009) 111,
  \href{http://dx.doi.org/10.1007/JHEP02(2010)011}{\doi{10.1007/JHEP02(2010)011}},
  \href{http://www.arXiv.org/abs/0907.4076}{\texttt{arXiv:0907.4076}}.
[Erratum: \DOI{10.1088/1126-6708/2009/09/111}].

\bibitem{Re:2010bp}
\hrefCMSnoop {}{E.~Re, ``Single-top {Wt-channel} production matched with parton
  showers using the {POWHEG} method'',} \textit{ Eur. Phys. J. C} \textbf{ 71}
  (2011) 1547,
  \href{http://dx.doi.org/10.1140/epjc/s10052-011-1547-z}{\doi{10.1140/epjc/s10052-011-1547-z}},
\href{http://www.arXiv.org/abs/1009.2450}{\texttt{arXiv:1009.2450}}.

\bibitem{Frederix:2012ps}
\hrefCMSnoop {}{R.~Frederix and S.~Frixione, ``Merging meets matching in
  {\MCATNLO}'',} \textit{ JHEP} \textbf{ 12} (2012) 061,
  \href{http://dx.doi.org/10.1007/JHEP12(2012)061}{\doi{10.1007/JHEP12(2012)061}},
\href{http://www.arXiv.org/abs/1209.6215}{\texttt{arXiv:1209.6215}}.

\bibitem{Nason:2013ydw}
\hrefCMSnoop {}{P.~Nason and G.~Zanderighi, ``{$\PW^+\PW^-$ , $\PW\PZ$ and
  $\PZ\PZ$ production in the {POWHEG-BOX-V2}}'',} \textit{ Eur. Phys. J. C}
  \textbf{ 74} (2014) 2702,
  \href{http://dx.doi.org/10.1140/epjc/s10052-013-2702-5}{\doi{10.1140/epjc/s10052-013-2702-5}},
\href{http://www.arXiv.org/abs/1311.1365}{\texttt{arXiv:1311.1365}}.

\bibitem{Sjostrand:2014zea}
T.~Sj{\"o}strand\hrefCMSnoop {}{ {et~al.}, ``{An introduction to PYTHIA
  8.2}'',} \textit{ Comput. Phys. Commun.} \textbf{ 191} (2015) 159,
  \href{http://dx.doi.org/10.1016/j.cpc.2015.01.024}{\doi{10.1016/j.cpc.2015.01.024}},
\href{http://www.arXiv.org/abs/1410.3012}{\texttt{arXiv:1410.3012}}.

\bibitem{Ball:2014uwa}
\hrefCMSnoop {}{{NNPDF} Collaboration, ``{Parton distributions for the LHC Run
  II}'',} \textit{ JHEP} \textbf{ 04} (2015) 040,
  \href{http://dx.doi.org/10.1007/JHEP04(2015)040}{\doi{10.1007/JHEP04(2015)040}},
\href{http://www.arXiv.org/abs/1410.8849}{\texttt{arXiv:1410.8849}}.

\bibitem{Khachatryan:2015pea}
\hrefCMSnoop {}{{CMS Collaboration}, ``Event generator tunes obtained from
  underlying event and multiparton scattering measurements'',} \textit{ Eur.
  Phys. J. C} \textbf{ 76} (2016) 155,
  \href{http://dx.doi.org/10.1140/epjc/s10052-016-3988-x}{\doi{10.1140/epjc/s10052-016-3988-x}},
\href{http://www.arXiv.org/abs/1512.00815}{\texttt{arXiv:1512.00815}}.

\bibitem{CMS-PAS-TOP-16-021}
\href {https://cds.cern.ch/record/2235192}{{CMS Collaboration},
  ``{Investigations of the impact of the parton shower tuning in Pythia 8 in
  the modelling of $\mathrm{t\overline{t}}$ at $\sqrt{s}=8$ and 13 TeV}'',} CMS
  Physics Analysis Summary CMS-PAS-TOP-16-021, 2016.

\bibitem{Ball:2017nwa}
\hrefCMSnoop {}{{NNPDF} Collaboration, ``Parton distributions from
  high-precision collider data'',} \textit{ Eur. Phys. J. C} \textbf{ 77}
  (2017) 663,
  \href{http://dx.doi.org/10.1140/epjc/s10052-017-5199-5}{\doi{10.1140/epjc/s10052-017-5199-5}},
\href{http://www.arXiv.org/abs/1706.00428}{\texttt{arXiv:1706.00428}}.

\bibitem{Sirunyan:2019dfx}
\hrefCMSnoop {}{{CMS Collaboration}, ``{Extraction and validation of a new set
  of {CMS} {PYTHIA8} tunes from underlying-event measurements}'',} \textit{
  Eur. Phys. J. C} \textbf{ 80} (2020) 4,
  \href{http://dx.doi.org/10.1140/epjc/s10052-019-7499-4}{\doi{10.1140/epjc/s10052-019-7499-4}},
\href{http://www.arXiv.org/abs/1903.12179}{\texttt{arXiv:1903.12179}}.

\bibitem{Agostinelli:2002hh}
\hrefCMSnoop {}{{GEANT4} Collaboration, ``{\GEANTfour} --- a simulation
  toolkit'',} \textit{ Nucl. Instrum. Meth. A} \textbf{ 506} (2003) 250,
\href{http://dx.doi.org/10.1016/S0168-9002(03)01368-8}{\doi{10.1016/S0168-9002(03)01368-8}}.

\bibitem{EGM-17-001}
\hrefCMSnoop {}{{CMS Collaboration}, ``{Electron and photon reconstruction and
  identification with the CMS experiment at the CERN LHC}'',} \textit{ JINST}
  \textbf{ 16} (2021) P05014,
  \href{http://dx.doi.org/10.1088/1748-0221/16/05/P05014}{\doi{10.1088/1748-0221/16/05/P05014}},
\href{http://www.arXiv.org/abs/2012.06888}{\texttt{arXiv:2012.06888}}.

\bibitem{MUO-17-001}
\hrefCMSnoop {}{{CMS Collaboration}, ``{Performance of the reconstruction and
  identification of high-momentum muons in proton-proton collisions at
  $\sqrt{s} = 13$\TeV}'',} \textit{ JINST} \textbf{ 15} (2020) P02027,
  \href{http://dx.doi.org/10.1088/1748-0221/15/02/P02027}{\doi{10.1088/1748-0221/15/02/P02027}},
\href{http://www.arXiv.org/abs/1912.03516}{\texttt{arXiv:1912.03516}}.

\bibitem{CMS-PAS-JME-16-003}
\href {https://cds.cern.ch/record/2256875}{{CMS Collaboration}, ``{Jet
  algorithms performance in 13{\TeV} data}'',} CMS Physics Analysis Summary
  CMS-PAS-JME-16-003, 2017.

\bibitem{Dasgupta:2013ihk}
\hrefCMSnoop {}{M.~Dasgupta, A.~Fregoso, S.~Marzani, and G.~P. Salam, ``Towards
  an understanding of jet substructure'',} \textit{ JHEP} \textbf{ 09} (2013)
  029,
  \href{http://dx.doi.org/10.1007/JHEP09(2013)029}{\doi{10.1007/JHEP09(2013)029}},
\href{http://www.arXiv.org/abs/1307.0007}{\texttt{arXiv:1307.0007}}.

\bibitem{Butterworth:2008iy}
\hrefCMSnoop {}{J.~M. Butterworth, A.~R. Davison, M.~Rubin, and G.~P. Salam,
  ``Jet substructure as a new {Higgs} search channel at the {LHC}'',} \textit{
  Phys. Rev. Lett.} \textbf{ 100} (2008) 242001,
  \href{http://dx.doi.org/10.1103/PhysRevLett.100.242001}{\doi{10.1103/PhysRevLett.100.242001}},
\href{http://www.arXiv.org/abs/0802.2470}{\texttt{arXiv:0802.2470}}.

\bibitem{Larkoski:2014wba}
\hrefCMSnoop {}{A.~J. Larkoski, S.~Marzani, G.~Soyez, and J.~Thaler, ``Soft
  drop'',} \textit{ JHEP} \textbf{ 05} (2014) 146,
  \href{http://dx.doi.org/10.1007/JHEP05(2014)146}{\doi{10.1007/JHEP05(2014)146}},
\href{http://www.arXiv.org/abs/1402.2657}{\texttt{arXiv:1402.2657}}.

\bibitem{Thaler:2010tr}
\hrefCMSnoop {}{J.~Thaler and K.~Van~Tilburg, ``Identifying boosted objects
  with {$N$}-subjettiness'',} \textit{ JHEP} \textbf{ 03} (2011) 015,
  \href{http://dx.doi.org/10.1007/JHEP03(2011)015}{\doi{10.1007/JHEP03(2011)015}},
\href{http://www.arXiv.org/abs/1011.2268}{\texttt{arXiv:1011.2268}}.

\bibitem{ddt}
J.~Dolen\hrefCMSnoop {}{ {et~al.}, ``Thinking outside the {ROC}s: Designing
  decorrelated taggers ({DDT}) for jet substructure'',} \textit{ JHEP} \textbf{
  05} (2016) 156,
  \href{http://dx.doi.org/10.1007/JHEP05(2016)156}{\doi{10.1007/JHEP05(2016)156}},
\href{http://www.arXiv.org/abs/1603.00027}{\texttt{arXiv:1603.00027}}.

\bibitem{BTV-16-002}
\hrefCMSnoop {}{{CMS Collaboration}, ``{Identification of heavy-flavour jets
  with the CMS detector in pp collisions at 13 TeV}'',} \textit{ JINST}
  \textbf{ 13} (2018) P05011,
  \href{http://dx.doi.org/10.1088/1748-0221/13/05/P05011}{\doi{10.1088/1748-0221/13/05/P05011}},
\href{http://www.arXiv.org/abs/1712.07158}{\texttt{arXiv:1712.07158}}.

\bibitem{Oreglia:1980cs}
\href {http://www.slac.stanford.edu/cgi-wrap/getdoc/slac-r-236.pdf}{M.~J.
  Oreglia, ``A study of the reactions $\psi^\prime \to \gamma \gamma \psi$''}.
\newblock PhD thesis, Stanford University, 1980.
\newblock
{SLAC} Report {SLAC-R-236}.

\bibitem{CMS:2021xjt}
\hrefCMSnoop {}{{CMS Collaboration}, ``Precision luminosity measurement in
  proton-proton collisions at $\sqrt{s} =$ 13 {TeV} in 2015 and 2016 at
  {CMS}'',} 2021.
  \href{http://www.arXiv.org/abs/2104.01927}{\texttt{arXiv:2104.01927}}.
  Accepted by \textit{Eur. Phys. J. C}.

\bibitem{CMS-PAS-LUM-17-004}
\href {https://cds.cern.ch/record/2621960/}{{{CMS}} Collaboration, ``{CMS}
  luminosity measurement for the 2017 data-taking period at $\sqrt{s}$ = 13
  {TeV}'',} CMS Physics Analysis Summary CMS-PAS-LUM-17-004, 2018.

\bibitem{CMS-PAS-LUM-18-002}
\href {https://cds.cern.ch/record/2676164/}{{{CMS}} Collaboration, ``{CMS}
  luminosity measurement for the 2018 data-taking period at $\sqrt{s}$ = 13
  {TeV}'',} CMS Physics Analysis Summary CMS-PAS-LUM-18-002, 2019.

\bibitem{Cacciari:2003fi}
M.~Cacciari\hrefCMSnoop {}{ {et~al.}, ``The $\ttbar$ cross-section at 1.8{\TeV}
  and 1.96{\TeV}: A study of the systematics due to parton densities and scale
  dependence'',} \textit{ JHEP} \textbf{ 04} (2004) 068,
  \href{http://dx.doi.org/10.1088/1126-6708/2004/04/068}{\doi{10.1088/1126-6708/2004/04/068}},
\href{http://www.arXiv.org/abs/hep-ph/0303085}{\texttt{arXiv:hep-ph/0303085}}.

\bibitem{Catani:2003zt}
\hrefCMSnoop {}{S.~Catani, D.~de~Florian, M.~Grazzini, and P.~Nason, ``{Soft
  gluon resummation for {Higgs} boson production at hadron colliders}'',}
  \textit{ JHEP} \textbf{ 07} (2003) 028,
  \href{http://dx.doi.org/10.1088/1126-6708/2003/07/028}{\doi{10.1088/1126-6708/2003/07/028}},
\href{http://www.arXiv.org/abs/hep-ph/0306211}{\texttt{arXiv:hep-ph/0306211}}.

\bibitem{Baker:1983tu}
\hrefCMSnoop {}{S.~Baker and R.~D. Cousins, ``Clarification of the use of chi
  square and likelihood functions in fits to histograms'',} \textit{ Nucl.
  Instrum. Meth.} \textbf{ 221} (1984) 437,
\href{http://dx.doi.org/10.1016/0167-5087(84)90016-4}{\doi{10.1016/0167-5087(84)90016-4}}.

\bibitem{Cowan:2010js}
\hrefCMSnoop {}{G.~Cowan, K.~Cranmer, E.~Gross, and O.~Vitells, ``Asymptotic
  formulae for likelihood-based tests of new physics'',} \textit{ Eur. Phys. J.
  C} \textbf{ 71} (2011) 1554,
  \href{http://dx.doi.org/10.1140/epjc/s10052-011-1554-0}{\doi{10.1140/epjc/s10052-011-1554-0}},
  \href{http://www.arXiv.org/abs/1007.1727}{\texttt{arXiv:1007.1727}}.
[Erratum: \DOI{10.1140/epjc/s10052-013-2501-z}].

\bibitem{Junk:1999kv}
\hrefCMSnoop {}{T.~Junk, ``Confidence level computation for combining searches
  with small statistics'',} \textit{ Nucl. Instrum. Meth. A} \textbf{ 434}
  (1999) 435,
  \href{http://dx.doi.org/10.1016/S0168-9002(99)00498-2}{\doi{10.1016/S0168-9002(99)00498-2}},
\href{http://www.arXiv.org/abs/hep-ex/9902006}{\texttt{arXiv:hep-ex/9902006}}.

\bibitem{Read:2002hq}
\hrefCMSnoop {}{A.~L. Read, ``Presentation of search results: The {\CLs}
  technique'',} \textit{ J. Phys. G} \textbf{ 28} (2002) 2693,
\href{http://dx.doi.org/10.1088/0954-3899/28/10/313}{\doi{10.1088/0954-3899/28/10/313}}.

\bibitem{hepdata}
\hrefCMSnoop {}{``{HEPD}ata record for this analysis'',} 2021.
\newblock
  \href{http://dx.doi.org/10.17182/hepdata.102465}{\doi{10.17182/hepdata.102465}}.

\end{thebibliography}\endgroup
